\documentclass{article}
\usepackage{graphicx}
\usepackage[margin=1in]{geometry}
\usepackage{amsmath}
\usepackage{amssymb}
\usepackage{amsthm}
\usepackage{accents}
\usepackage[maxfloats=128]{morefloats}
\usepackage[breaklinks,hidelinks]{hyperref}

\newtheorem{theorem}{Theorem}[]

\newtheorem{remark1}[theorem]{Remark}
\newenvironment{remark}{\begin{remark1} \rm}{\end{remark1}}

\title{Compressed sensing with a jackknife and a bootstrap}
\author{Mark Tygert, Rachel Ward, and Jure Zbontar \medskip\\
Facebook Artificial Intelligence Research}

\begin{document}

\maketitle

\begin{abstract}
Compressed sensing proposes to reconstruct more degrees of freedom in a signal
than the number of values actually measured.
Compressed sensing therefore risks introducing errors ---
inserting spurious artifacts or masking the abnormalities that
medical imaging seeks to discover.
The present case study of estimating errors using the standard statistical
tools of a jackknife and a bootstrap yields error ``bars'' in the form
of full images that are remarkably representative of the actual errors
(at least when evaluated and validated on data sets for which the ground truth
and hence the actual error is available).
These images show the structure of possible errors --- without recourse
to measuring the entire ground truth directly --- and build confidence
in regions of the images where the estimated errors are small.
\end{abstract}

\section{Introduction}

Compressed sensing is the concept that many interesting signals are recoverable
from undersampled measurements of the representations of those signals
in a special basis. A widely touted potential application
is to the acceleration of magnetic resonance imaging (MRI).
In MRI, the special basis for representations of signals is the Fourier basis,
and the goal of compressed sensing is to recover high-resolution images
from relatively sparse measurements of the Fourier components of those images.
Here, ``sparse'' means substantially fewer measurements of values
in the Fourier domain than the numbers of pixels in the reconstructed images.
Of course, recovering more degrees of freedom than the number of measured
values is an ill-posed problem, yet has been rigorously proven to be solvable
when the gradients of the images being recovered are known to be small
except at a few pixels, for instance, when edges dominate the images.
This recovery is still non-trivial,
as the small number of pixels where the gradients are non-trivial
very well may vary from image to image, while the same reconstruction procedure
works irrespective of where the gradients concentrate (so long as they
concentrate on sparse subsets of all pixels in the reconstructed domain).
The requirement that gradients be concentrated on sparse subsets
is sufficient but may not be necessary, and much recent research aims
to generalize beyond this requirement by applying machine learning
to representative data sets. Indeed, the literature on compressed sensing
is vast and growing rapidly; see, for example,
the recent review of~\cite{tropp} for explication of all this and more.

Needless to say, compressed sensing risks introducing errors
into the resulting reconstructions, especially if the assumption of sparsity
is unfounded for the real data at hand. The works
of~\cite{malioutov-sanghavi-willsky} and~\cite{ward}
quantify these errors via a single scalar estimate of confidence
in the reconstruction, namely, an estimate of the mean-square error.
The present paper extends these methods, producing estimates
of the entire image displaying the discrepancy between the reconstruction
in compressed sensing and the actual ground truth.
Of course, compressed sensing takes too few measurements to ascertain
the actual ground truth, so only an estimate of the discrepancy
--- an error ``bar'' in the form of an image --- is possible.
However, the examples of the present paper show that ``jackknife''
and ``bootstrap'' estimates of the errors are reasonably representative
of the reality, at least for the cases in MRI tested here,
in which the ground truth is available for comparison and evaluation.
Those unfamiliar with the jackknife and the bootstrap may wish
to consult~\cite{efron-tibshirani};
that said, the presentation below is completely self-contained,
not presuming any prior knowledge of either the jackknife or the bootstrap.
The jackknife and bootstrap images highlight when, where, and what errors
may have arisen in each reconstruction from compressed sensing for MRI,
tailored to the specific object being imaged.

The jackknife is similar to standard {\it a posteriori} tests for convergence
of numerical methods; such numerical tests for convergence
often serve as proxies for estimates of accuracy.
The bootstrap leverages more extensive computation, simulating measurements
that could have been taken but were not in fact
(recall that compressed sensing involves taking fewer measurements
than the number of degrees of freedom being reconstructed).
The bootstrap simulates plausible alternative reconstructions
from hypothetical measurements that are consistent
with the reconstruction from the measurements actually made.
The alternative reconstructions fluctuate around the reconstruction
from the measurements actually made;
the fluctuation is an estimate of the error,
when averaged over various sampling patterns
for the measurements being considered.

The structure of the remainder of the present paper is as follows:
First, Section~\ref{methods} introduces the jackknife and the bootstrap
for compressed sensing.
Then, Section~\ref{numex} illustrates the performance of the methods
on data sets from MRI, with copious additional examples provided
in the appendix.

\section{Methods}
\label{methods}

We denote by $X$ a data set $(x_i)_{i \in I}$,
where each $x_i$ is a scalar or a vector and $I$ is a set of indices.
We consider a vector-valued (or image-valued) function $f = f(X, S)$
of both $X$ and a subset $S$ of the index set $I$
such that the value of $f$ depends only on $(x_i)_{i \in S}$.
Compressed sensing approximates the full $f(X, I)$ with $f(X, S)$,
where $S$ is a subset of $I$ collecting together independent uniformly random
draws from $I$, perhaps plus some fixed subset $T$ of $I$.
(Obviously, this construction makes $T$ a subset of $S$. However,
$T$ need not be disjoint from the set of independent uniformly random draws.)

In compressed sensing for MRI, measured observations in the Fourier domain
of the object being imaged are $(x_i)_{i \in I}$, and $f(X, I)$ uses those
measurements to reconstruct the object in the original domain
(hence involving an inverse Fourier transform to map from $X$ to $f(X, I)$).
The reconstruction $f(X, S)$ from a subset $S$ of $I$ commonly involves
minimizing a total-variation objective function or deep learning of some sort,
as discussed by~\cite{tao-yang}, \cite{yang-zhang}, and their references.

With such undersampled measurements, the reconstruction is oblivious to much
of the Fourier domain, sampling fewer measurements
than at the usual Nyquist rate.
We will tacitly be assuming that the procedure for reconstruction works
not only for the set $S$ specifying the measurements actually used,
but also for other sets of random observations, that is,
for other random realizations of $S$.
For machine-learned reconstructions, the model for reconstruction must train
on measurements taken from many different possible samplings, not just one,
as otherwise the model will be blind to parts of the Fourier domain.
If we can simulate on a computer what could have happened with measurements
that we do not take in reality, then we can construct error ``bars''
highlighting when, where, and what might have gone wrong in a reconstruction
from actual measurements taken with only one realized sampling set $S$.
The computational simulation allows us to gauge what could have happened
with unseen measurements.
While seeing the unseen (at least in part) may seem counterintuitive,
in fact the field of statistics is all about what might have occurred
given observations of what actually did happen.
The bootstrap defined below follows this prescription literally.
The jackknife is a somewhat simpler formulation.

The goal of both the jackknife and the bootstrap is to provide an estimated
bound on $f(X, S) - f(X, I)$, without having access to the full reconstruction
$f(X, I)$. (The full reconstruction depends on all measurements
$(x_i)_{i \in I}$ --- the whole $I$ --- so is unavailable when performing
compressed sensing.)

First, we define the jackknife error ``bar'' for $f$ on $S$ to be
\begin{equation}
\label{jackknife}
d = 2 \sum_{i \in S \backslash T}
    \Bigl( f(X, S \backslash \{i\}) - f(X, S) \Bigr),
\end{equation}
where the sum ranges over every index $i \in S$ such that $i \notin T$,
and $S \backslash \{i\}$ is just $S$ after removing $i$.
The jackknife $d$ defined in~(\ref{jackknife}) characterizes
what would happen to the output of $f$ if the input $S$ were slightly smaller;
if $f(X, S)$ is close to converging on $f(X, I)$,
then $f(X, S \backslash \{i\})$ in~(\ref{jackknife}) should also be close
to $f(X, I)$, so $f(X, S \backslash \{i\})$ should be close to $f(X, S)$,
aside from errors.
We refer to $f(X, S \backslash \{i\}) - f(X, S)$ in the right-hand side
of~(\ref{jackknife}) as a ``leave-one-out'' difference,
as in ``leave-one-out'' cross-validation.
We could empirically (or semi-empirically) determine a calibration constant $c$
such that $cd$ becomes of the same size as the actual discrepancy
$f(X, S) - f(X, I)$ on average for a training set of exemplars
(the training set could consist of many different $X$ together
with the corresponding $f(X, S)$ and $f(X,I)$).
However, we find that $c = 1$ works exceptionally well;
$d$ is typically of about the same size as $f(X, S) - f(X, I)$.

Next, we define the bootstrap, assuming that $S$ is the union of the set $T$
and a set of $\ell$ independent uniformly random samples from $I$
(where $\ell$ is a parameter, and the number of distinct members
of the latter set may be less than $\ell$ due to repetition
in the $\ell$ samples):
First, having already computed $f(X, S)$,
we solve for $\tilde{X} = (\tilde{x}_i)_{i \in I}$ such that
\begin{equation}
\label{bootstrapped}
f(\tilde{X}, I) = f(X, S).
\end{equation}
Then, we form the set $R$ of $\ell$ independent uniformly random draws from $I$
(not all $\ell$ of which need be distinct),
plus the fixed subset $T$ of $I$ (in so-called parallel MRI,
as described by~\cite{brown-cheng-haacke-thompson-venkatesan},
$T$ would naturally contain all the lowest frequencies).
We select a positive integer $k$ and repeat this resampling independently
$k$ times, thus obtaining sets $R_1$, $R_2$, \dots, $R_k$.
We define the bootstrap error ``bar'' to be
\begin{equation}
\label{difference}
e = \frac{3}{k} \sum_{j = 1}^k
    \Bigl( f(\tilde{X}, R_j) - f(\tilde{X}, I) \Bigr).
\end{equation}
We could say that $f(\tilde{X}, R_j)$ arises from $f(\tilde{X}, I)$
in the same way as $f(X, S)$ arises from $f(X, I)$,
having constructed $\tilde{X}$ assuming that $f(X, S)$ is ``correct''
in the sense of~(\ref{bootstrapped}); so the summand in~(\ref{difference})
is a proxy for the actual error $f(X, S) - f(X, I)$
(and the averaging over independent realizations reduces noise).

\begin{remark}
Both $\bigl( f(X, S \backslash \{i\}) - f(X, S) \bigr)_{i \in S \backslash T}$
from~(\ref{jackknife})
and $\bigl( f(\tilde{X}, R_j) - f(\tilde{X}, I) \bigr)_{j = 1}^k$
from~(\ref{difference})
span whole spaces of errors that potentially could have happened
given the actually observed measurements.
While the sum and average in~(\ref{jackknife}) and~(\ref{difference}),
respectively, of these sets of differences characterize the leading modes
of these spaces, principal component analysis can characterize all modes.
However, looking at even just the leading modes seems somewhat overwhelming
already; having to investigate more modes could really try the patience
of a physician interpreting MRI scans, for instance.
The present paper focuses on the leading modes.
\end{remark}

\section{Numerical examples}
\label{numex}

The examples of this section illustrate
the most commonly discussed compressed sensing for MRI,
in which we reconstruct an image from measured observations of some
of its values in the Fourier domain --- ``some'' meaning significantly less
than usually required by the Nyquist-Shannon-Whittaker sampling theory.
To reconstruct an image from measurements taken in the Fourier domain
(with independent and identically distributed centered complex Gaussian noise
of standard deviation $0.02$ added to mimic machine imprecision),
we minimize the sum of deviations from the measurements
plus a total-variation regularizer
via Algorithm~1 at the end of Section~2.2 of~\cite{tao-yang}
(which is based on the work of~\cite{yang-zhang}), with 100 iterations,
using the typical parameter settings $\mu = 10^{12}$ and $\beta = 10$
($\mu$ governs the fidelity to the measurements taken in the Fourier domain,
and $\beta$ is the strength of the coupling in the operator splitting
for the alternating-direction method of multipliers).
As discussed by~\cite{tropp}, this is the canonical setting
for compressed sensing.
All computations take place in IEEE standard double-precision arithmetic.
We use $k =$ 1,000 resamplings for the bootstrap in~(\ref{difference}).

We consider two kinds of sampling patterns,
radially retained and horizontally retained.
All sampling takes place on an $m \times n$ Cartesian grid,
allowing direct use of the fast Fourier transform
for acceleration of the reconstruction (as described by~\cite{tao-yang}).
Future implementations could consider sampling off the grid, too.

\begin{figure}
\parbox{\textwidth}{\includegraphics[width=\textwidth]{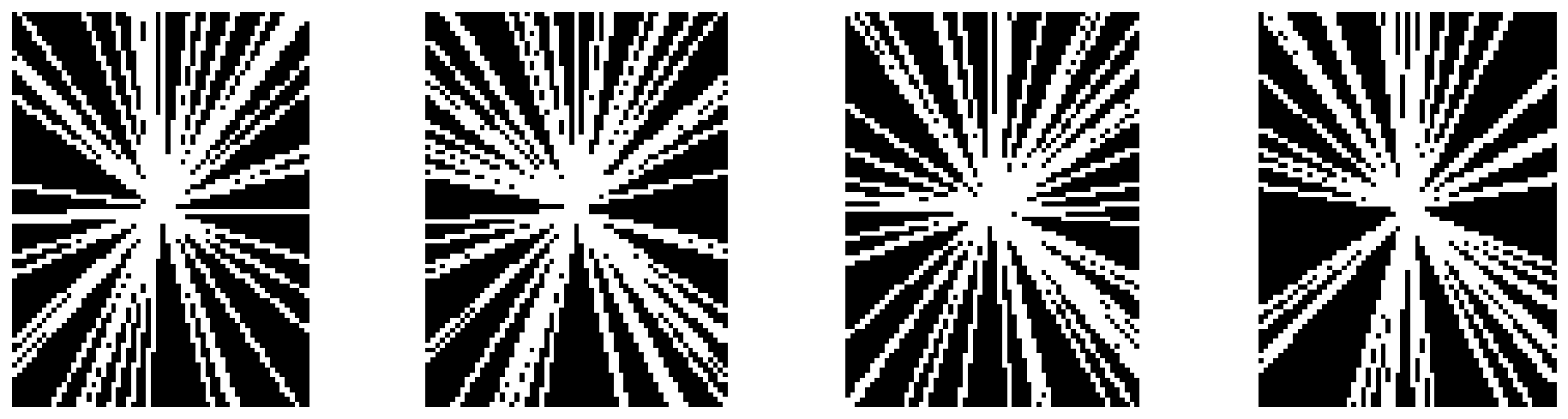}}
\caption{radially retained sampling ---
         sampling on a Cartesian grid along rays emanating from the origin}
\label{radialines}
\end{figure}

With radially retained sampling,
each $x_i$ in our data set $X = (x_i)_{i \in I}$
consists of all pixels on an $m \times n$ Cartesian grid in the Fourier domain
that intersect a ray emanating from the origin (each angle corresponds
to $x_i$ for a different index $i$). Figure~\ref{radialines} displays
four examples of uniformly random subsets of $X$, sampling the angles
of the rays uniformly at random.
For radially retained sampling, we refrain from supplementing
the subsampled set $S$ with any fixed subset; that is, the set $T$ is empty.
To construct $S$, we generate $\frac{m + n}{5}$ angles uniformly at random
(rounding $\frac{m + n}{5}$ to the nearest integer), which makes the errors
easy to see in the coming figures, yet not too extreme.

With horizontally retained sampling,
each $x_i$ in our data set $X = (x_i)_{i \in I}$
consists of a horizontal line $n$ pixels wide on an $m \times n$ Cartesian grid
in the Fourier domain, with $I$ consisting of the $m$ integers
from $-\frac{m}{2}$ to $\frac{m-2}{2}$.
The subsampled set $S$ always includes
all horizontal lines ranging from the $-\sqrt{2m}\,$th lowest frequency
to the $\sqrt{2m}\,$th lowest frequency
(rounding $\sqrt{2m}$ to the nearest integer);
that is, the set $T$ consists of these low-frequency indices.
To construct the remainder of $S$, we generate $\frac{m}{4}$ integers
from $-\frac{m}{2}$ to $\frac{m-2}{2}$ uniformly at random
(rounding $\frac{m}{4}$ to the nearest integer), which makes the errors
easy to see in the coming figures, yet not too extreme.
Recall that $S$ is a set: each member $i$ of $S$
occurs only once irrespective of how many times the sampling procedure
just described chooses to include the index $i$.

\begin{figure}
\begin{centering}

\parbox{.65\textwidth}{\includegraphics[width=.65\textwidth]{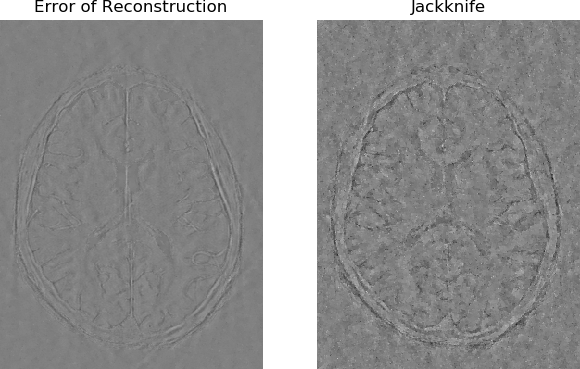}}

\vspace{.25in}

\parbox{.65\textwidth}{\includegraphics[width=.65\textwidth]{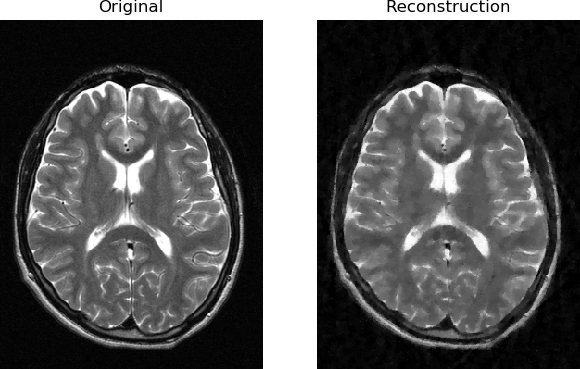}}

\vspace{.25in}

\parbox{.65\textwidth}{\includegraphics[width=.65\textwidth]{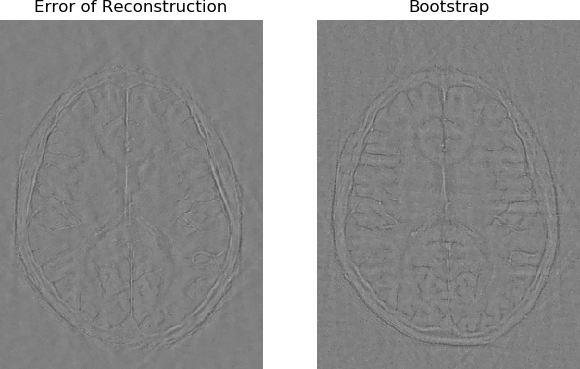}}

\end{centering}
\caption{radially retained sampling}
\label{bigradial}
\end{figure}

\begin{figure}
\begin{centering}

\parbox{.65\textwidth}{\includegraphics[width=.65\textwidth]{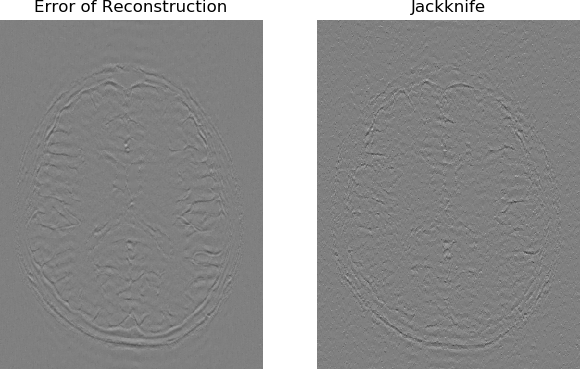}}

\vspace{.25in}

\parbox{.65\textwidth}{\includegraphics[width=.65\textwidth]{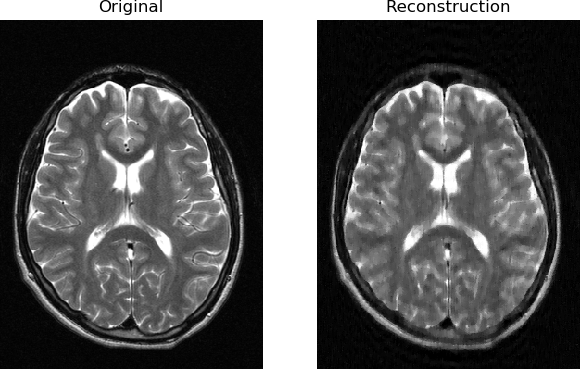}}

\vspace{.25in}

\parbox{.65\textwidth}{\includegraphics[width=.65\textwidth]{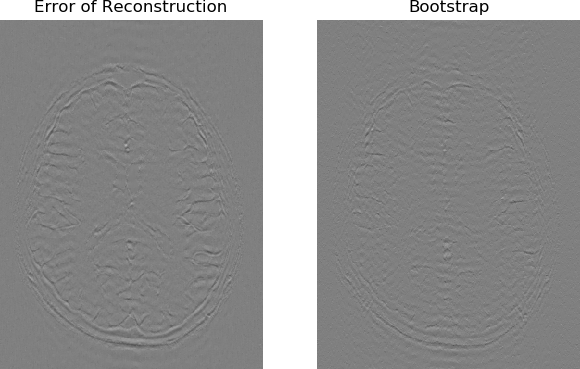}}

\end{centering}
\caption{horizontally retained sampling}
\label{bighorizontal}
\end{figure}

Figures~\ref{bigradial} and~\ref{bighorizontal} display results for retaining
radial and horizontal lines, respectively, using the same original image.
Further examples are available in the appendix.
The figures whose captions specify ``$2\times$'' use
$\frac{2(m + n)}{5}$ random angles for radially retained sampling
and $\frac{m}{2}$ random integers for horizontally retained sampling
instead of the $\frac{m + n}{5}$ random angles
and $\frac{m}{4}$ random integers used in all other figures.
All figures concern MRI scans of patients' heads from the data
of~\cite{mri2}, \cite{mri1}, \cite{mri3}, and~\cite{mri4}.
The resolution in pixels of the original image for Figures~\ref{bigradial}
and~\ref{bighorizontal} is $378 \times 284$.
The resolutions in pixels of the original images for the appendix range
from $376 \times 286$ to $456 \times 371$.

In the figures,
``Original'' displays the original image,
``Reconstruction'' displays the reconstruction $f(X, S)$,
``Error of Reconstruction'' displays the difference between the original image
and the reconstruction,
``Jackknife'' displays the jackknife $d$ from~(\ref{jackknife}),
and ``Bootstrap'' displays the bootstrap $e$ from~(\ref{difference}).
The values of the original pixels are normalized to range from $0$ to $1$.
In the images ``Original'' and ``Reconstruction,''
pure black corresponds to 0 while pure white corresponds to 1.
In the images ``Error of Reconstruction,'' ``Jackknife,'' and ``Bootstrap,''
pure white and pure black correspond to the extreme values $\pm 1$,
whereas 50\% gray (halfway to black or to white) corresponds to 0.
Thus, in the images displaying errors and potential errors,
middling halftone grays correspond to little or no error, while
extreme pure white and pure black correspond to more substantial errors.

The jackknife images are generally noisier than the bootstrap images.
The bootstrap directly explores parts of the Fourier domain
outside the observed measurements, whereas the jackknife is more
like a convergence test or a differential approximation to the bootstrap ---
see, for example, the review of~\cite{efron-tibshirani}.
Both the jackknife and the bootstrap occasionally display artifacts
where in fact the reconstruction was accurate.
Moreover, they miss some anomalies;
if the reconstruction completely washes out a feature of the original image,
then neither the jackknife nor the bootstrap can detect the washed-out feature.
That said, in most cases they show the actual errors nicely.
The estimates bear an uncanny resemblance to the actual errors.
Using both the jackknife and the bootstrap may be somewhat conservative,
but if the jackknife misses an error, then the bootstrap usually catches it,
and vice versa.

\section*{Acknowledgements}

We would like to thank Florian Knoll, Jerry Ma, Jitendra Malik, Matt Muckley,
Mike Rabbat, Dan Sodickson, and Larry Zitnick.

\begin{figure}
{\Large \bf Appendix: Supplementary figures}

\bigskip

This appendix supplements the examples of Section~\ref{numex}
with analogous figures for twenty cross-sectional slices
through the head of another patient.
\end{figure}

\newlength{\vertsep}
\setlength{\vertsep}{.25in}
\newlength{\imsize}
\setlength{\imsize}{.65\textwidth}
\newlength{\vertseps}
\setlength{\vertseps}{.3in}
\newlength{\imsizes}
\setlength{\imsizes}{.46\textwidth}

\begin{figure}
\begin{centering}

\parbox{\imsize}{\includegraphics[width=\imsize]{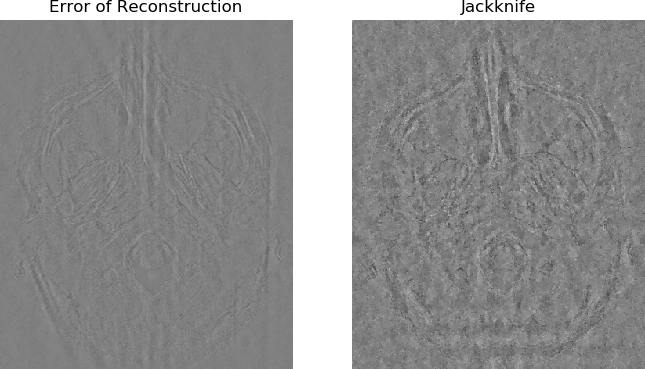}}

\vspace{\vertsep}

\parbox{\imsize}{\includegraphics[width=\imsize]{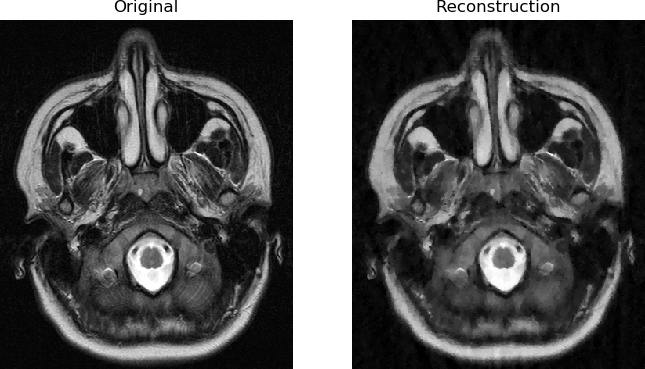}}

\vspace{\vertsep}

\parbox{\imsize}{\includegraphics[width=\imsize]{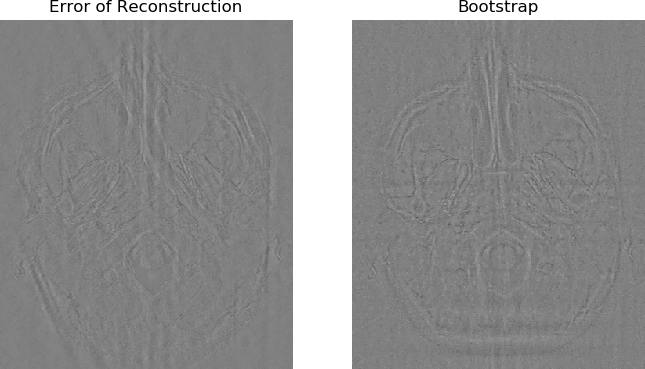}}

\end{centering}
\caption{radially retained sampling --- slice 1}
\end{figure}

\begin{figure}
\begin{centering}

\parbox{\imsize}{\includegraphics[width=\imsize]{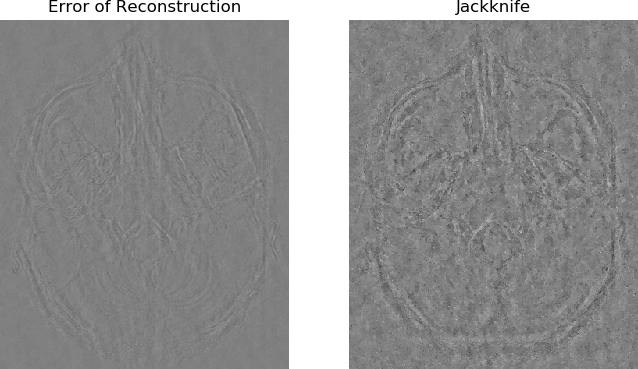}}

\vspace{\vertsep}

\parbox{\imsize}{\includegraphics[width=\imsize]{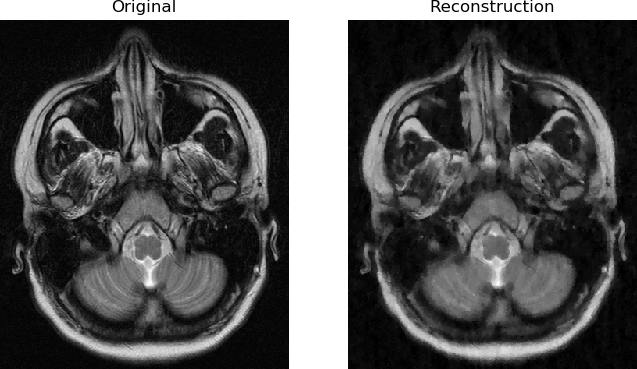}}

\vspace{\vertsep}

\parbox{\imsize}{\includegraphics[width=\imsize]{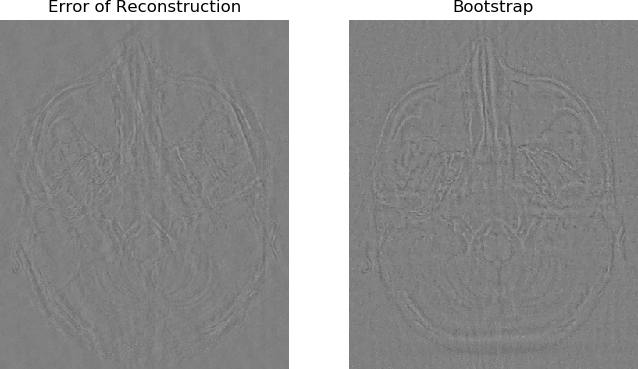}}

\end{centering}
\caption{radially retained sampling --- slice 2}
\end{figure}

\begin{figure}
\begin{centering}

\parbox{\imsize}{\includegraphics[width=\imsize]{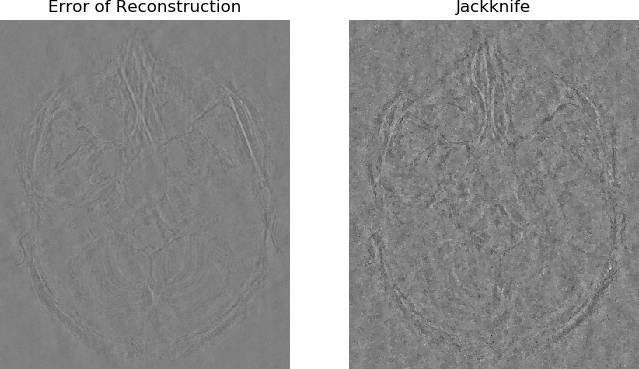}}

\vspace{\vertsep}

\parbox{\imsize}{\includegraphics[width=\imsize]{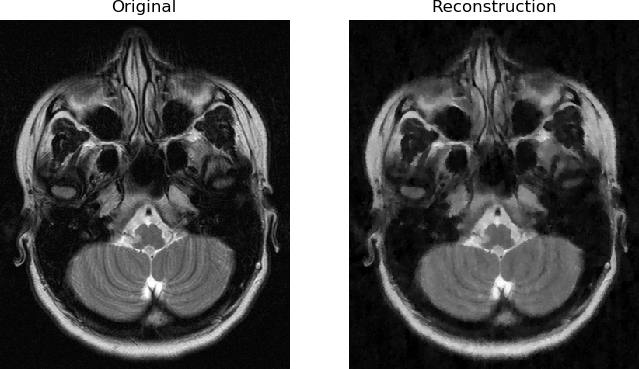}}

\vspace{\vertsep}

\parbox{\imsize}{\includegraphics[width=\imsize]{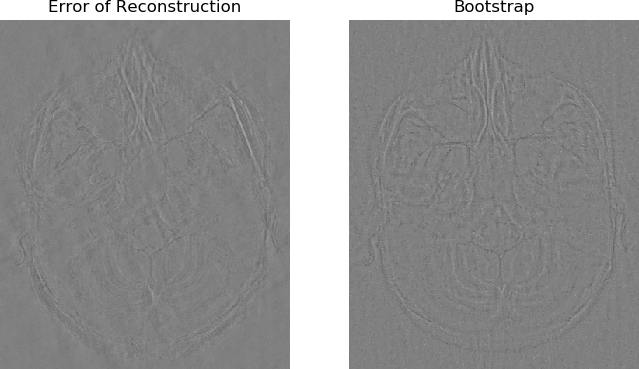}}

\end{centering}
\caption{radially retained sampling --- slice 3}
\end{figure}

\begin{figure}
\begin{centering}

\parbox{\imsize}{\includegraphics[width=\imsize]{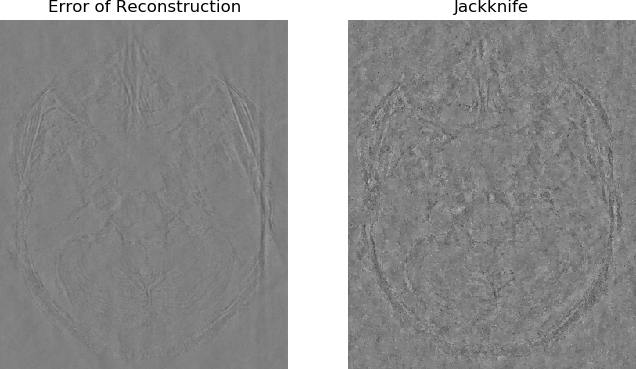}}

\vspace{\vertsep}

\parbox{\imsize}{\includegraphics[width=\imsize]{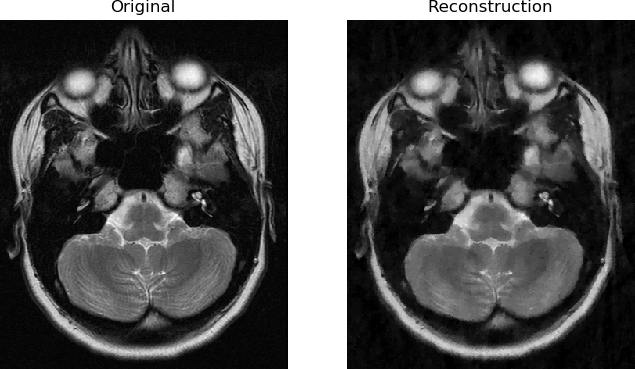}}

\vspace{\vertsep}

\parbox{\imsize}{\includegraphics[width=\imsize]{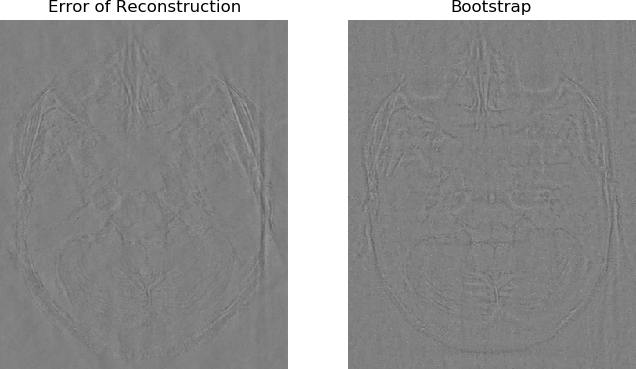}}

\end{centering}
\caption{radially retained sampling --- slice 4}
\end{figure}

\begin{figure}
\begin{centering}

\parbox{\imsize}{\includegraphics[width=\imsize]{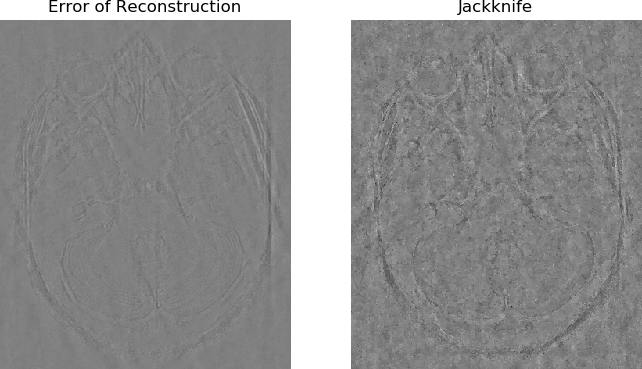}}

\vspace{\vertsep}

\parbox{\imsize}{\includegraphics[width=\imsize]{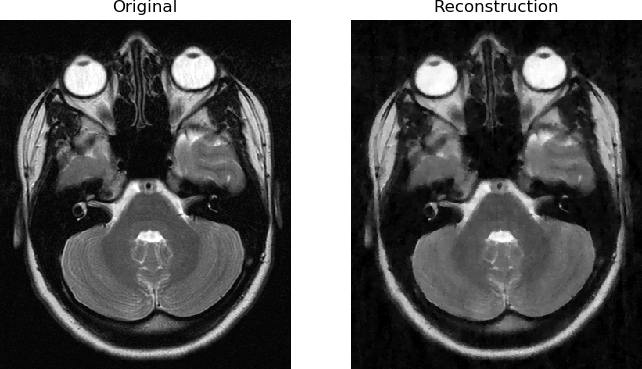}}

\vspace{\vertsep}

\parbox{\imsize}{\includegraphics[width=\imsize]{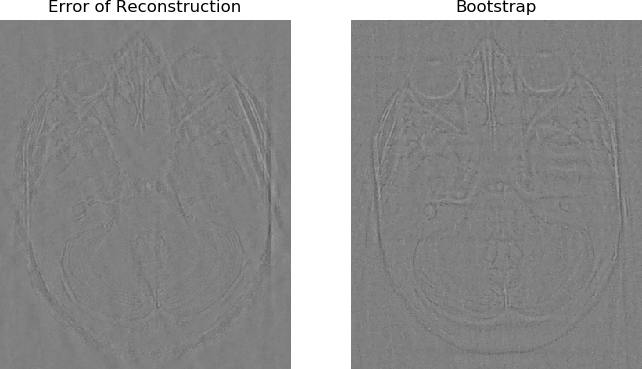}}

\end{centering}
\caption{radially retained sampling --- slice 5}
\end{figure}

\begin{figure}
\begin{centering}

\parbox{\imsize}{\includegraphics[width=\imsize]{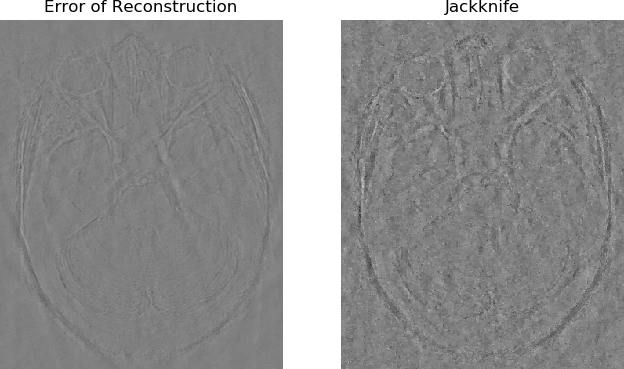}}

\vspace{\vertsep}

\parbox{\imsize}{\includegraphics[width=\imsize]{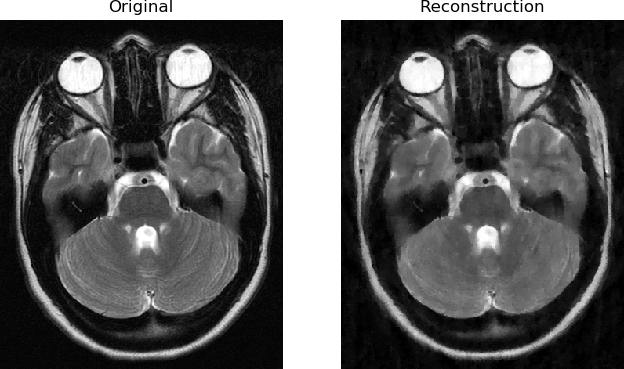}}

\vspace{\vertsep}

\parbox{\imsize}{\includegraphics[width=\imsize]{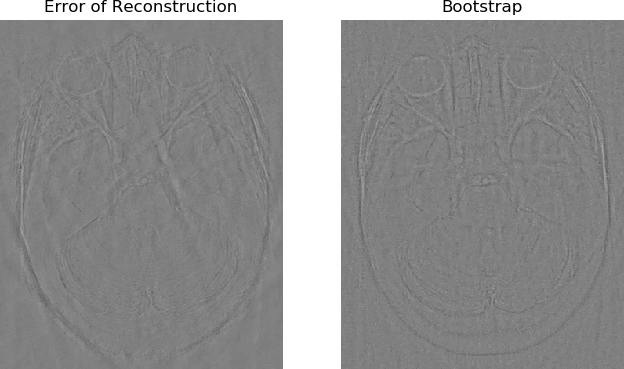}}

\end{centering}
\caption{radially retained sampling --- slice 6}
\end{figure}

\begin{figure}
\begin{centering}

\parbox{\imsize}{\includegraphics[width=\imsize]{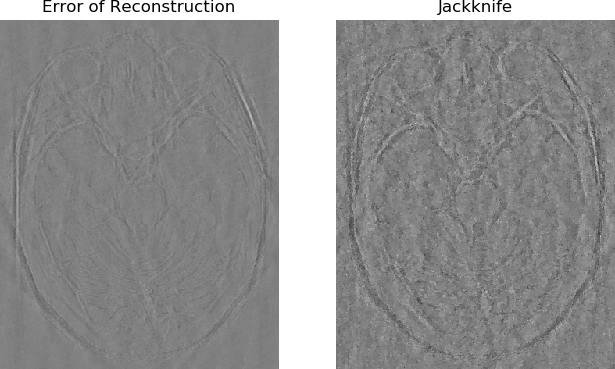}}

\vspace{\vertsep}

\parbox{\imsize}{\includegraphics[width=\imsize]{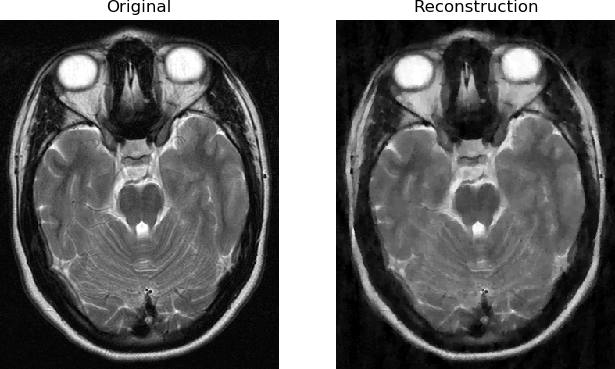}}

\vspace{\vertsep}

\parbox{\imsize}{\includegraphics[width=\imsize]{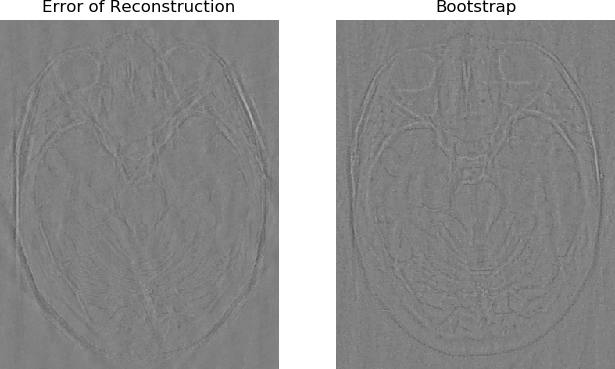}}

\end{centering}
\caption{radially retained sampling --- slice 7}
\end{figure}

\begin{figure}
\begin{centering}

\parbox{\imsize}{\includegraphics[width=\imsize]{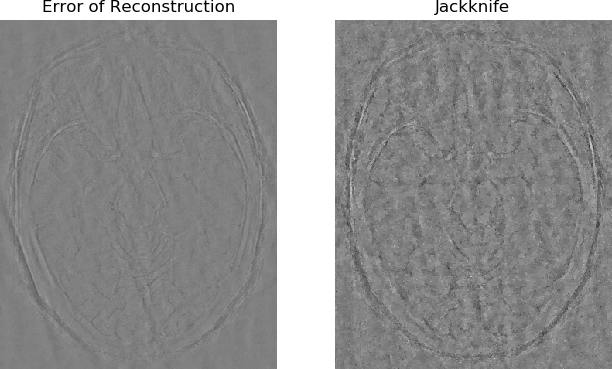}}

\vspace{\vertsep}

\parbox{\imsize}{\includegraphics[width=\imsize]{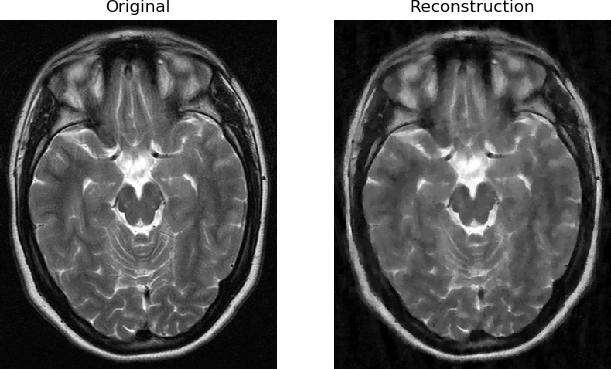}}

\vspace{\vertsep}

\parbox{\imsize}{\includegraphics[width=\imsize]{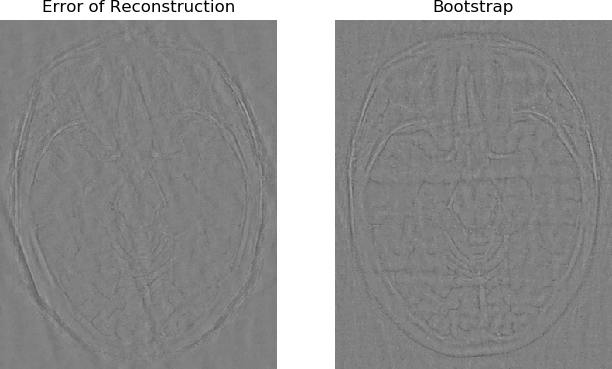}}

\end{centering}
\caption{radially retained sampling --- slice 8}
\end{figure}

\begin{figure}
\begin{centering}

\parbox{\imsize}{\includegraphics[width=\imsize]{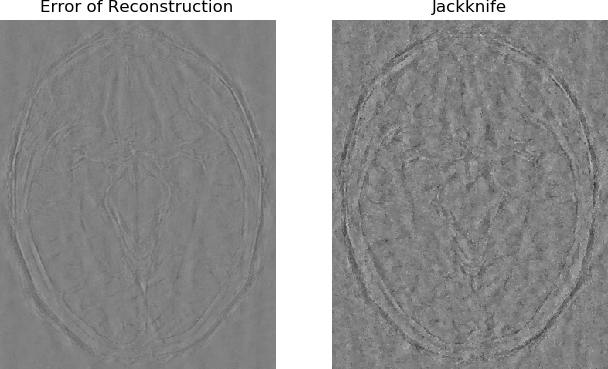}}

\vspace{\vertsep}

\parbox{\imsize}{\includegraphics[width=\imsize]{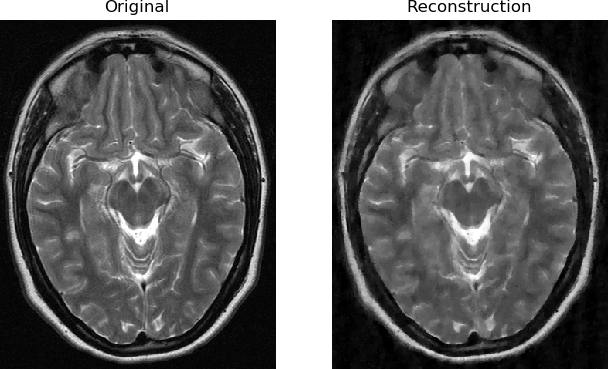}}

\vspace{\vertsep}

\parbox{\imsize}{\includegraphics[width=\imsize]{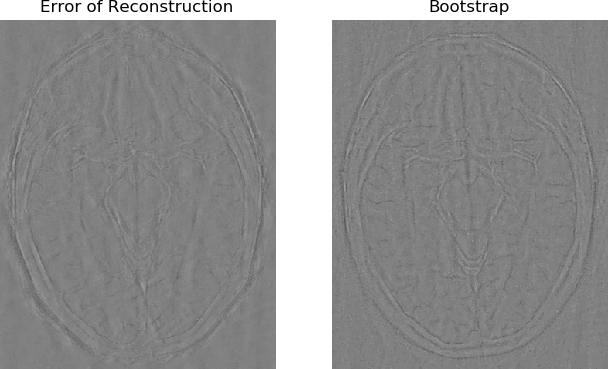}}

\end{centering}
\caption{radially retained sampling --- slice 9}
\end{figure}

\begin{figure}
\begin{centering}

\parbox{\imsize}{\includegraphics[width=\imsize]{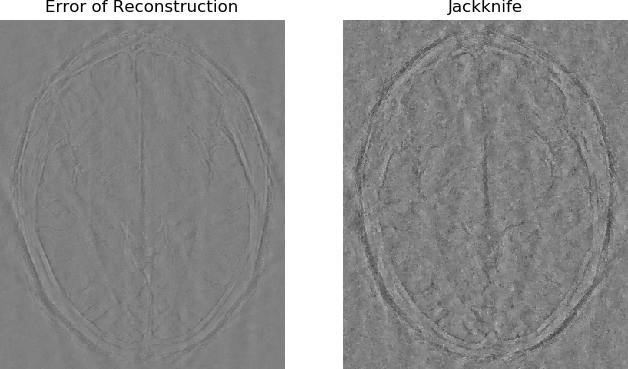}}

\vspace{\vertsep}

\parbox{\imsize}{\includegraphics[width=\imsize]{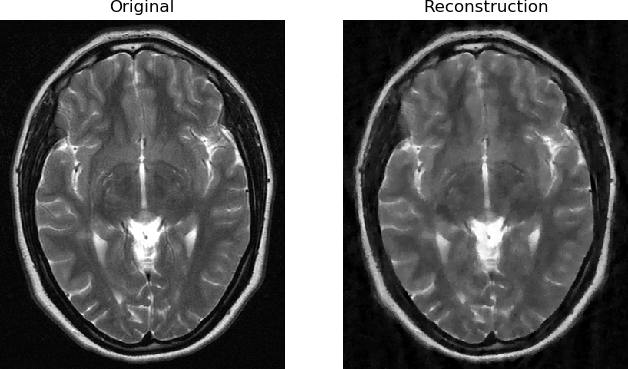}}

\vspace{\vertsep}

\parbox{\imsize}{\includegraphics[width=\imsize]{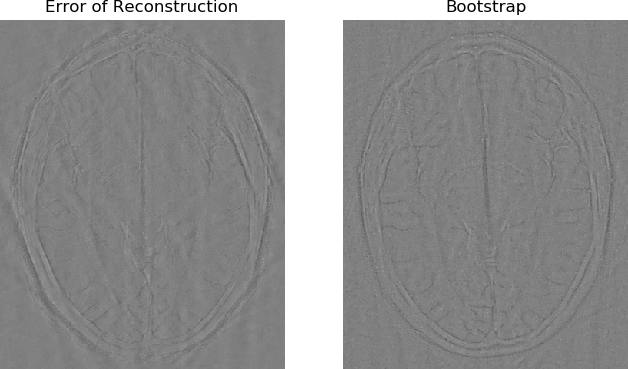}}

\end{centering}
\caption{radially retained sampling --- slice 10}
\end{figure}

\begin{figure}
\begin{centering}

\parbox{\imsize}{\includegraphics[width=\imsize]{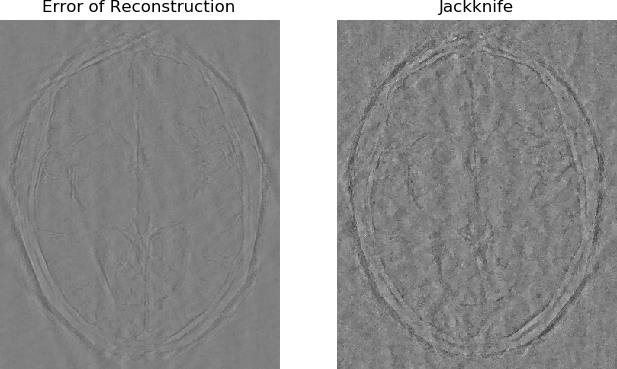}}

\vspace{\vertsep}

\parbox{\imsize}{\includegraphics[width=\imsize]{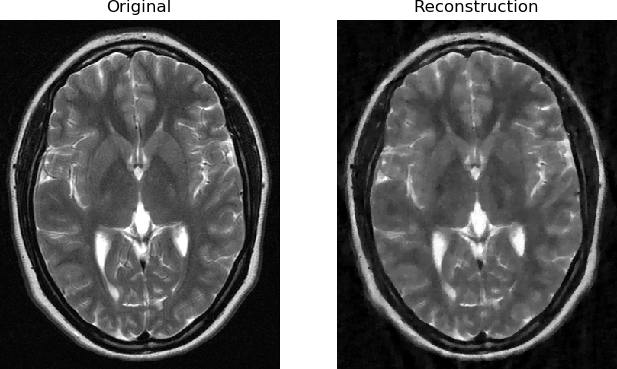}}

\vspace{\vertsep}

\parbox{\imsize}{\includegraphics[width=\imsize]{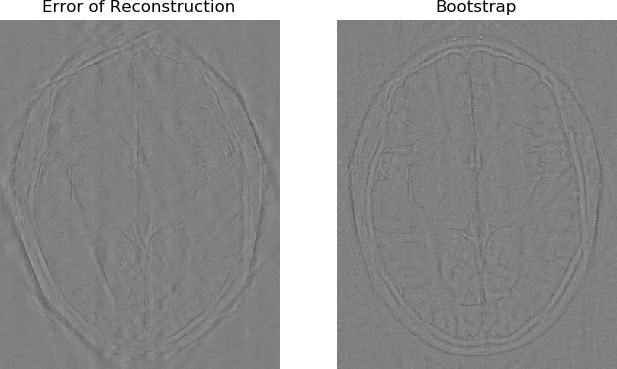}}

\end{centering}
\caption{radially retained sampling --- slice 11}
\end{figure}

\begin{figure}
\begin{centering}

\parbox{\imsize}{\includegraphics[width=\imsize]{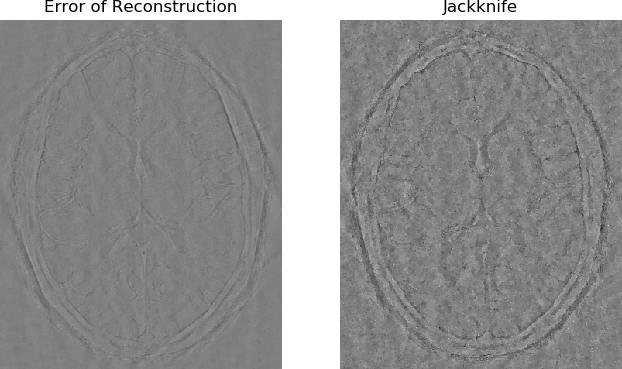}}

\vspace{\vertsep}

\parbox{\imsize}{\includegraphics[width=\imsize]{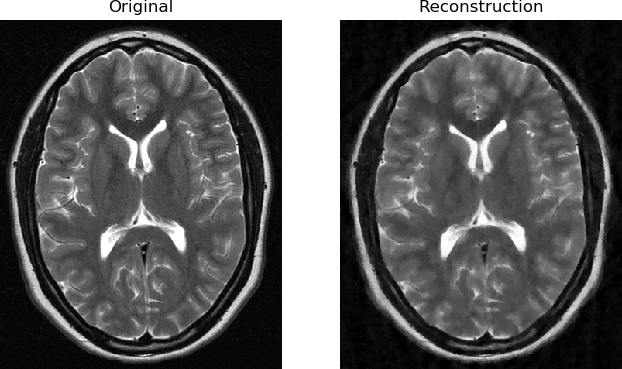}}

\vspace{\vertsep}

\parbox{\imsize}{\includegraphics[width=\imsize]{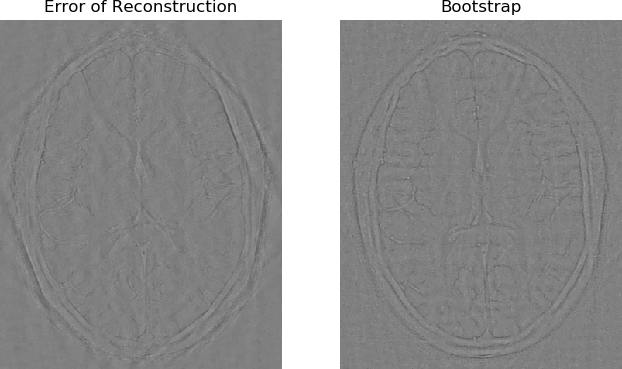}}

\end{centering}
\caption{radially retained sampling --- slice 12}
\end{figure}

\begin{figure}
\begin{centering}

\parbox{\imsize}{\includegraphics[width=\imsize]{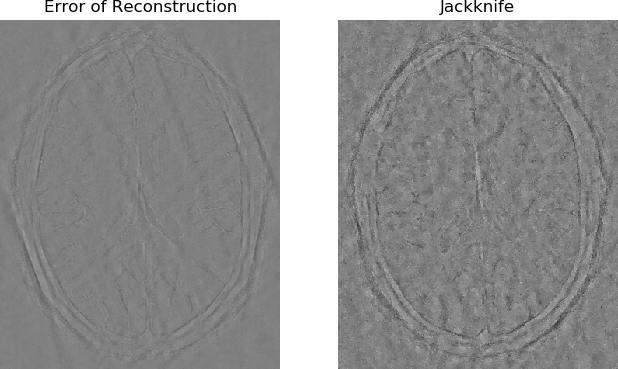}}

\vspace{\vertsep}

\parbox{\imsize}{\includegraphics[width=\imsize]{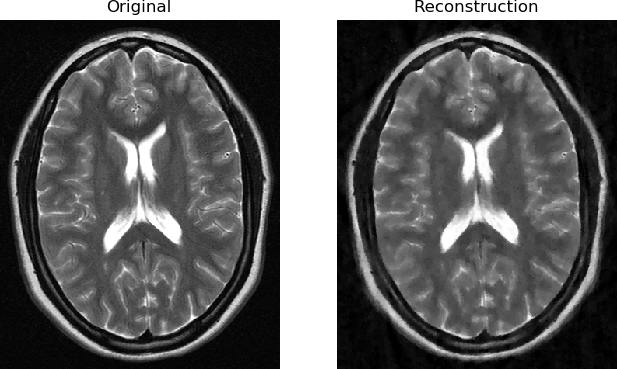}}

\vspace{\vertsep}

\parbox{\imsize}{\includegraphics[width=\imsize]{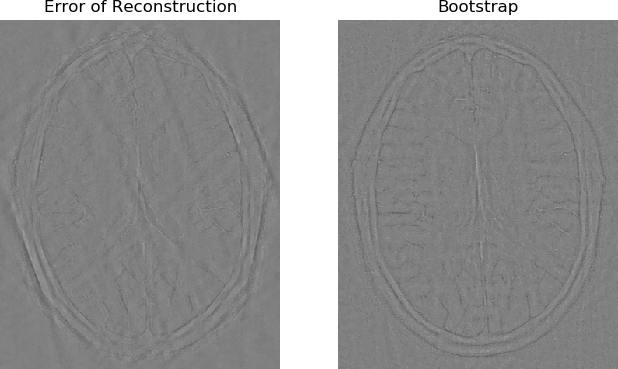}}

\end{centering}
\caption{radially retained sampling --- slice 13}
\end{figure}

\begin{figure}
\begin{centering}

\parbox{\imsize}{\includegraphics[width=\imsize]{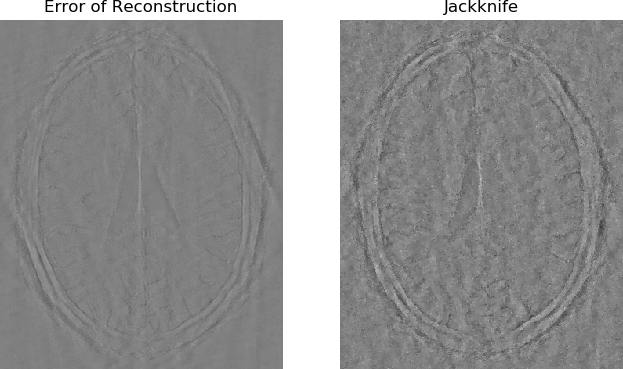}}

\vspace{\vertsep}

\parbox{\imsize}{\includegraphics[width=\imsize]{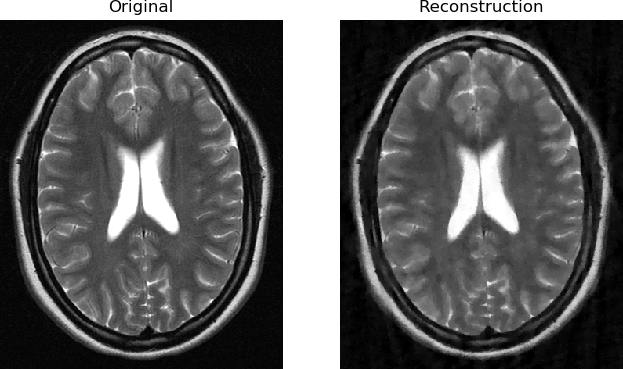}}

\vspace{\vertsep}

\parbox{\imsize}{\includegraphics[width=\imsize]{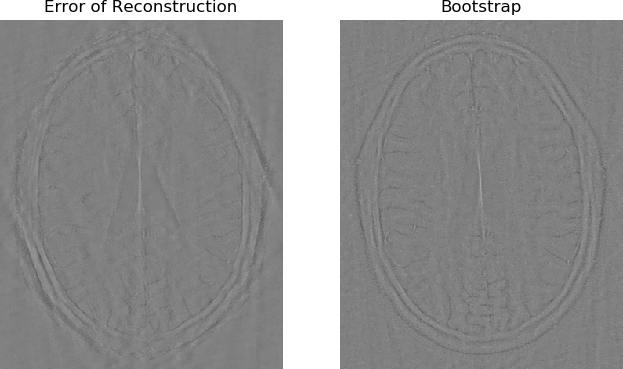}}

\end{centering}
\caption{radially retained sampling --- slice 14}
\end{figure}

\begin{figure}
\begin{centering}

\parbox{\imsize}{\includegraphics[width=\imsize]{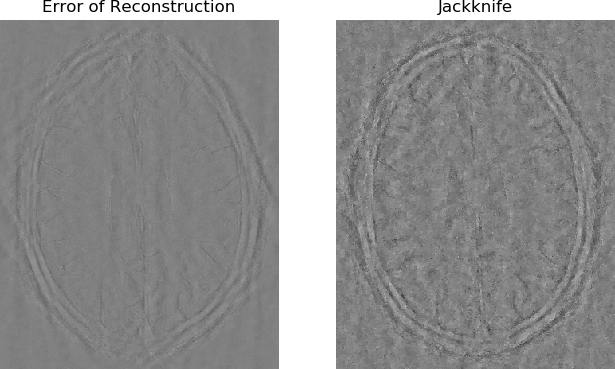}}

\vspace{\vertsep}

\parbox{\imsize}{\includegraphics[width=\imsize]{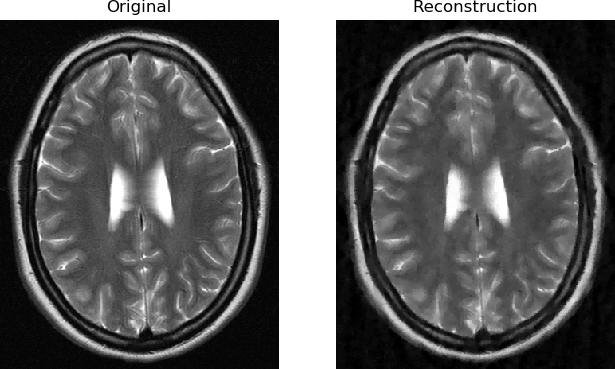}}

\vspace{\vertsep}

\parbox{\imsize}{\includegraphics[width=\imsize]{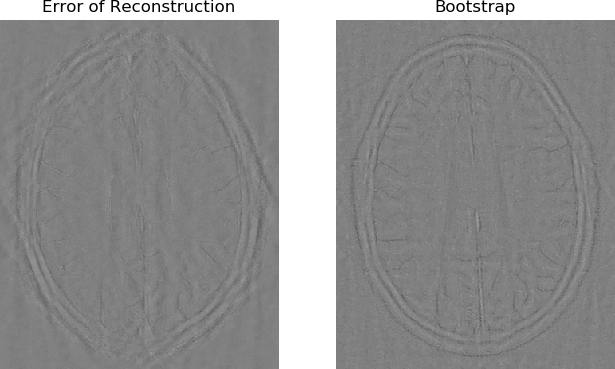}}

\end{centering}
\caption{radially retained sampling --- slice 15}
\end{figure}

\begin{figure}
\begin{centering}

\parbox{\imsize}{\includegraphics[width=\imsize]{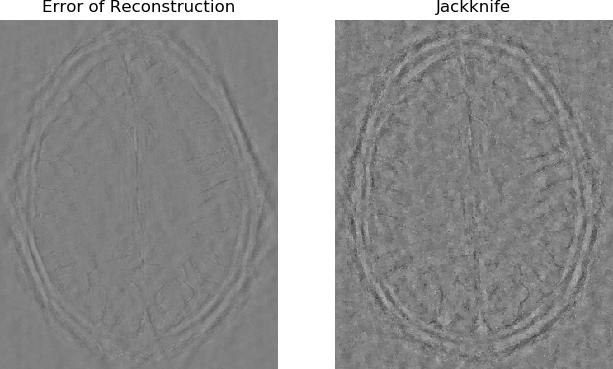}}

\vspace{\vertsep}

\parbox{\imsize}{\includegraphics[width=\imsize]{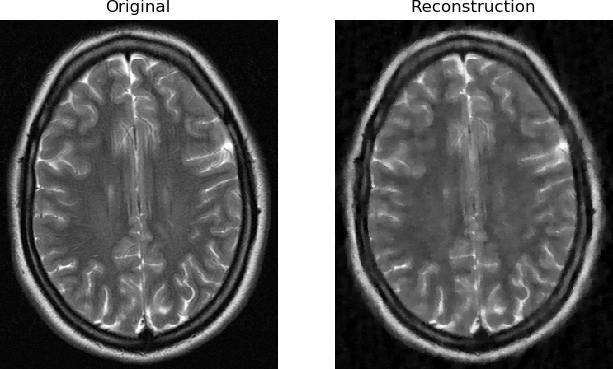}}

\vspace{\vertsep}

\parbox{\imsize}{\includegraphics[width=\imsize]{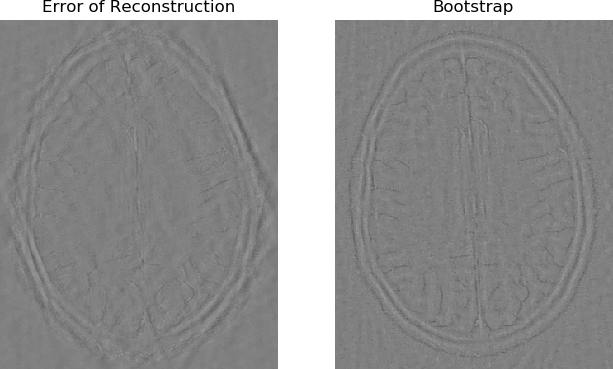}}

\end{centering}
\caption{radially retained sampling --- slice 16}
\end{figure}

\begin{figure}
\begin{centering}

\parbox{\imsize}{\includegraphics[width=\imsize]{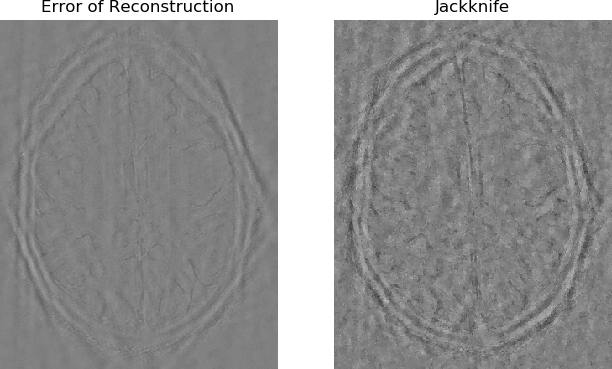}}

\vspace{\vertsep}

\parbox{\imsize}{\includegraphics[width=\imsize]{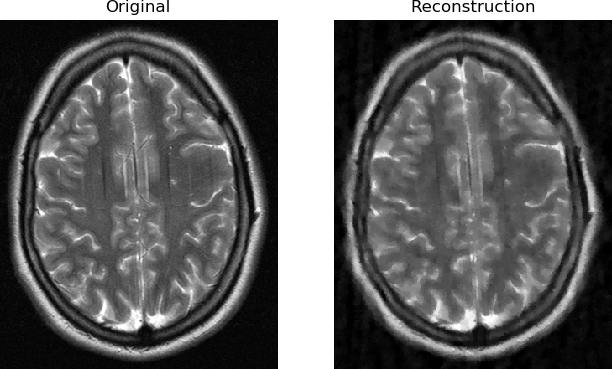}}

\vspace{\vertsep}

\parbox{\imsize}{\includegraphics[width=\imsize]{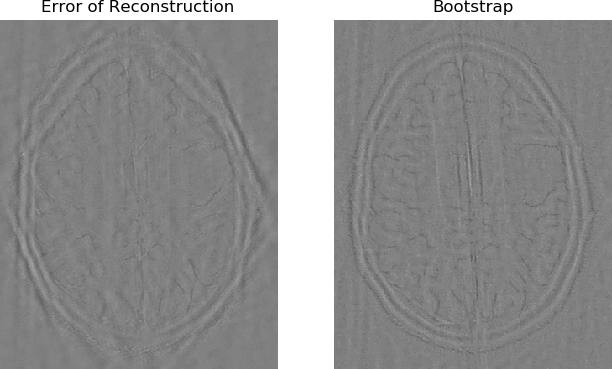}}

\end{centering}
\caption{radially retained sampling --- slice 17}
\end{figure}

\begin{figure}
\begin{centering}

\parbox{\imsize}{\includegraphics[width=\imsize]{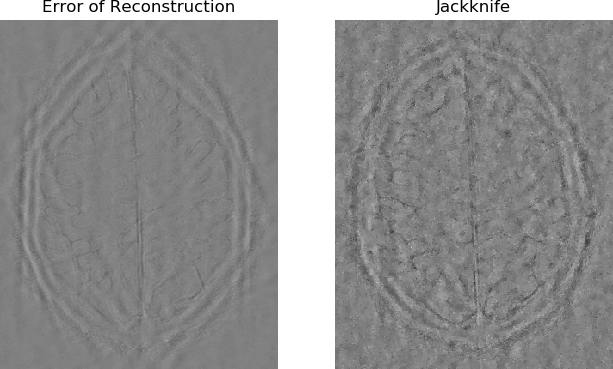}}

\vspace{\vertsep}

\parbox{\imsize}{\includegraphics[width=\imsize]{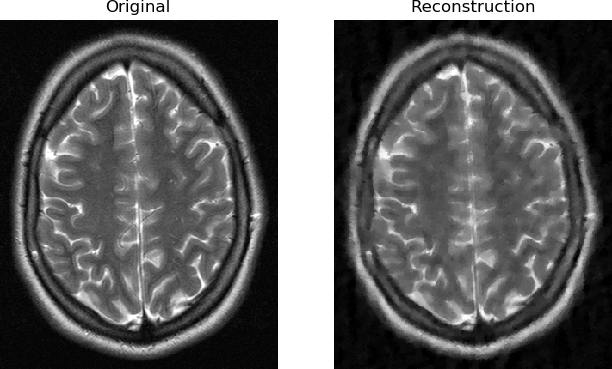}}

\vspace{\vertsep}

\parbox{\imsize}{\includegraphics[width=\imsize]{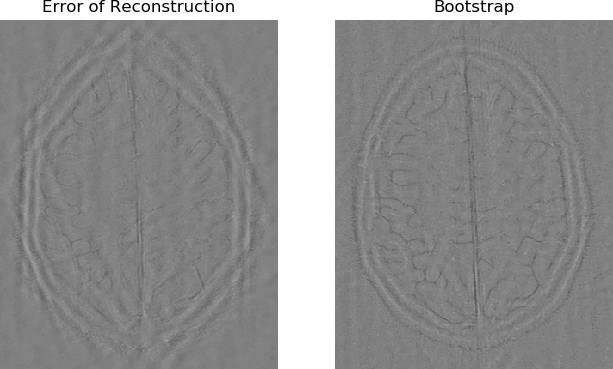}}

\end{centering}
\caption{radially retained sampling --- slice 18}
\end{figure}

\begin{figure}
\begin{centering}

\parbox{\imsize}{\includegraphics[width=\imsize]{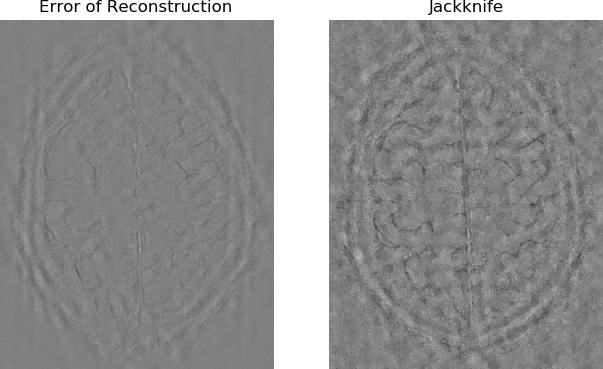}}

\vspace{\vertsep}

\parbox{\imsize}{\includegraphics[width=\imsize]{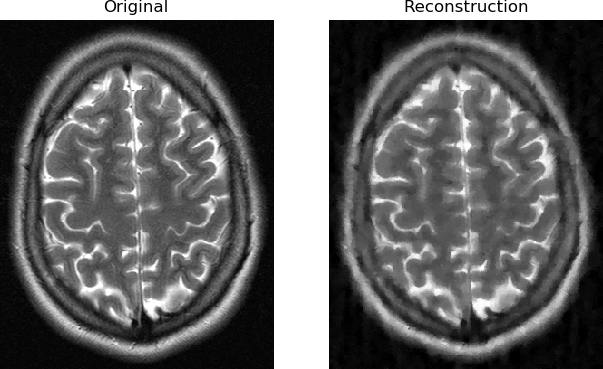}}

\vspace{\vertsep}

\parbox{\imsize}{\includegraphics[width=\imsize]{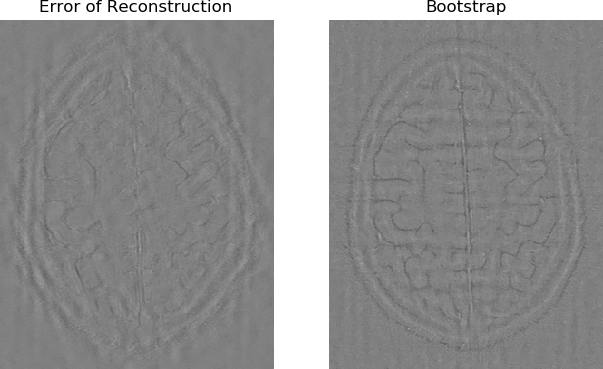}}

\end{centering}
\caption{radially retained sampling --- slice 19}
\end{figure}

\begin{figure}
\begin{centering}

\parbox{\imsize}{\includegraphics[width=\imsize]{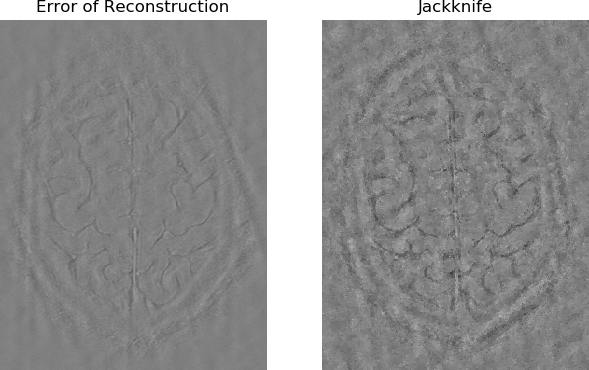}}

\vspace{\vertsep}

\parbox{\imsize}{\includegraphics[width=\imsize]{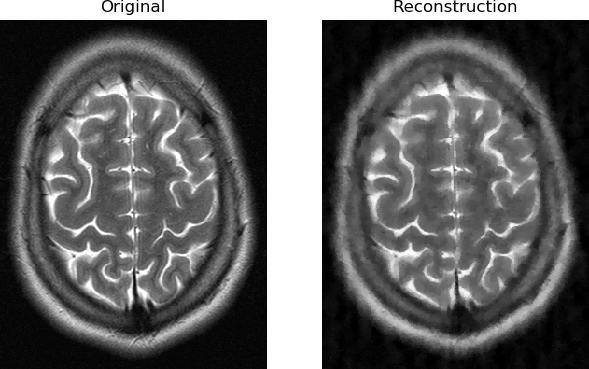}}

\vspace{\vertsep}

\parbox{\imsize}{\includegraphics[width=\imsize]{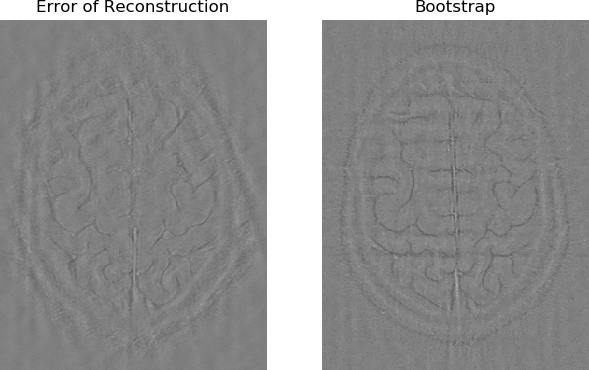}}

\end{centering}
\caption{radially retained sampling --- slice 20}
\end{figure}

\begin{figure}
\begin{centering}

\parbox{\imsize}{\includegraphics[width=\imsize]{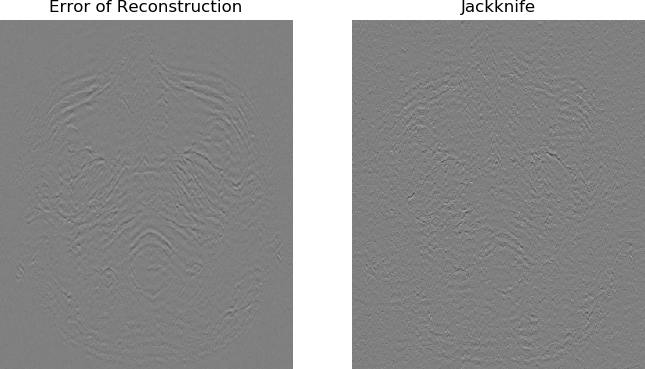}}

\vspace{\vertsep}

\parbox{\imsize}{\includegraphics[width=\imsize]{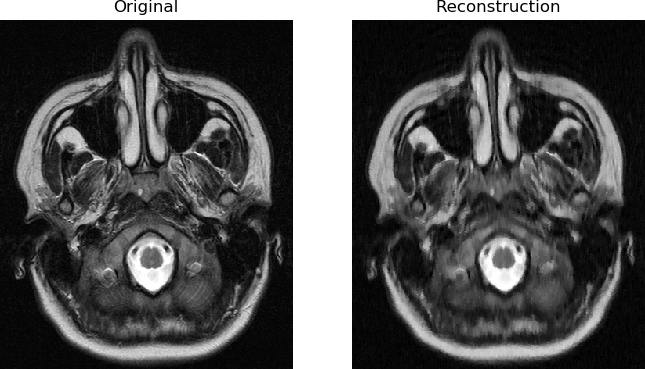}}

\vspace{\vertsep}

\parbox{\imsize}{\includegraphics[width=\imsize]{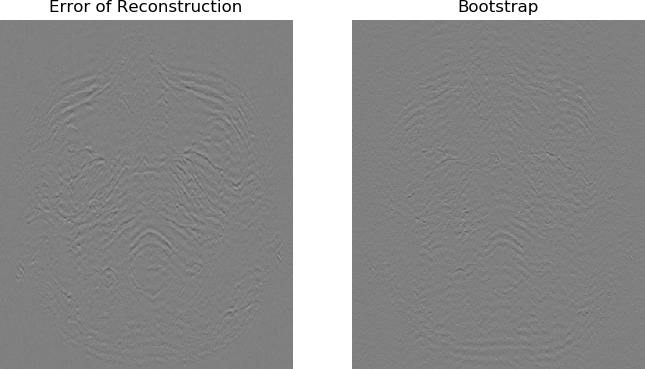}}

\end{centering}
\caption{horizontally retained sampling --- slice 1}
\end{figure}

\begin{figure}
\begin{centering}

\parbox{\imsize}{\includegraphics[width=\imsize]{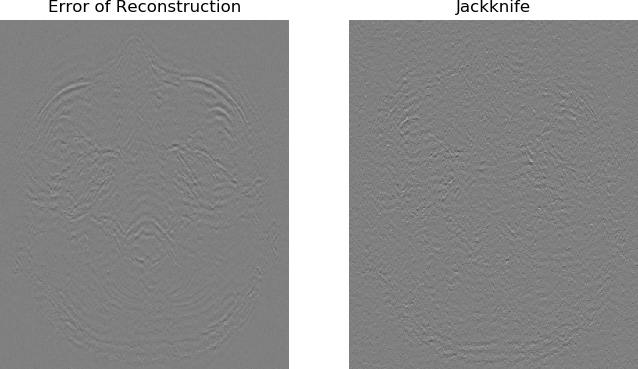}}

\vspace{\vertsep}

\parbox{\imsize}{\includegraphics[width=\imsize]{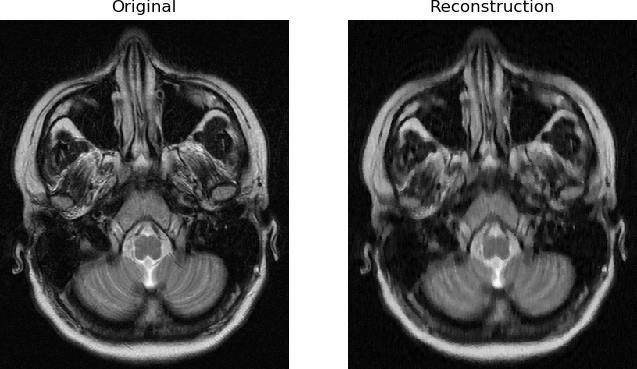}}

\vspace{\vertsep}

\parbox{\imsize}{\includegraphics[width=\imsize]{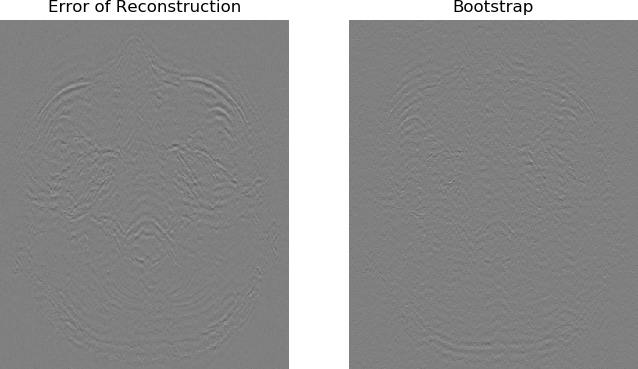}}

\end{centering}
\caption{horizontally retained sampling --- slice 2}
\end{figure}

\begin{figure}
\begin{centering}

\parbox{\imsize}{\includegraphics[width=\imsize]{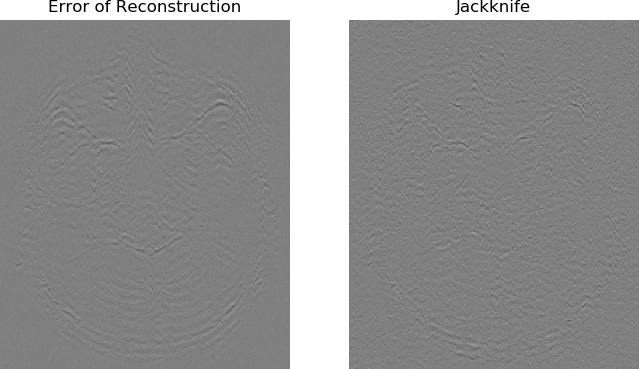}}

\vspace{\vertsep}

\parbox{\imsize}{\includegraphics[width=\imsize]{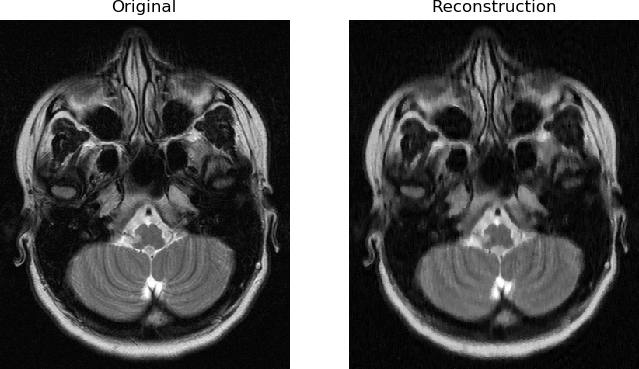}}

\vspace{\vertsep}

\parbox{\imsize}{\includegraphics[width=\imsize]{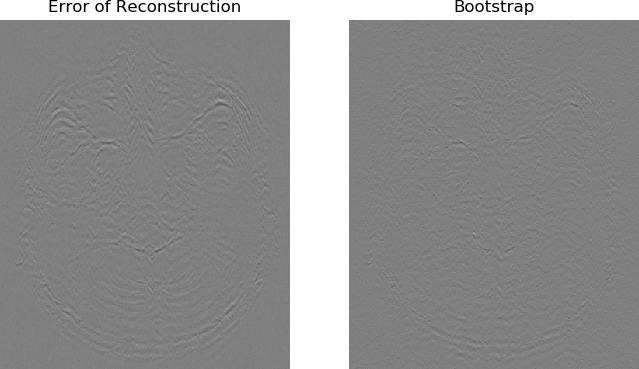}}

\end{centering}
\caption{horizontally retained sampling --- slice 3}
\end{figure}

\begin{figure}
\begin{centering}

\parbox{\imsize}{\includegraphics[width=\imsize]{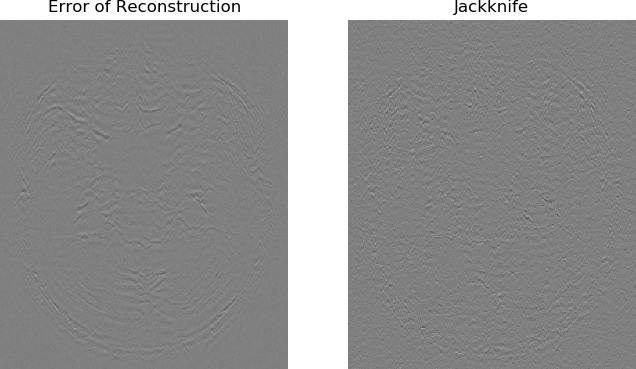}}

\vspace{\vertsep}

\parbox{\imsize}{\includegraphics[width=\imsize]{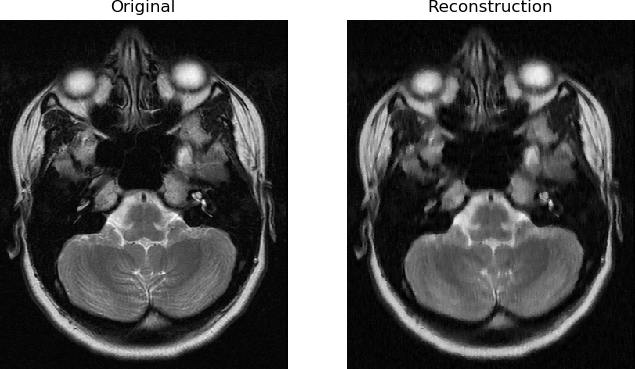}}

\vspace{\vertsep}

\parbox{\imsize}{\includegraphics[width=\imsize]{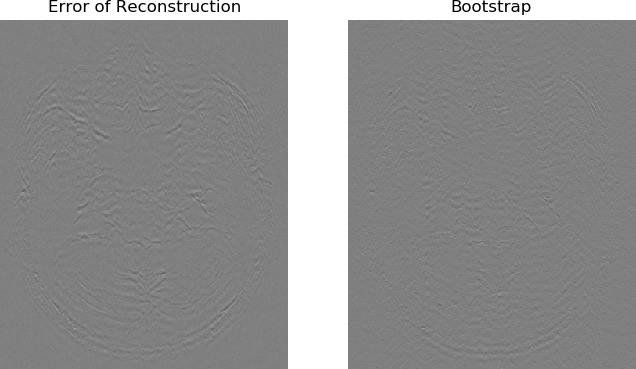}}

\end{centering}
\caption{horizontally retained sampling --- slice 4}
\end{figure}

\begin{figure}
\begin{centering}

\parbox{\imsize}{\includegraphics[width=\imsize]{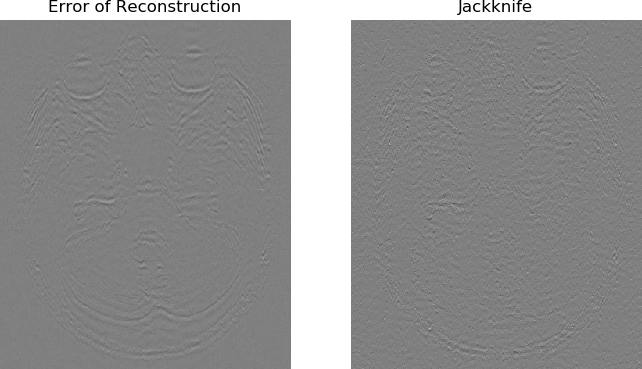}}

\vspace{\vertsep}

\parbox{\imsize}{\includegraphics[width=\imsize]{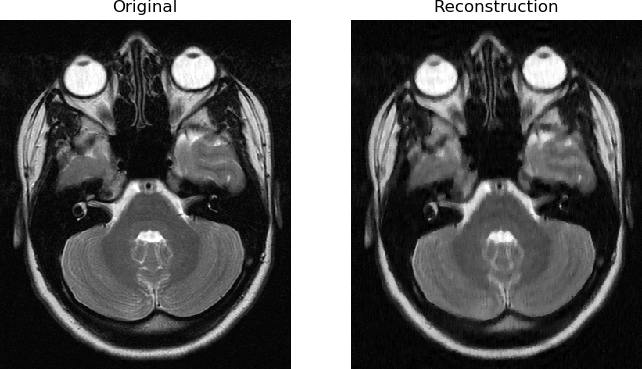}}

\vspace{\vertsep}

\parbox{\imsize}{\includegraphics[width=\imsize]{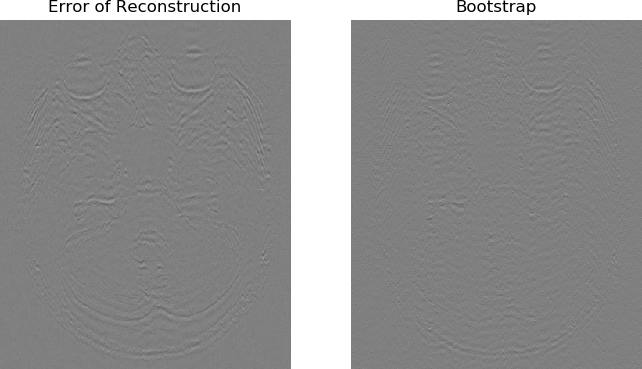}}

\end{centering}
\caption{horizontally retained sampling --- slice 5}
\end{figure}

\begin{figure}
\begin{centering}

\parbox{\imsize}{\includegraphics[width=\imsize]{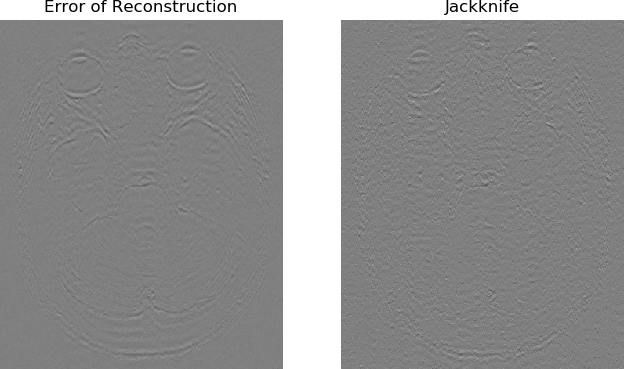}}

\vspace{\vertsep}

\parbox{\imsize}{\includegraphics[width=\imsize]{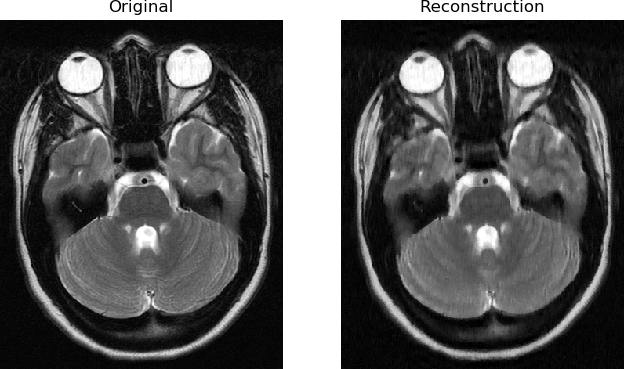}}

\vspace{\vertsep}

\parbox{\imsize}{\includegraphics[width=\imsize]{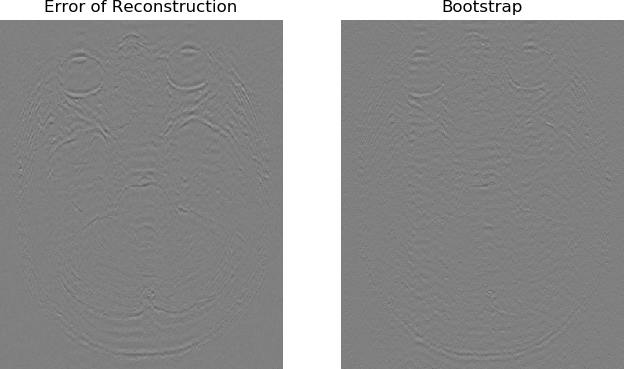}}

\end{centering}
\caption{horizontally retained sampling --- slice 6}
\end{figure}

\begin{figure}
\begin{centering}

\parbox{\imsize}{\includegraphics[width=\imsize]{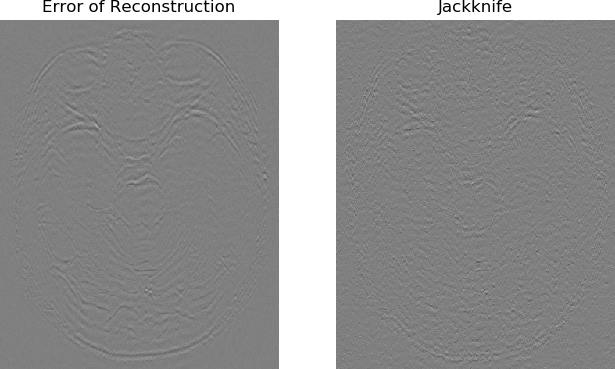}}

\vspace{\vertsep}

\parbox{\imsize}{\includegraphics[width=\imsize]{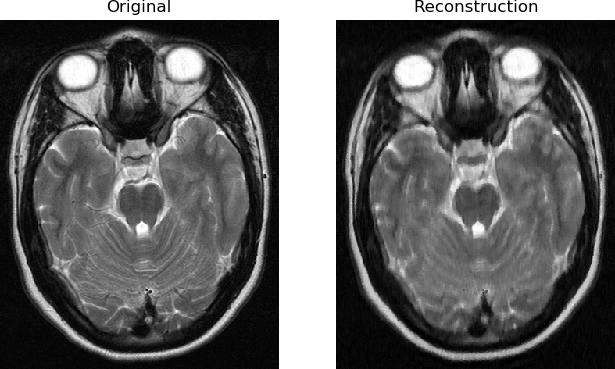}}

\vspace{\vertsep}

\parbox{\imsize}{\includegraphics[width=\imsize]{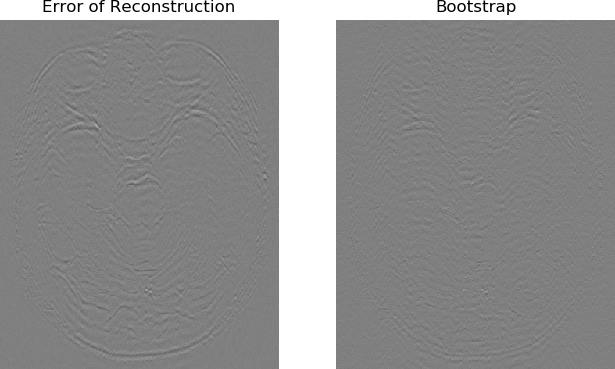}}

\end{centering}
\caption{horizontally retained sampling --- slice 7}
\end{figure}

\begin{figure}
\begin{centering}

\parbox{\imsize}{\includegraphics[width=\imsize]{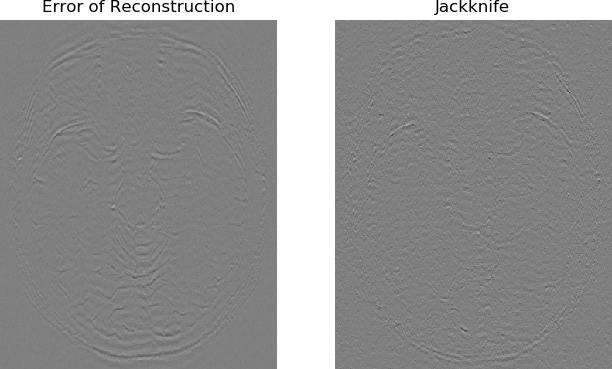}}

\vspace{\vertsep}

\parbox{\imsize}{\includegraphics[width=\imsize]{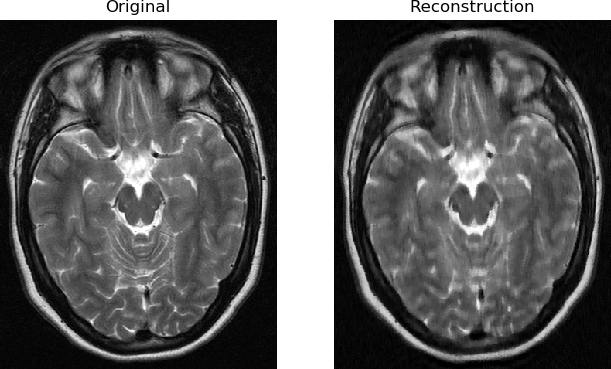}}

\vspace{\vertsep}

\parbox{\imsize}{\includegraphics[width=\imsize]{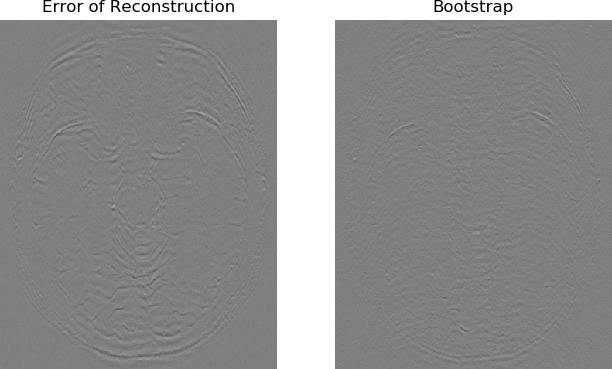}}

\end{centering}
\caption{horizontally retained sampling --- slice 8}
\end{figure}

\begin{figure}
\begin{centering}

\parbox{\imsize}{\includegraphics[width=\imsize]{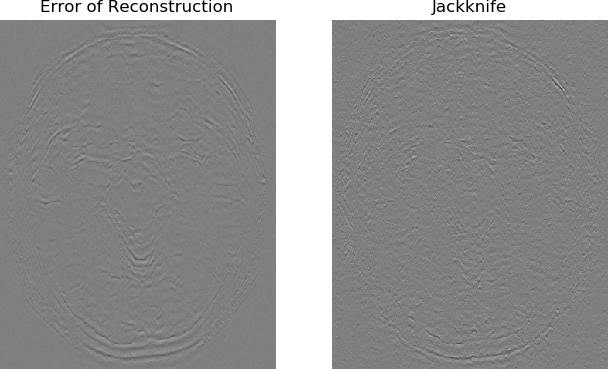}}

\vspace{\vertsep}

\parbox{\imsize}{\includegraphics[width=\imsize]{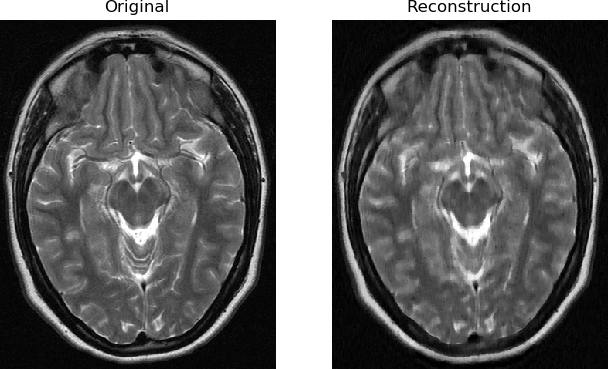}}

\vspace{\vertsep}

\parbox{\imsize}{\includegraphics[width=\imsize]{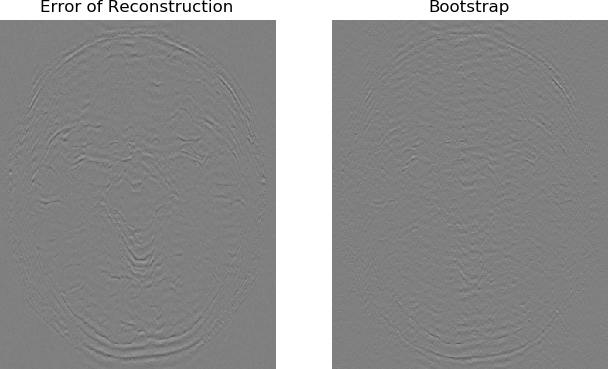}}

\end{centering}
\caption{horizontally retained sampling --- slice 9}
\end{figure}

\begin{figure}
\begin{centering}

\parbox{\imsize}{\includegraphics[width=\imsize]{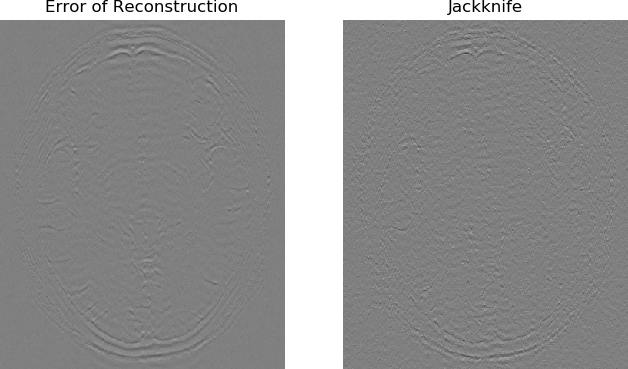}}

\vspace{\vertsep}

\parbox{\imsize}{\includegraphics[width=\imsize]{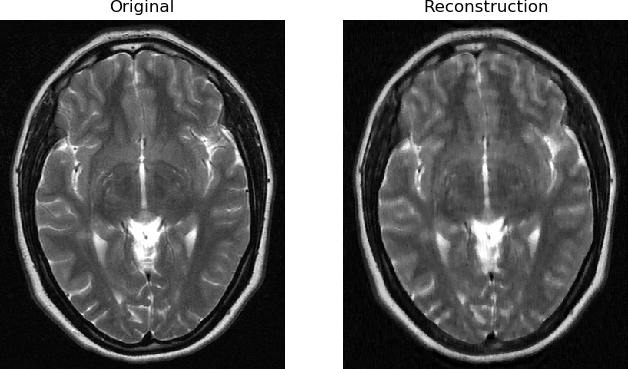}}

\vspace{\vertsep}

\parbox{\imsize}{\includegraphics[width=\imsize]{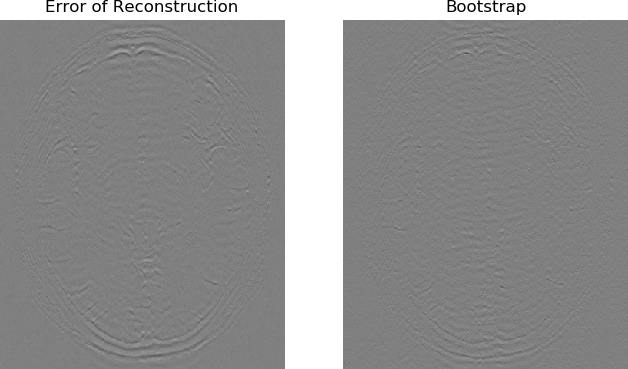}}

\end{centering}
\caption{horizontally retained sampling --- slice 10}
\end{figure}

\begin{figure}
\begin{centering}

\parbox{\imsize}{\includegraphics[width=\imsize]{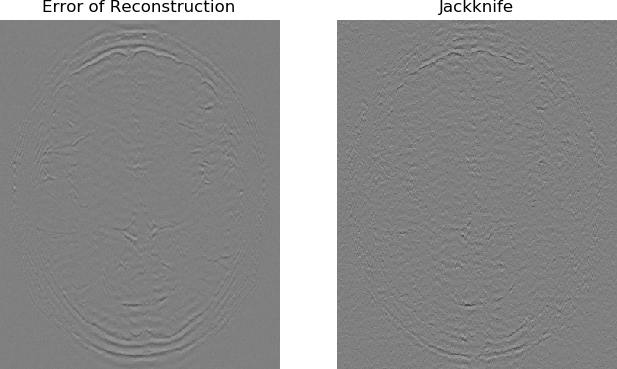}}

\vspace{\vertsep}

\parbox{\imsize}{\includegraphics[width=\imsize]{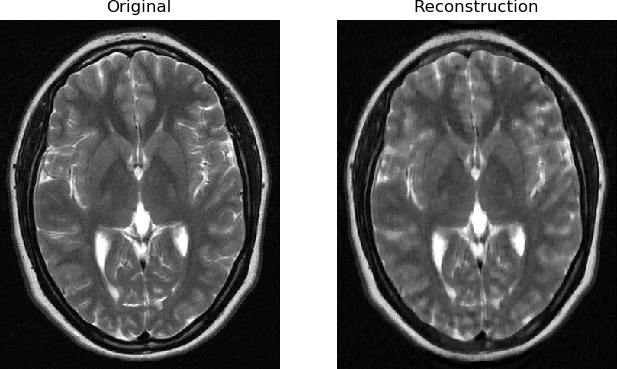}}

\vspace{\vertsep}

\parbox{\imsize}{\includegraphics[width=\imsize]{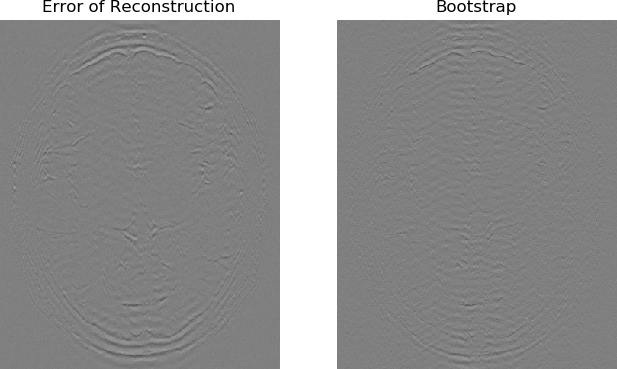}}

\end{centering}
\caption{horizontally retained sampling --- slice 11}
\end{figure}

\begin{figure}
\begin{centering}

\parbox{\imsize}{\includegraphics[width=\imsize]{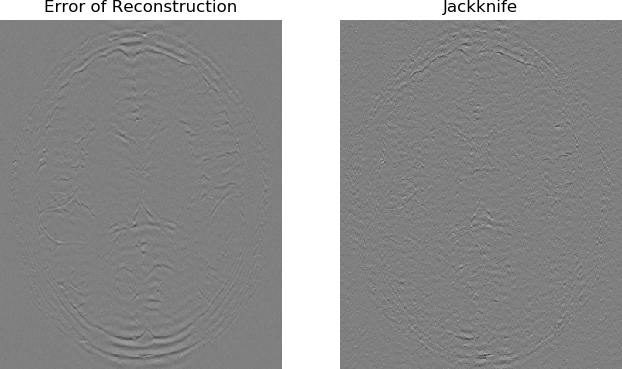}}

\vspace{\vertsep}

\parbox{\imsize}{\includegraphics[width=\imsize]{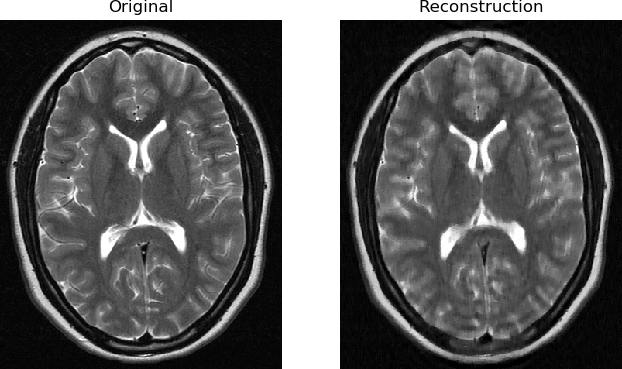}}

\vspace{\vertsep}

\parbox{\imsize}{\includegraphics[width=\imsize]{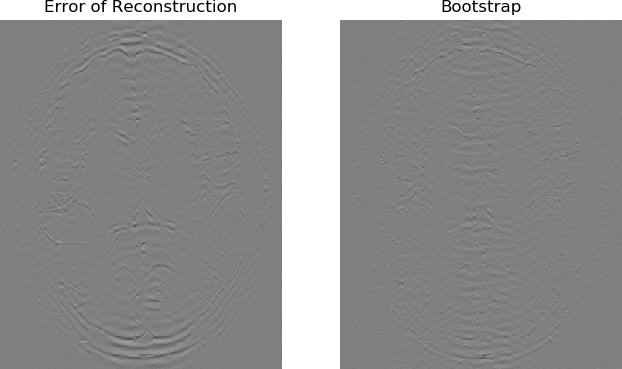}}

\end{centering}
\caption{horizontally retained sampling --- slice 12}
\end{figure}

\begin{figure}
\begin{centering}

\parbox{\imsize}{\includegraphics[width=\imsize]{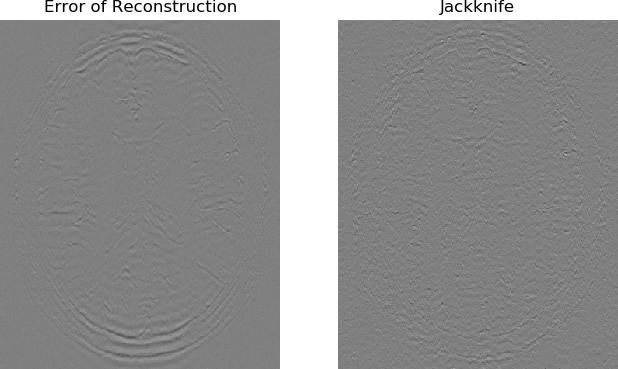}}

\vspace{\vertsep}

\parbox{\imsize}{\includegraphics[width=\imsize]{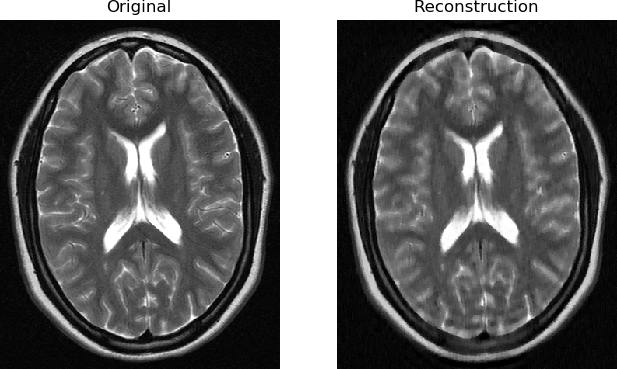}}

\vspace{\vertsep}

\parbox{\imsize}{\includegraphics[width=\imsize]{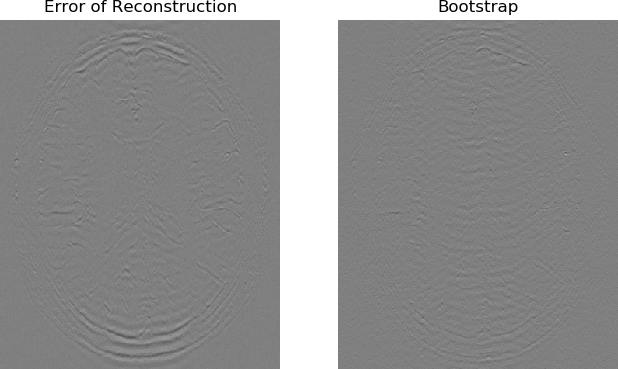}}

\end{centering}
\caption{horizontally retained sampling --- slice 13}
\end{figure}

\begin{figure}
\begin{centering}

\parbox{\imsize}{\includegraphics[width=\imsize]{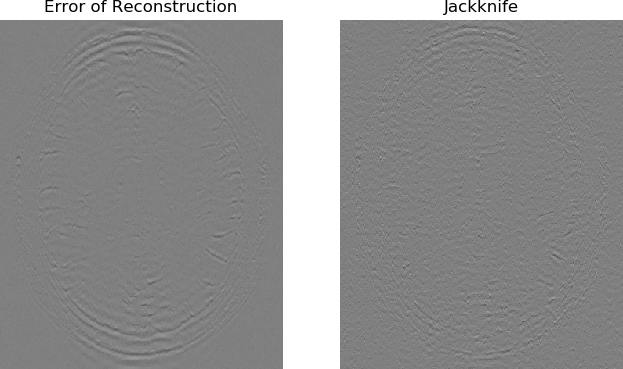}}

\vspace{\vertsep}

\parbox{\imsize}{\includegraphics[width=\imsize]{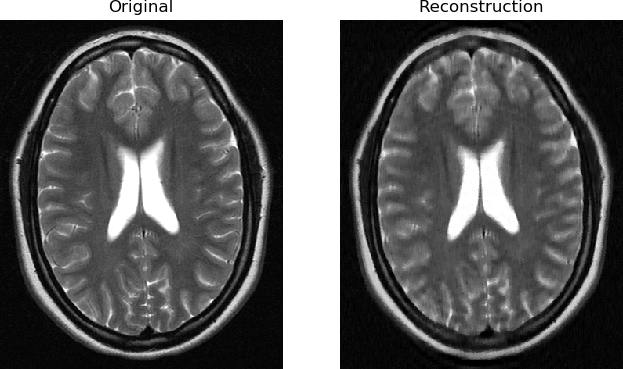}}

\vspace{\vertsep}

\parbox{\imsize}{\includegraphics[width=\imsize]{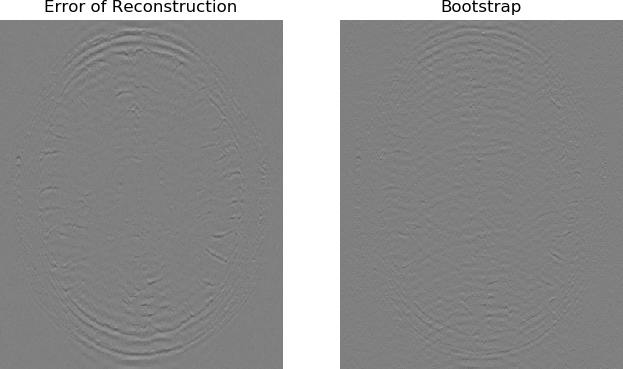}}

\end{centering}
\caption{horizontally retained sampling --- slice 14}
\end{figure}

\begin{figure}
\begin{centering}

\parbox{\imsize}{\includegraphics[width=\imsize]{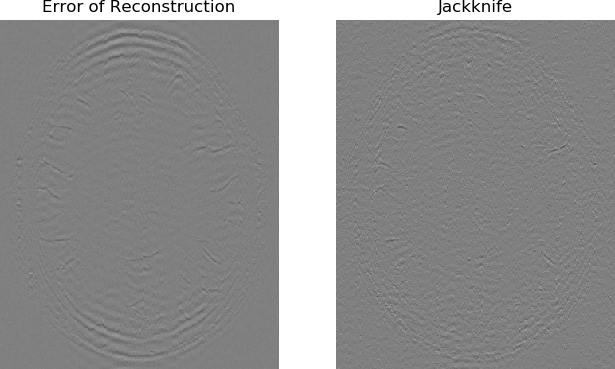}}

\vspace{\vertsep}

\parbox{\imsize}{\includegraphics[width=\imsize]{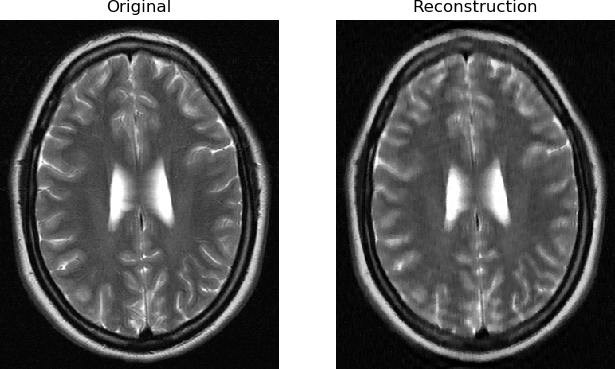}}

\vspace{\vertsep}

\parbox{\imsize}{\includegraphics[width=\imsize]{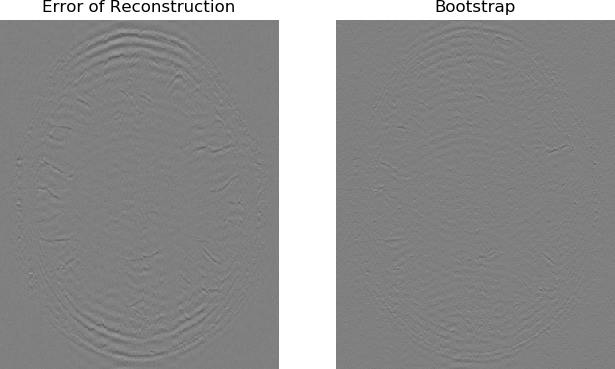}}

\end{centering}
\caption{horizontally retained sampling --- slice 15}
\end{figure}

\begin{figure}
\begin{centering}

\parbox{\imsize}{\includegraphics[width=\imsize]{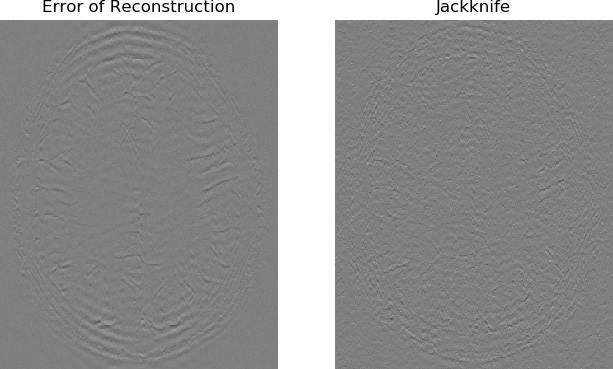}}

\vspace{\vertsep}

\parbox{\imsize}{\includegraphics[width=\imsize]{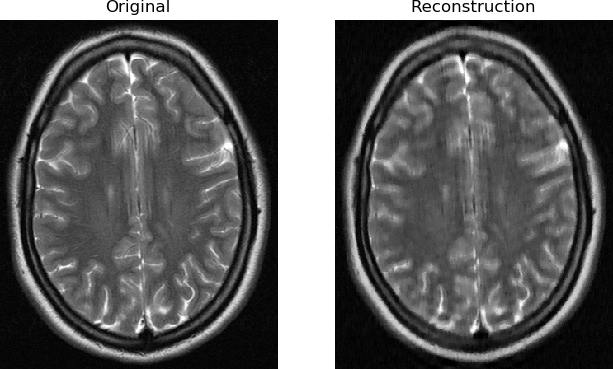}}

\vspace{\vertsep}

\parbox{\imsize}{\includegraphics[width=\imsize]{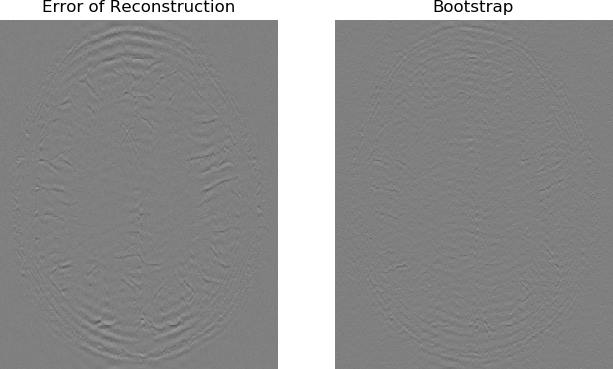}}

\end{centering}
\caption{horizontally retained sampling --- slice 16}
\end{figure}

\begin{figure}
\begin{centering}

\parbox{\imsize}{\includegraphics[width=\imsize]{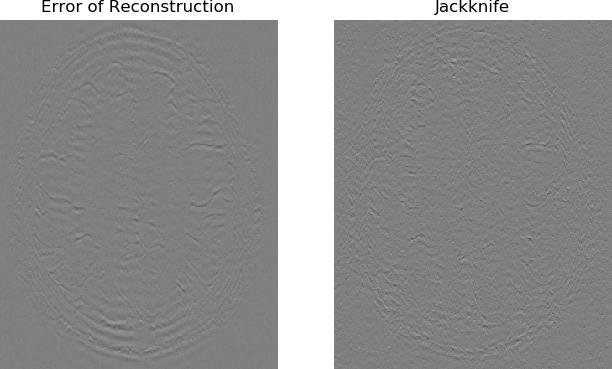}}

\vspace{\vertsep}

\parbox{\imsize}{\includegraphics[width=\imsize]{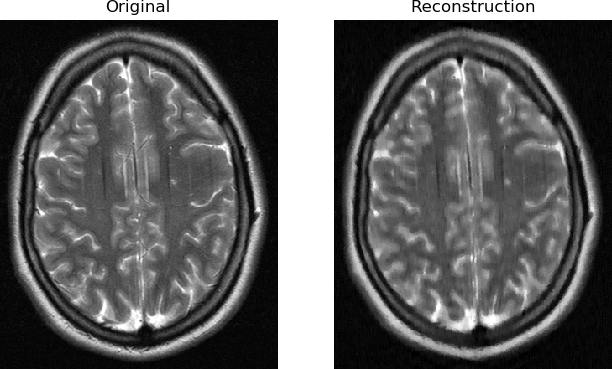}}

\vspace{\vertsep}

\parbox{\imsize}{\includegraphics[width=\imsize]{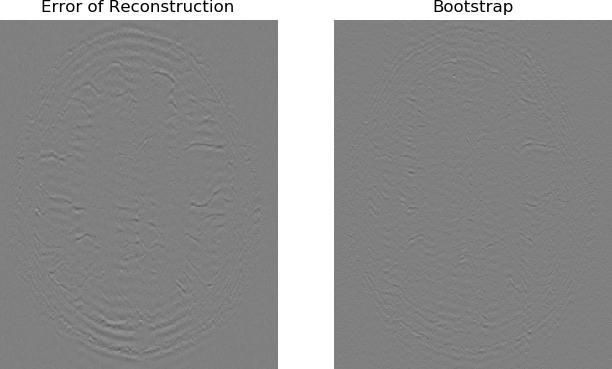}}

\end{centering}
\caption{horizontally retained sampling --- slice 17}
\end{figure}

\begin{figure}
\begin{centering}

\parbox{\imsize}{\includegraphics[width=\imsize]{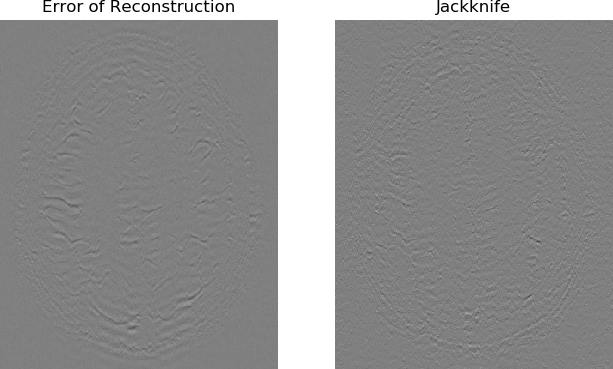}}

\vspace{\vertsep}

\parbox{\imsize}{\includegraphics[width=\imsize]{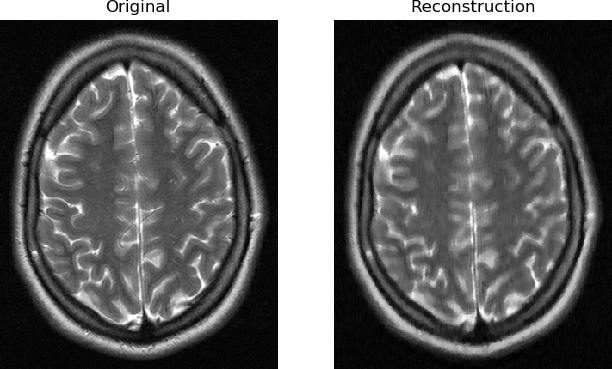}}

\vspace{\vertsep}

\parbox{\imsize}{\includegraphics[width=\imsize]{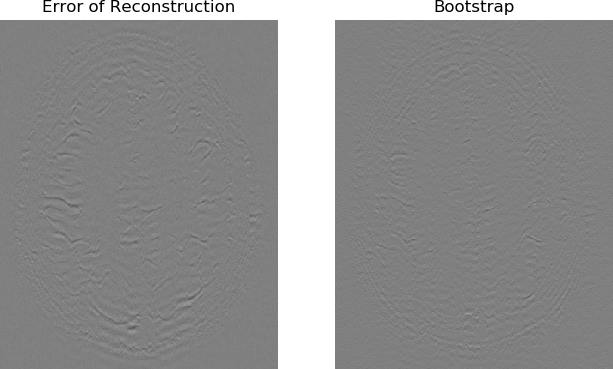}}

\end{centering}
\caption{horizontally retained sampling --- slice 18}
\end{figure}

\begin{figure}
\begin{centering}

\parbox{\imsize}{\includegraphics[width=\imsize]{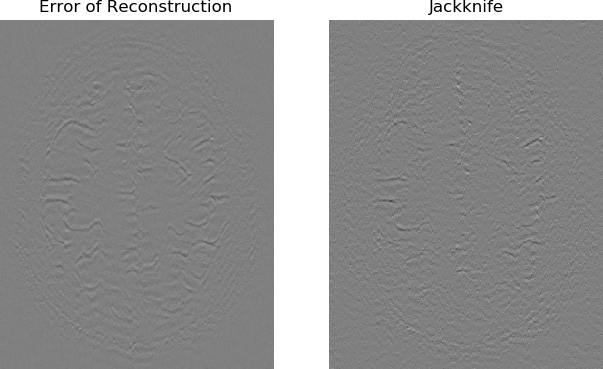}}

\vspace{\vertsep}

\parbox{\imsize}{\includegraphics[width=\imsize]{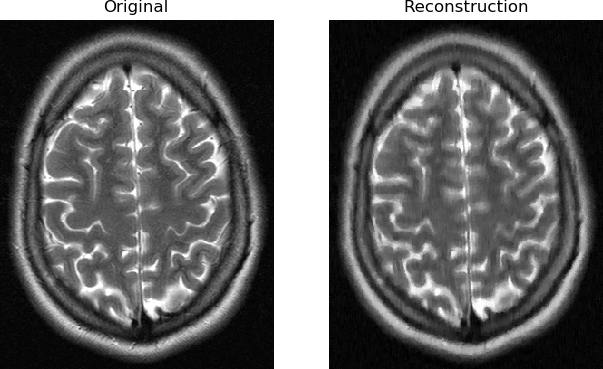}}

\vspace{\vertsep}

\parbox{\imsize}{\includegraphics[width=\imsize]{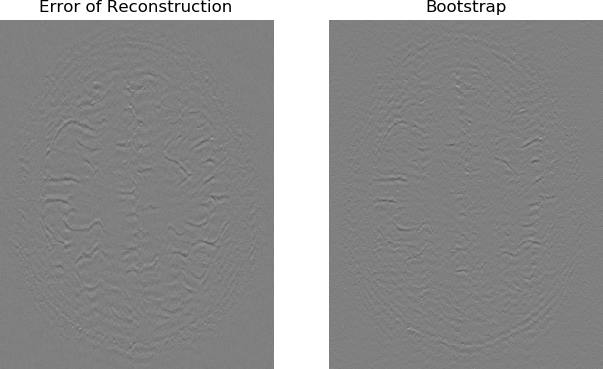}}

\end{centering}
\caption{horizontally retained sampling --- slice 19}
\end{figure}

\begin{figure}
\begin{centering}

\parbox{\imsize}{\includegraphics[width=\imsize]{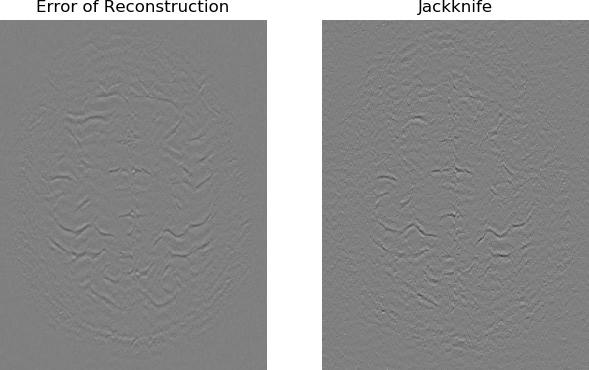}}

\vspace{\vertsep}

\parbox{\imsize}{\includegraphics[width=\imsize]{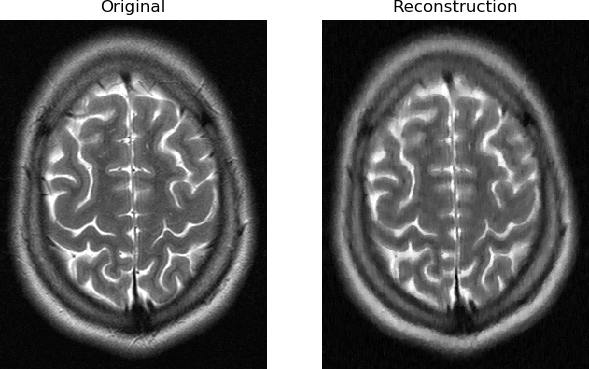}}

\vspace{\vertsep}

\parbox{\imsize}{\includegraphics[width=\imsize]{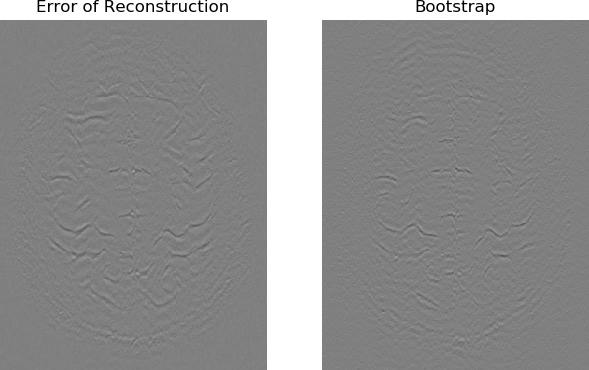}}

\end{centering}
\caption{horizontally retained sampling --- slice 20}
\end{figure}

\begin{figure}
\begin{centering}

\parbox{\imsizes}{\includegraphics[width=\imsizes]{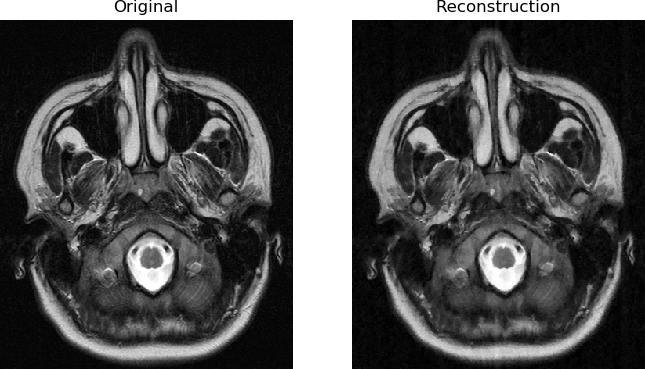}}

\vspace{\vertseps}

\parbox{\imsizes}{\includegraphics[width=\imsizes]{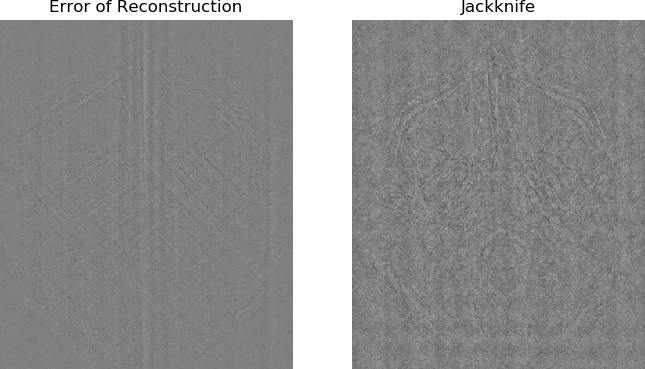}}
\hfill
\parbox{\imsizes}{\includegraphics[width=\imsizes]{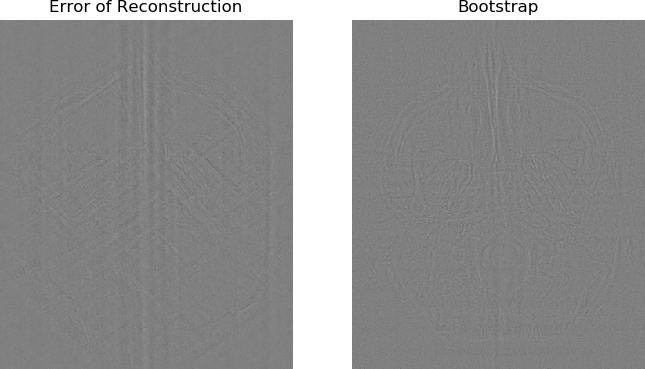}}

\end{centering}
\caption{$2\times$ radially retained sampling --- slice 1}
\end{figure}

\begin{figure}
\begin{centering}

\parbox{\imsizes}{\includegraphics[width=\imsizes]{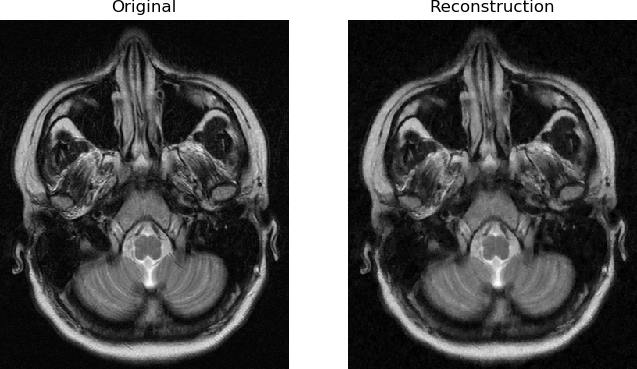}}

\vspace{\vertseps}

\parbox{\imsizes}{\includegraphics[width=\imsizes]{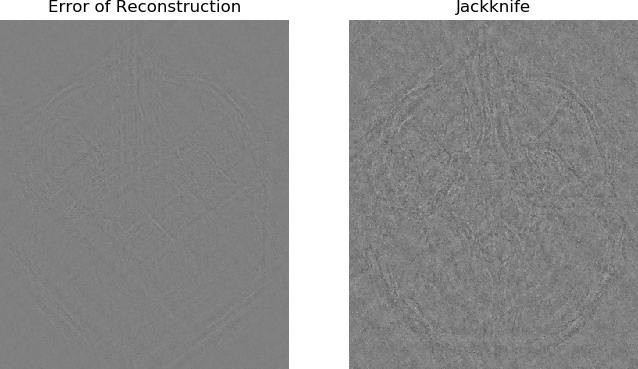}}
\hfill
\parbox{\imsizes}{\includegraphics[width=\imsizes]{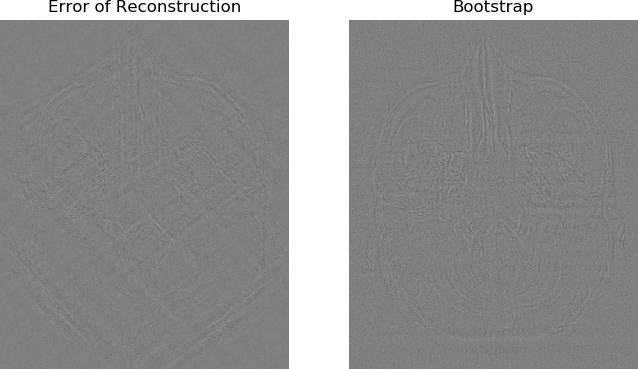}}

\end{centering}
\caption{$2\times$ radially retained sampling --- slice 2}
\end{figure}

\begin{figure}
\begin{centering}

\parbox{\imsizes}{\includegraphics[width=\imsizes]{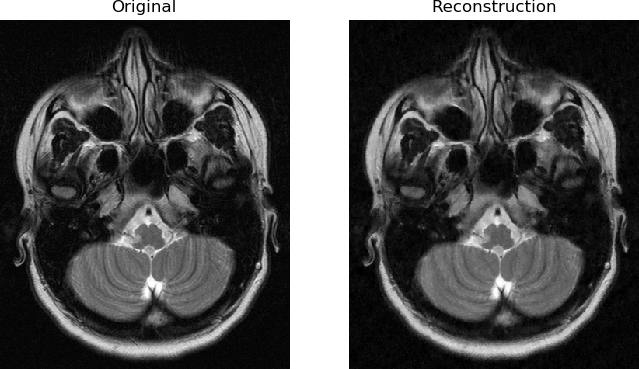}}

\vspace{\vertseps}

\parbox{\imsizes}{\includegraphics[width=\imsizes]{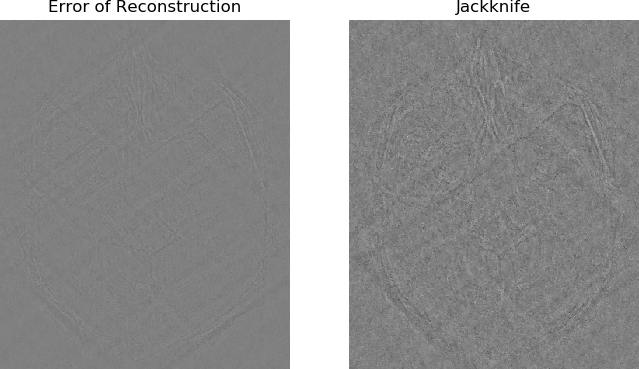}}
\hfill
\parbox{\imsizes}{\includegraphics[width=\imsizes]{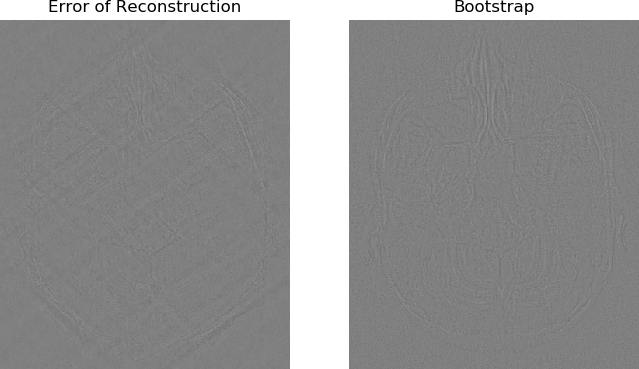}}

\end{centering}
\caption{$2\times$ radially retained sampling --- slice 3}
\end{figure}

\begin{figure}
\begin{centering}

\parbox{\imsizes}{\includegraphics[width=\imsizes]{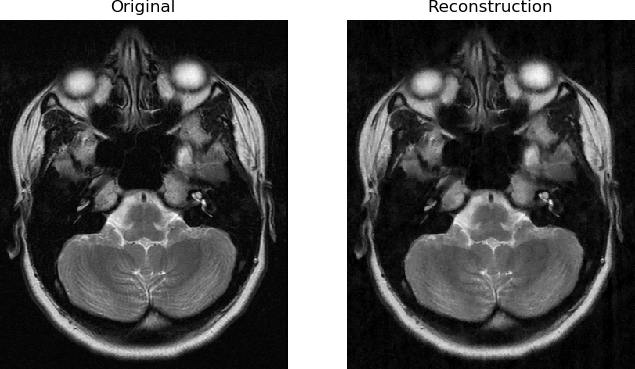}}

\vspace{\vertseps}

\parbox{\imsizes}{\includegraphics[width=\imsizes]{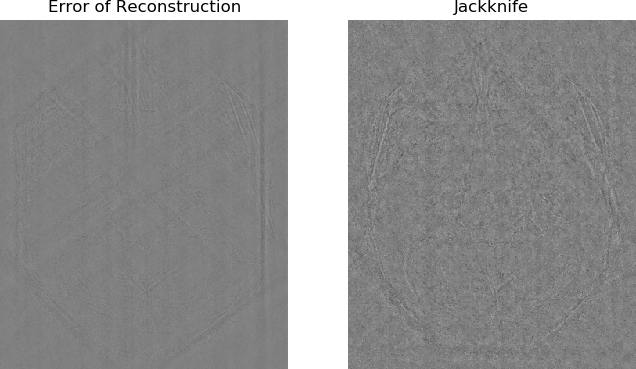}}
\hfill
\parbox{\imsizes}{\includegraphics[width=\imsizes]{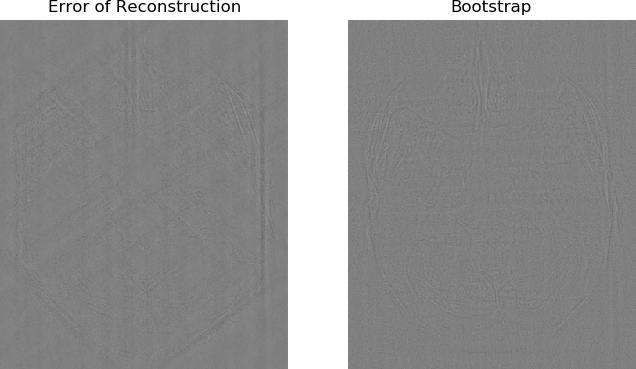}}

\end{centering}
\caption{$2\times$ radially retained sampling --- slice 4}
\end{figure}

\begin{figure}
\begin{centering}

\parbox{\imsizes}{\includegraphics[width=\imsizes]{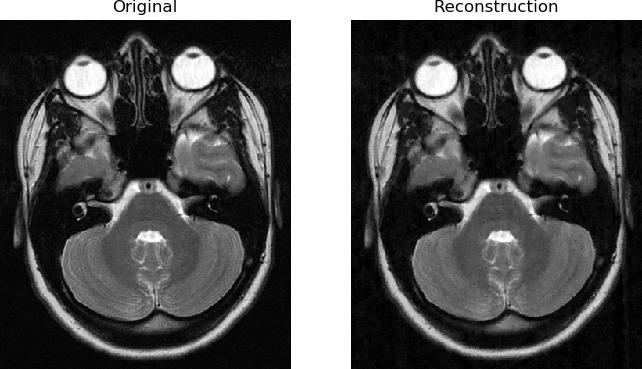}}

\vspace{\vertseps}

\parbox{\imsizes}{\includegraphics[width=\imsizes]{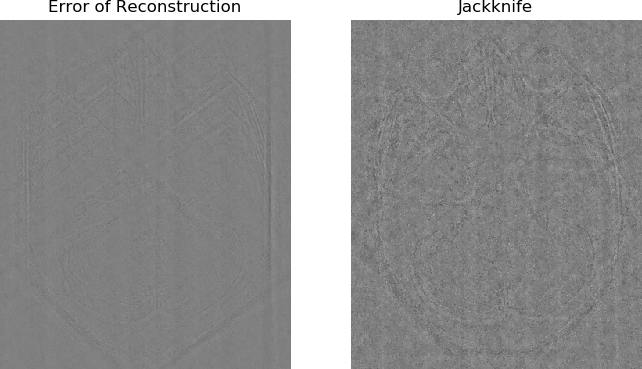}}
\hfill
\parbox{\imsizes}{\includegraphics[width=\imsizes]{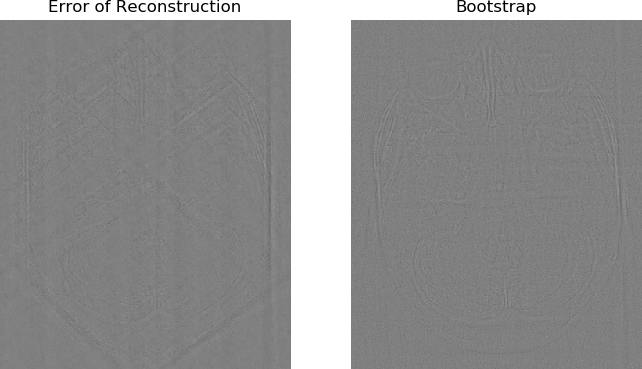}}

\end{centering}
\caption{$2\times$ radially retained sampling --- slice 5}
\end{figure}

\begin{figure}
\begin{centering}

\parbox{\imsizes}{\includegraphics[width=\imsizes]{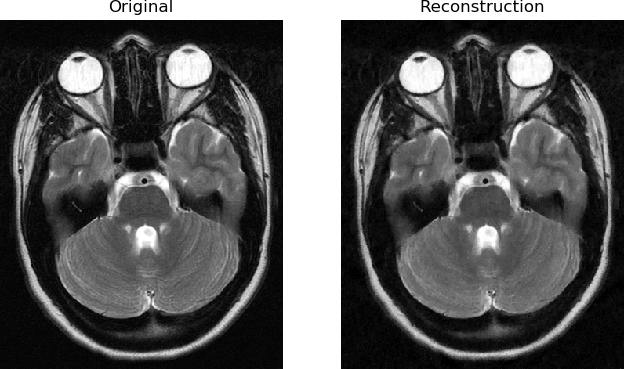}}

\vspace{\vertseps}

\parbox{\imsizes}{\includegraphics[width=\imsizes]{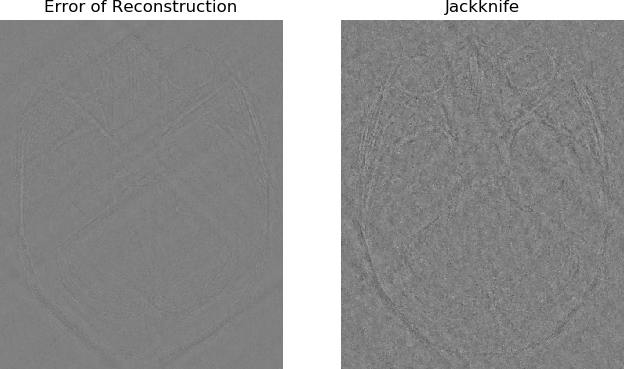}}
\hfill
\parbox{\imsizes}{\includegraphics[width=\imsizes]{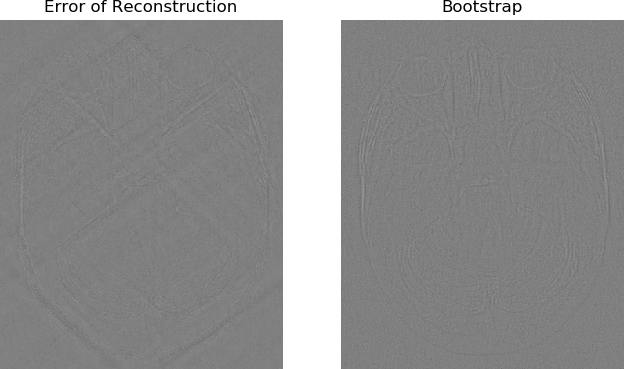}}

\end{centering}
\caption{$2\times$ radially retained sampling --- slice 6}
\end{figure}

\begin{figure}
\begin{centering}

\parbox{\imsizes}{\includegraphics[width=\imsizes]{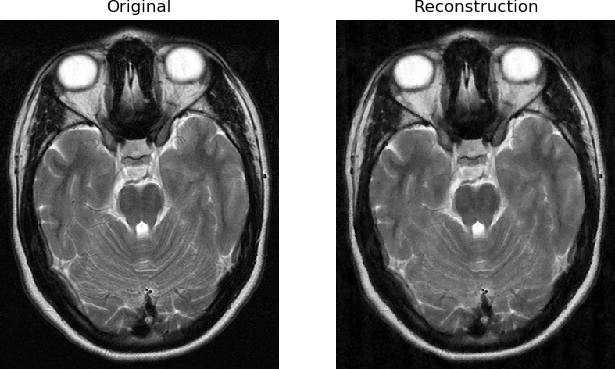}}

\vspace{\vertseps}

\parbox{\imsizes}{\includegraphics[width=\imsizes]{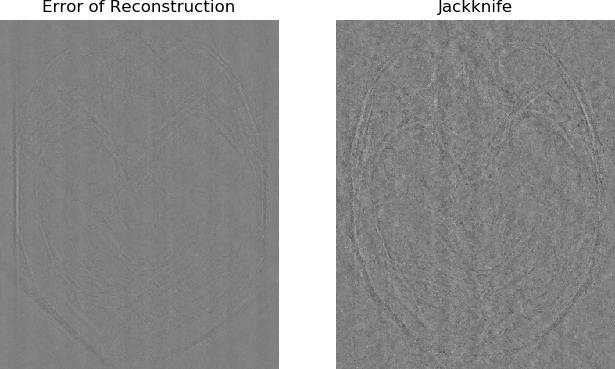}}
\hfill
\parbox{\imsizes}{\includegraphics[width=\imsizes]{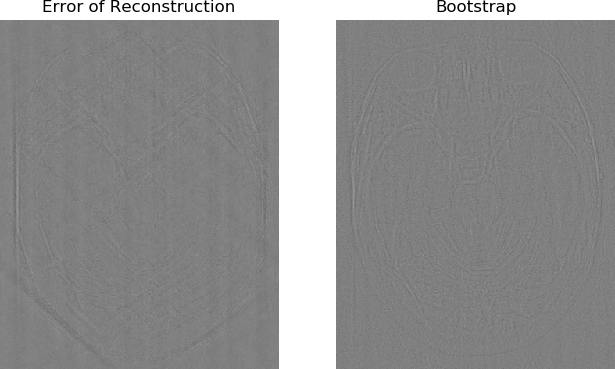}}

\end{centering}
\caption{$2\times$ radially retained sampling --- slice 7}
\end{figure}

\begin{figure}
\begin{centering}

\parbox{\imsizes}{\includegraphics[width=\imsizes]{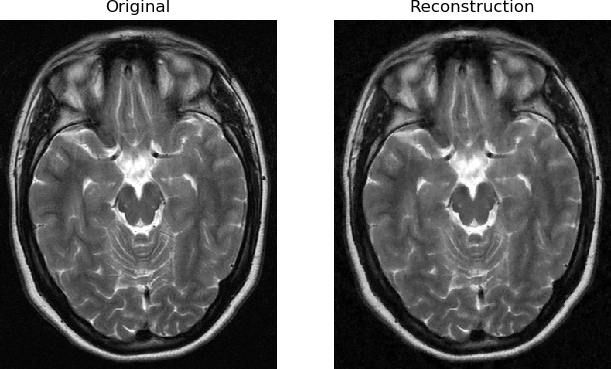}}

\vspace{\vertseps}

\parbox{\imsizes}{\includegraphics[width=\imsizes]{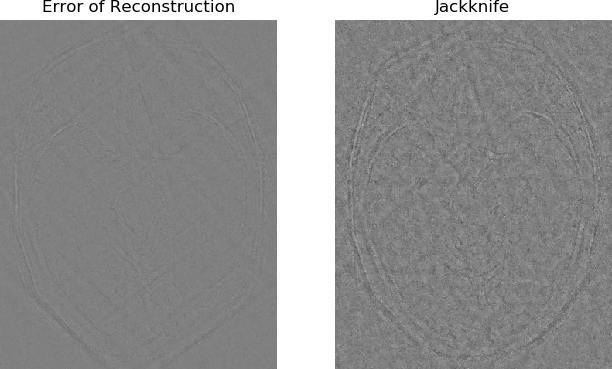}}
\hfill
\parbox{\imsizes}{\includegraphics[width=\imsizes]{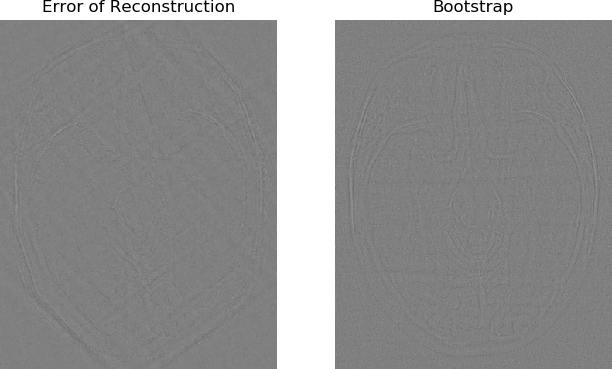}}

\end{centering}
\caption{$2\times$ radially retained sampling --- slice 8}
\end{figure}

\begin{figure}
\begin{centering}

\parbox{\imsizes}{\includegraphics[width=\imsizes]{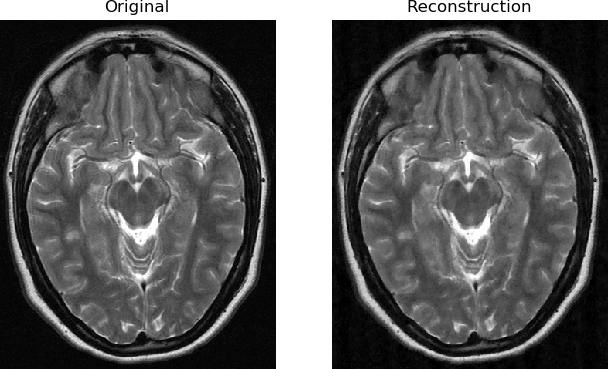}}

\vspace{\vertseps}

\parbox{\imsizes}{\includegraphics[width=\imsizes]{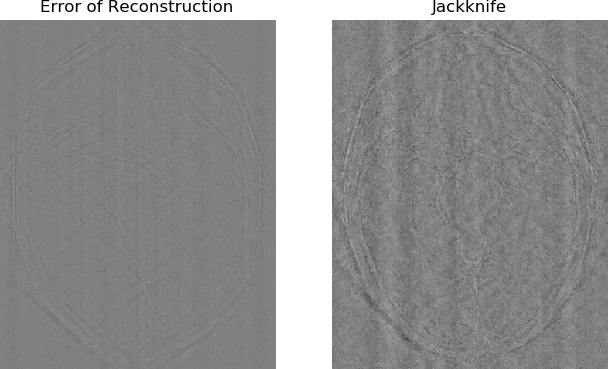}}
\hfill
\parbox{\imsizes}{\includegraphics[width=\imsizes]{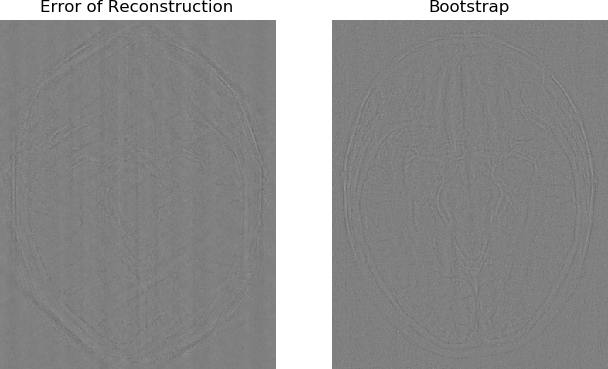}}

\end{centering}
\caption{$2\times$ radially retained sampling --- slice 9}
\end{figure}

\begin{figure}
\begin{centering}

\parbox{\imsizes}{\includegraphics[width=\imsizes]{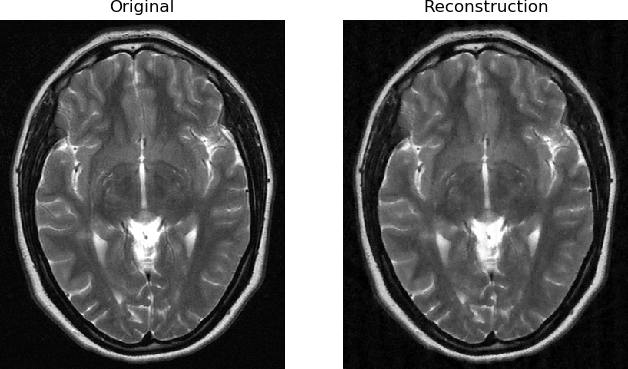}}

\vspace{\vertseps}

\parbox{\imsizes}{\includegraphics[width=\imsizes]{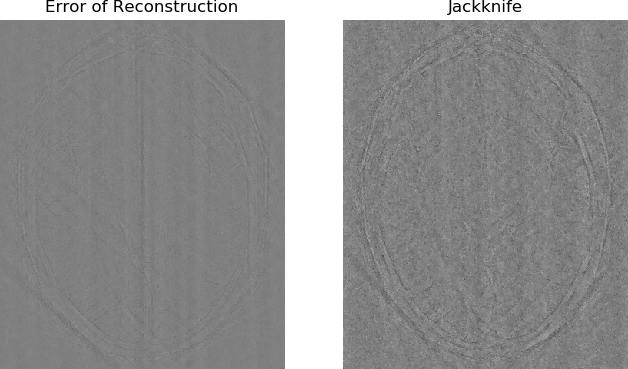}}
\hfill
\parbox{\imsizes}{\includegraphics[width=\imsizes]{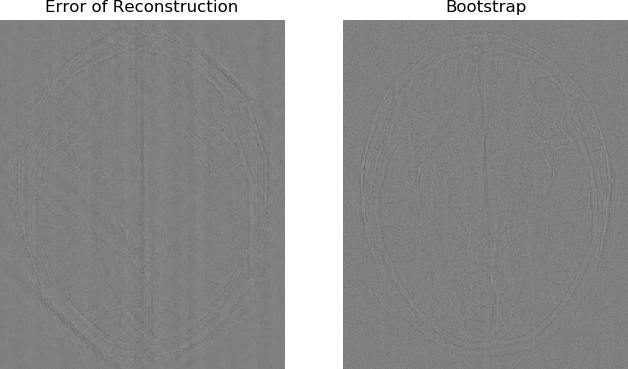}}

\end{centering}
\caption{$2\times$ radially retained sampling --- slice 10}
\end{figure}

\begin{figure}
\begin{centering}

\parbox{\imsizes}{\includegraphics[width=\imsizes]{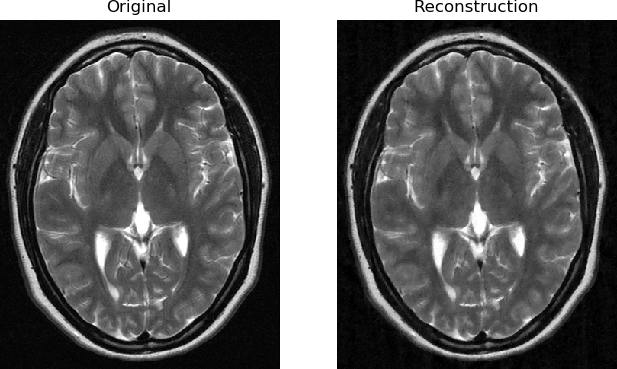}}

\vspace{\vertseps}

\parbox{\imsizes}{\includegraphics[width=\imsizes]{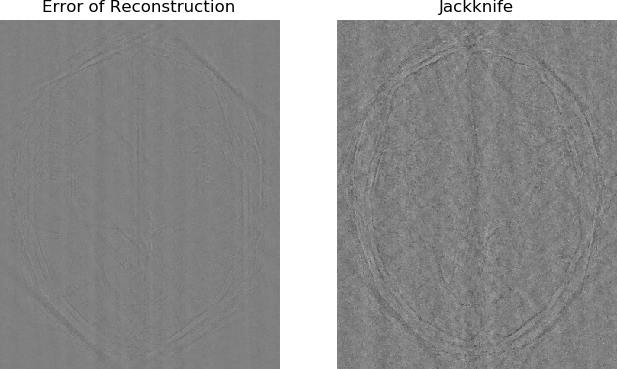}}
\hfill
\parbox{\imsizes}{\includegraphics[width=\imsizes]{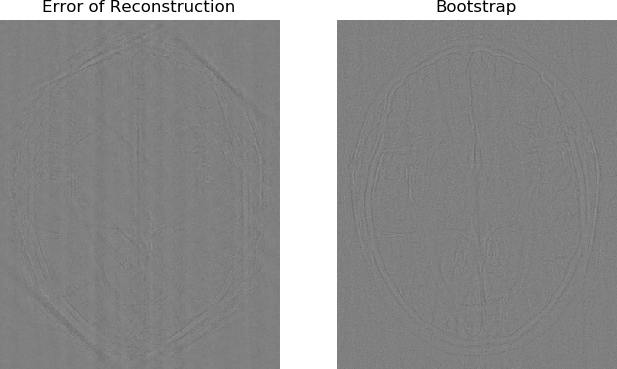}}

\end{centering}
\caption{$2\times$ radially retained sampling --- slice 11}
\end{figure}

\begin{figure}
\begin{centering}

\parbox{\imsizes}{\includegraphics[width=\imsizes]{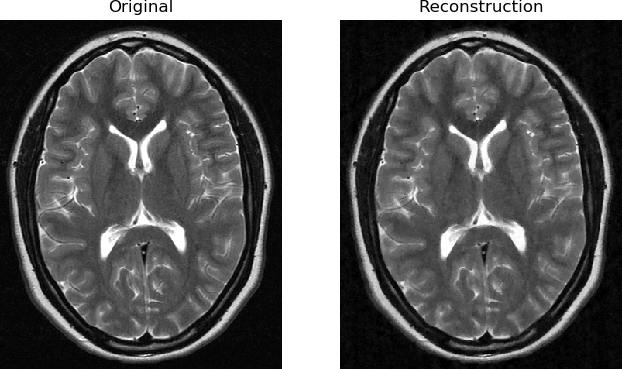}}

\vspace{\vertseps}

\parbox{\imsizes}{\includegraphics[width=\imsizes]{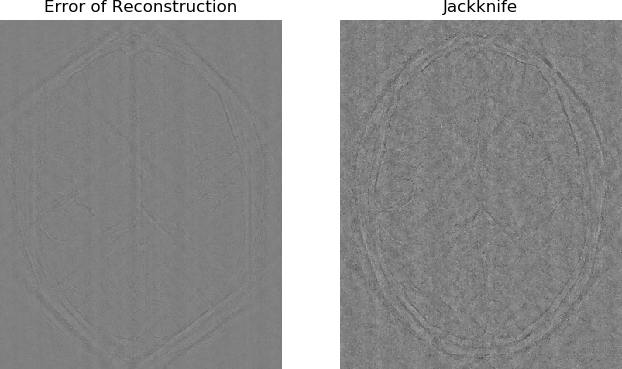}}
\hfill
\parbox{\imsizes}{\includegraphics[width=\imsizes]{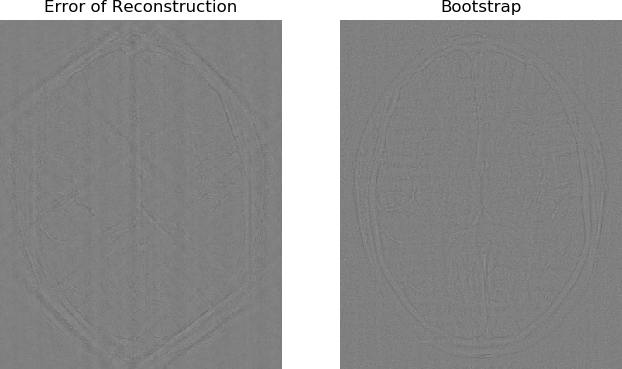}}

\end{centering}
\caption{$2\times$ radially retained sampling --- slice 12}
\end{figure}

\begin{figure}
\begin{centering}

\parbox{\imsizes}{\includegraphics[width=\imsizes]{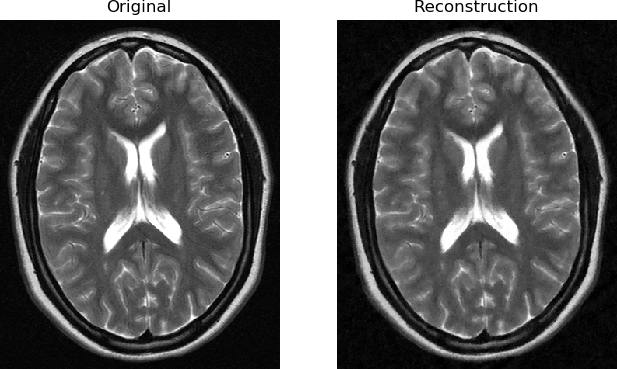}}

\vspace{\vertseps}

\parbox{\imsizes}{\includegraphics[width=\imsizes]{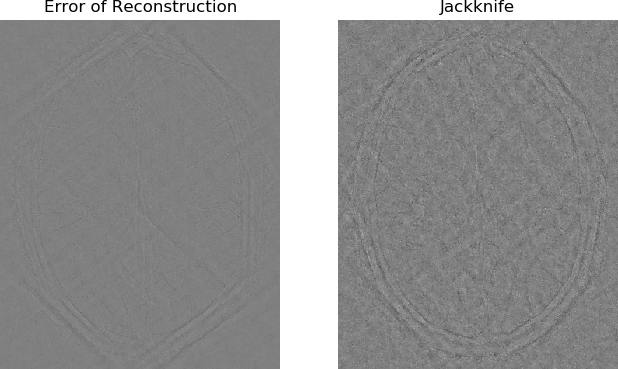}}
\hfill
\parbox{\imsizes}{\includegraphics[width=\imsizes]{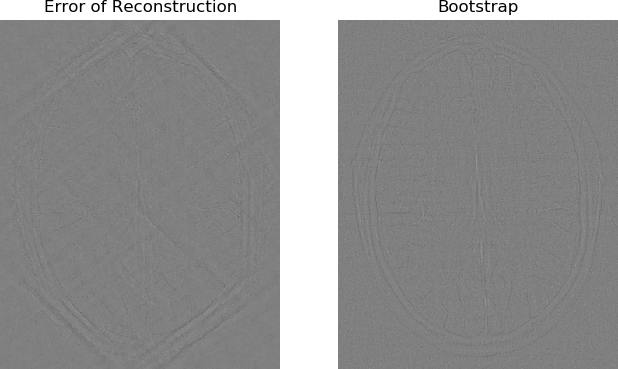}}

\end{centering}
\caption{$2\times$ radially retained sampling --- slice 13}
\end{figure}

\begin{figure}
\begin{centering}

\parbox{\imsizes}{\includegraphics[width=\imsizes]{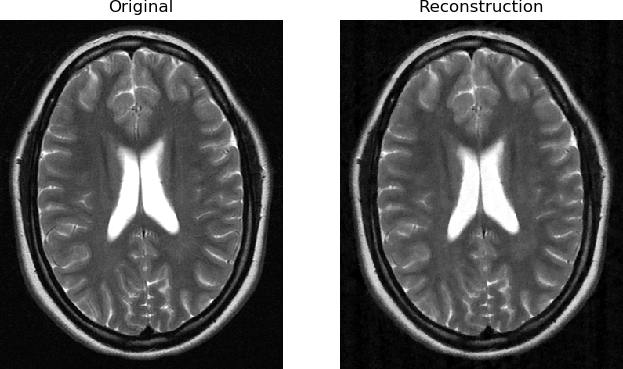}}

\vspace{\vertseps}

\parbox{\imsizes}{\includegraphics[width=\imsizes]{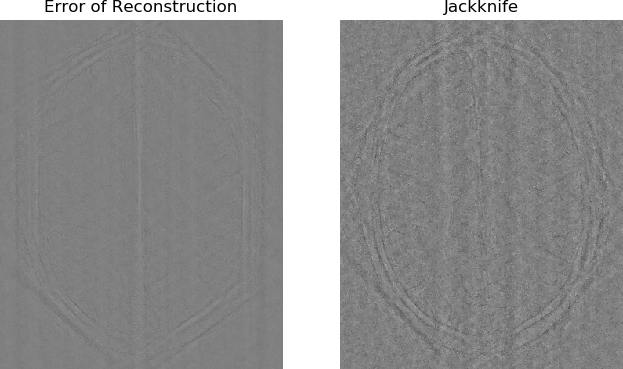}}
\hfill
\parbox{\imsizes}{\includegraphics[width=\imsizes]{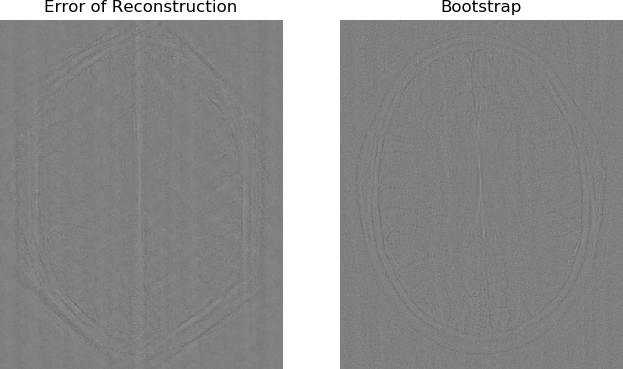}}

\end{centering}
\caption{$2\times$ radially retained sampling --- slice 14}
\end{figure}

\begin{figure}
\begin{centering}

\parbox{\imsizes}{\includegraphics[width=\imsizes]{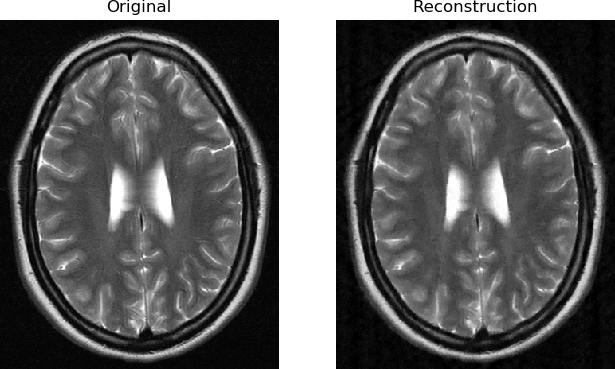}}

\vspace{\vertseps}

\parbox{\imsizes}{\includegraphics[width=\imsizes]{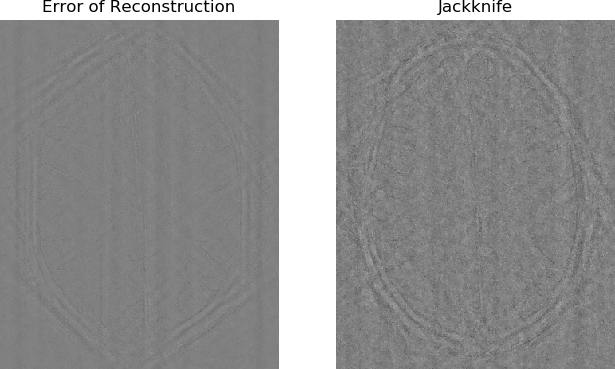}}
\hfill
\parbox{\imsizes}{\includegraphics[width=\imsizes]{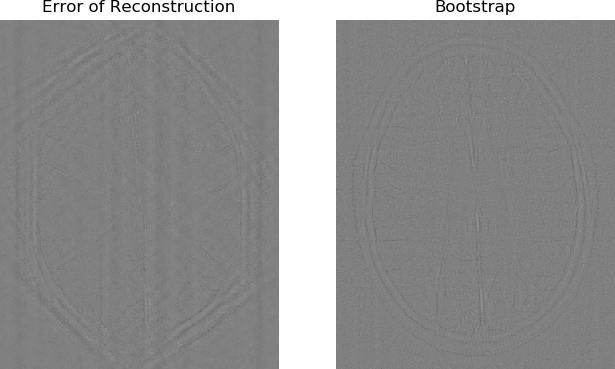}}

\end{centering}
\caption{$2\times$ radially retained sampling --- slice 15}
\end{figure}

\begin{figure}
\begin{centering}

\parbox{\imsizes}{\includegraphics[width=\imsizes]{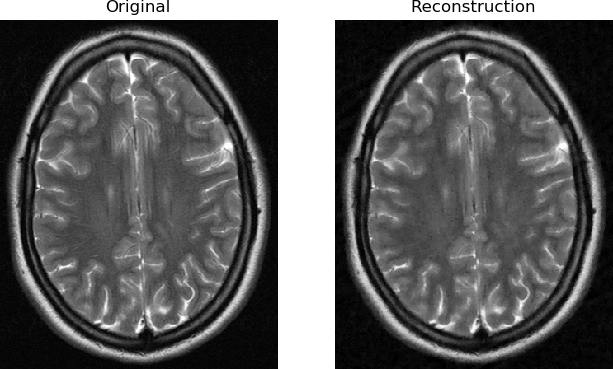}}

\vspace{\vertseps}

\parbox{\imsizes}{\includegraphics[width=\imsizes]{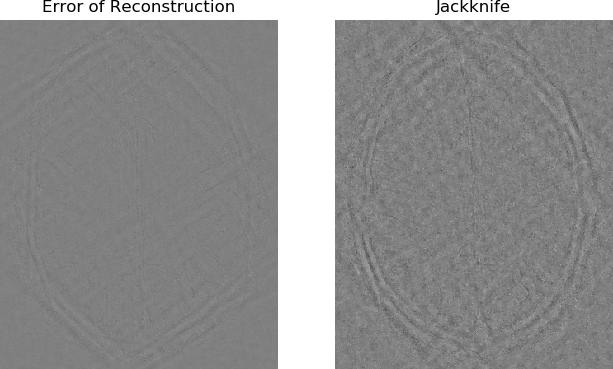}}
\hfill
\parbox{\imsizes}{\includegraphics[width=\imsizes]{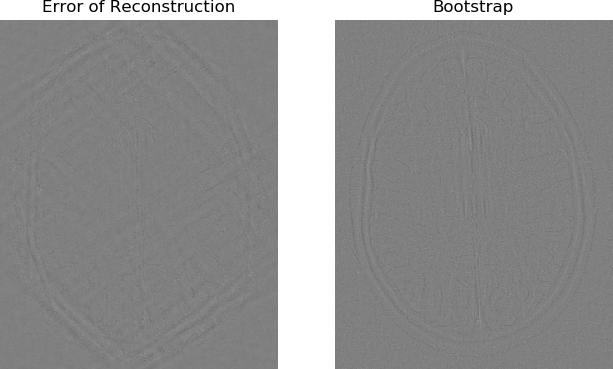}}

\end{centering}
\caption{$2\times$ radially retained sampling --- slice 16}
\end{figure}

\begin{figure}
\begin{centering}

\parbox{\imsizes}{\includegraphics[width=\imsizes]{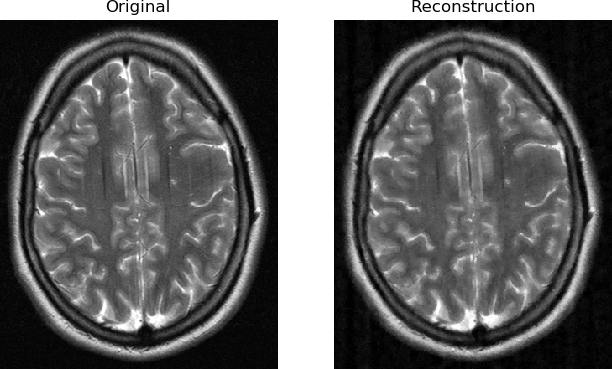}}

\vspace{\vertseps}

\parbox{\imsizes}{\includegraphics[width=\imsizes]{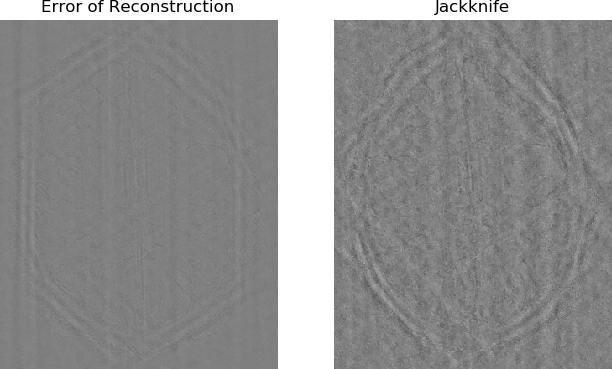}}
\hfill
\parbox{\imsizes}{\includegraphics[width=\imsizes]{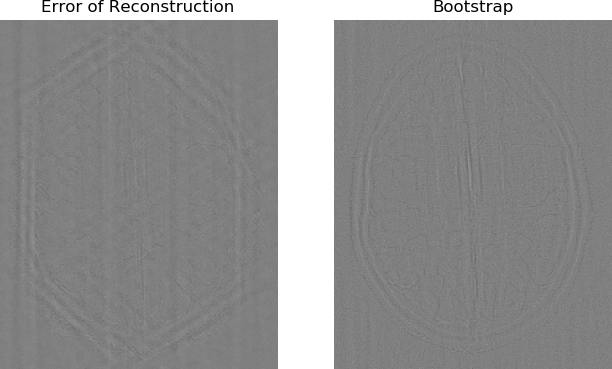}}

\end{centering}
\caption{$2\times$ radially retained sampling --- slice 17}
\end{figure}

\begin{figure}
\begin{centering}

\parbox{\imsizes}{\includegraphics[width=\imsizes]{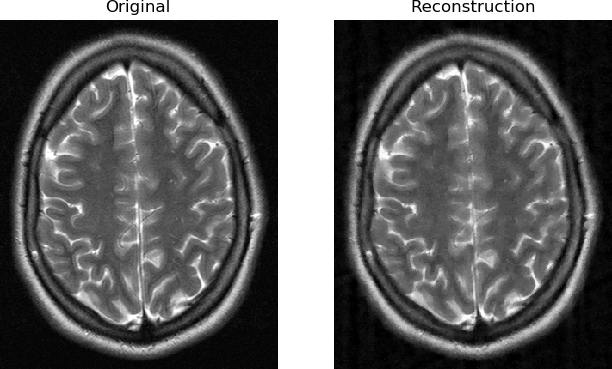}}

\vspace{\vertseps}

\parbox{\imsizes}{\includegraphics[width=\imsizes]{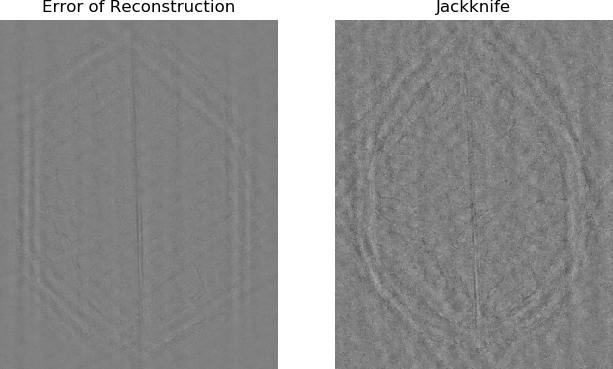}}
\hfill
\parbox{\imsizes}{\includegraphics[width=\imsizes]{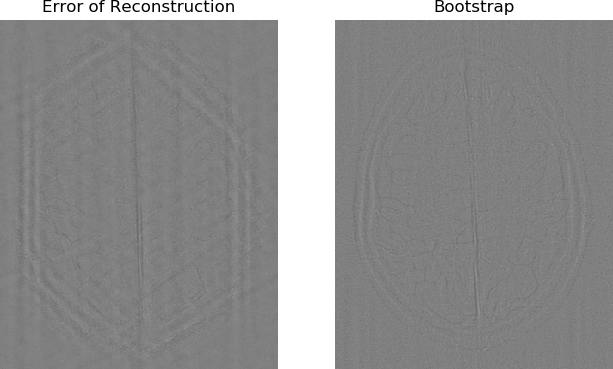}}

\end{centering}
\caption{$2\times$ radially retained sampling --- slice 18}
\end{figure}

\begin{figure}
\begin{centering}

\parbox{\imsizes}{\includegraphics[width=\imsizes]{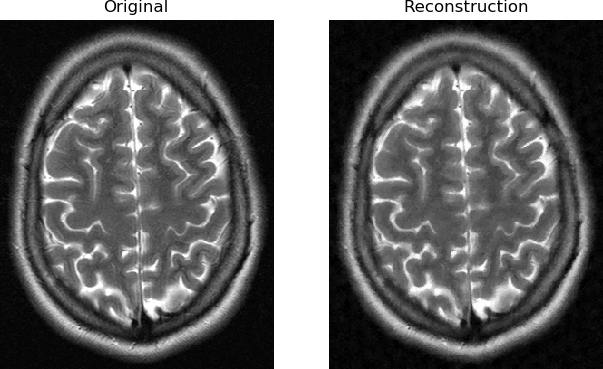}}

\vspace{\vertseps}

\parbox{\imsizes}{\includegraphics[width=\imsizes]{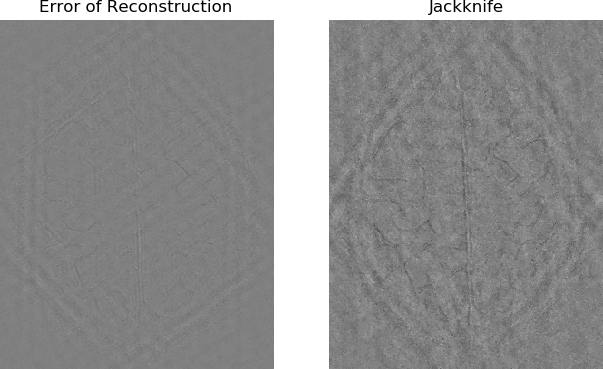}}
\hfill
\parbox{\imsizes}{\includegraphics[width=\imsizes]{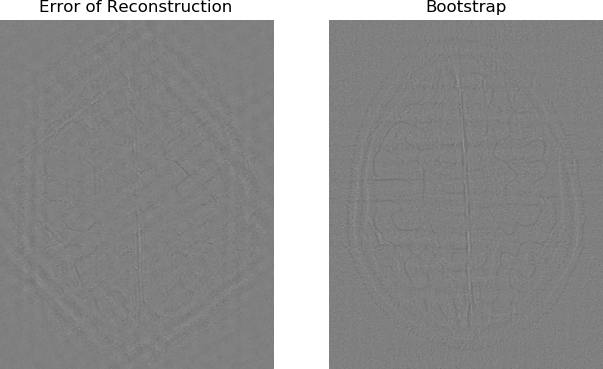}}

\end{centering}
\caption{$2\times$ radially retained sampling --- slice 19}
\end{figure}

\begin{figure}
\begin{centering}

\parbox{\imsizes}{\includegraphics[width=\imsizes]{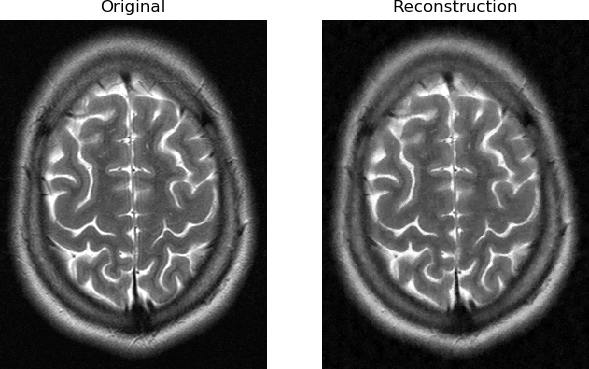}}

\vspace{\vertseps}

\parbox{\imsizes}{\includegraphics[width=\imsizes]{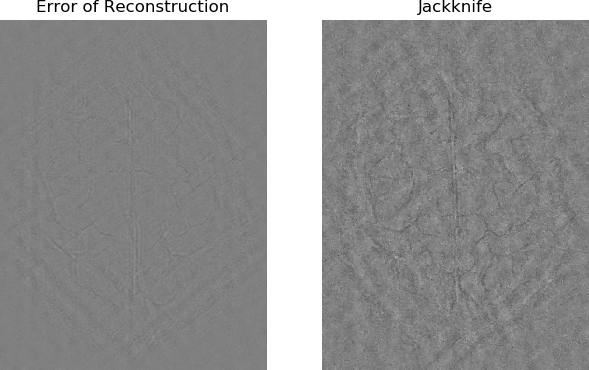}}
\hfill
\parbox{\imsizes}{\includegraphics[width=\imsizes]{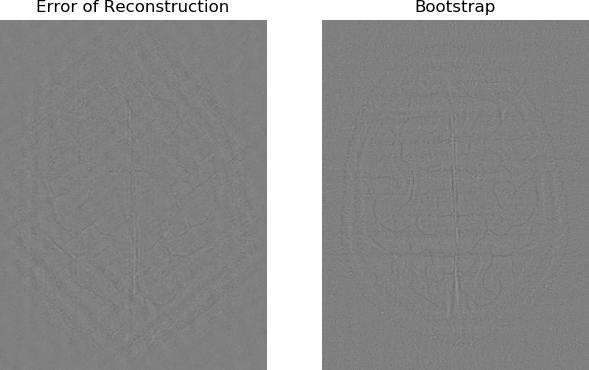}}

\end{centering}
\caption{$2\times$ radially retained sampling --- slice 20}
\end{figure}

\begin{figure}
\begin{centering}

\parbox{\imsizes}{\includegraphics[width=\imsizes]{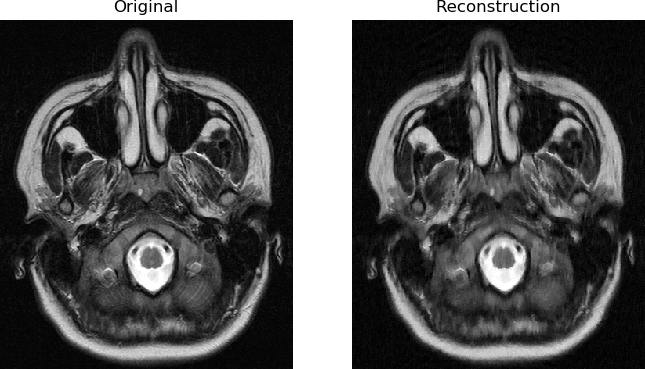}}

\vspace{\vertseps}

\parbox{\imsizes}{\includegraphics[width=\imsizes]{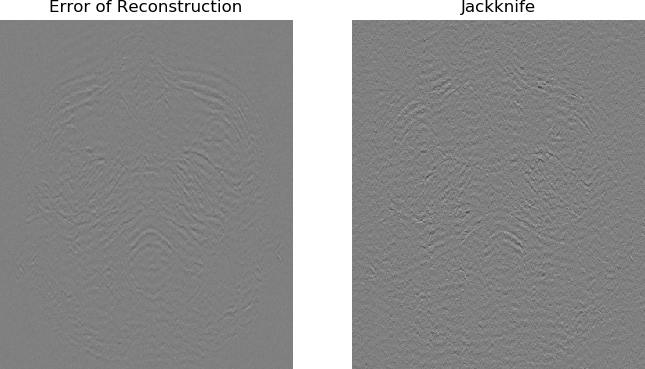}}
\hfill
\parbox{\imsizes}{\includegraphics[width=\imsizes]{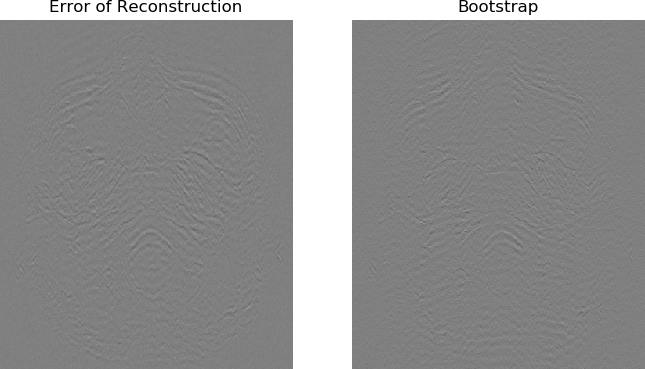}}

\end{centering}
\caption{$2\times$ horizontally retained sampling --- slice 1}
\end{figure}

\begin{figure}
\begin{centering}

\parbox{\imsizes}{\includegraphics[width=\imsizes]{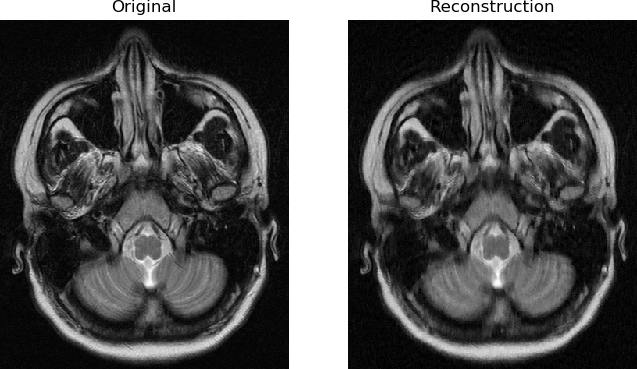}}

\vspace{\vertseps}

\parbox{\imsizes}{\includegraphics[width=\imsizes]{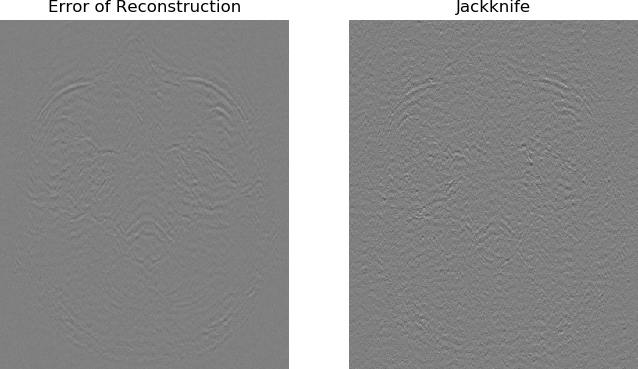}}
\hfill
\parbox{\imsizes}{\includegraphics[width=\imsizes]{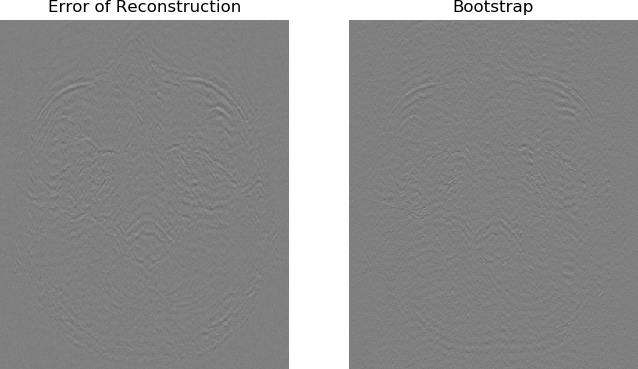}}

\end{centering}
\caption{$2\times$ horizontally retained sampling --- slice 2}
\end{figure}

\begin{figure}
\begin{centering}

\parbox{\imsizes}{\includegraphics[width=\imsizes]{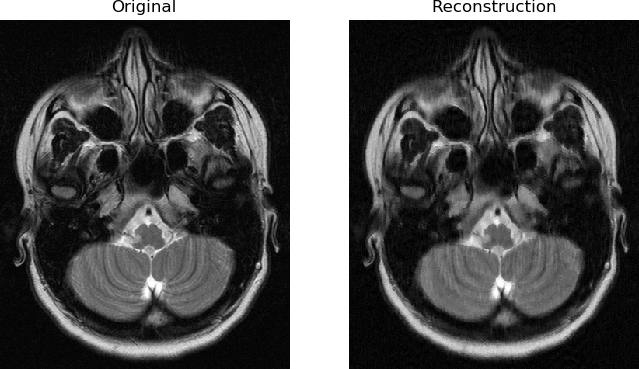}}

\vspace{\vertseps}

\parbox{\imsizes}{\includegraphics[width=\imsizes]{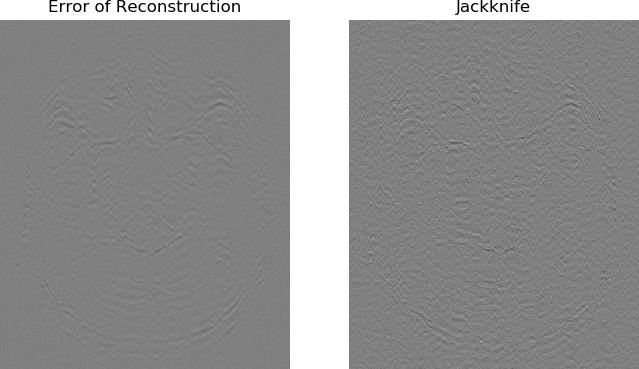}}
\hfill
\parbox{\imsizes}{\includegraphics[width=\imsizes]{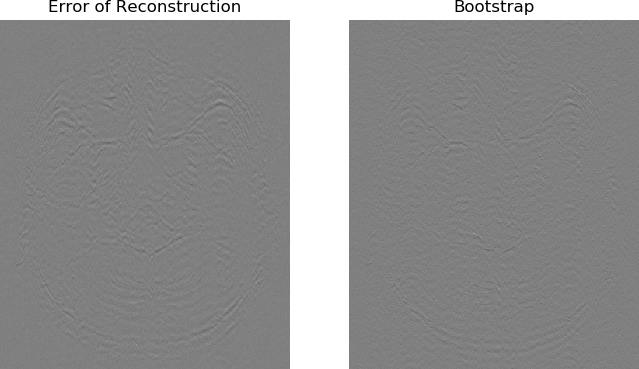}}

\end{centering}
\caption{$2\times$ horizontally retained sampling --- slice 3}
\end{figure}

\begin{figure}
\begin{centering}

\parbox{\imsizes}{\includegraphics[width=\imsizes]{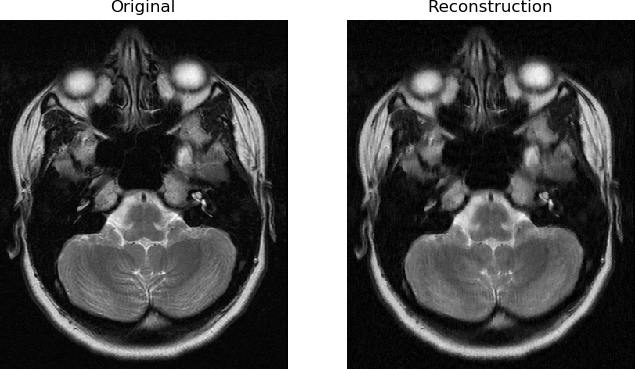}}

\vspace{\vertseps}

\parbox{\imsizes}{\includegraphics[width=\imsizes]{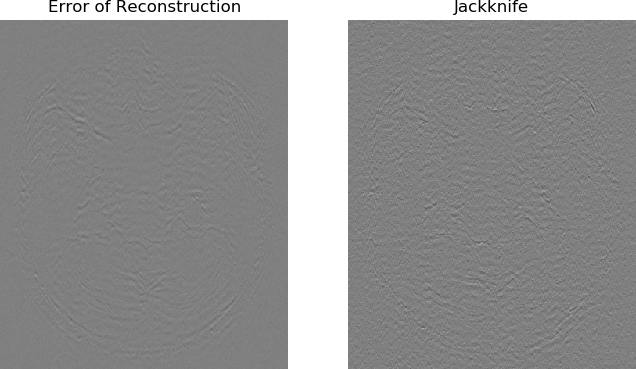}}
\hfill
\parbox{\imsizes}{\includegraphics[width=\imsizes]{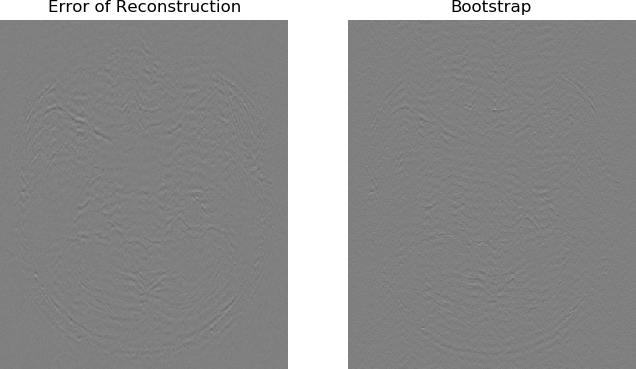}}

\end{centering}
\caption{$2\times$ horizontally retained sampling --- slice 4}
\end{figure}

\begin{figure}
\begin{centering}

\parbox{\imsizes}{\includegraphics[width=\imsizes]{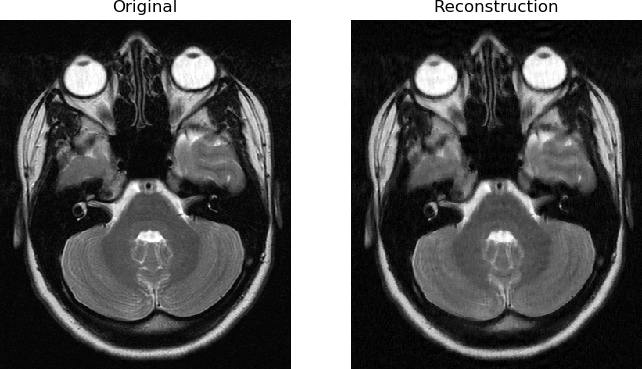}}

\vspace{\vertseps}

\parbox{\imsizes}{\includegraphics[width=\imsizes]{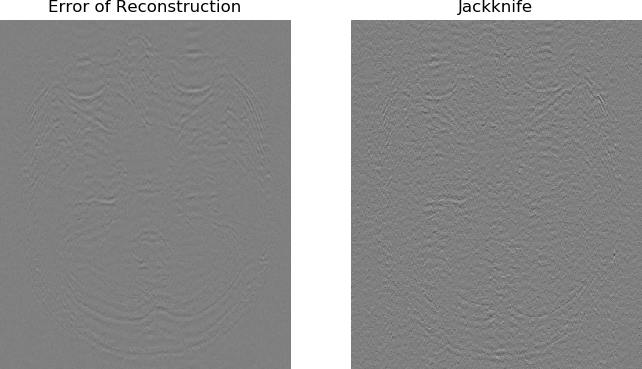}}
\hfill
\parbox{\imsizes}{\includegraphics[width=\imsizes]{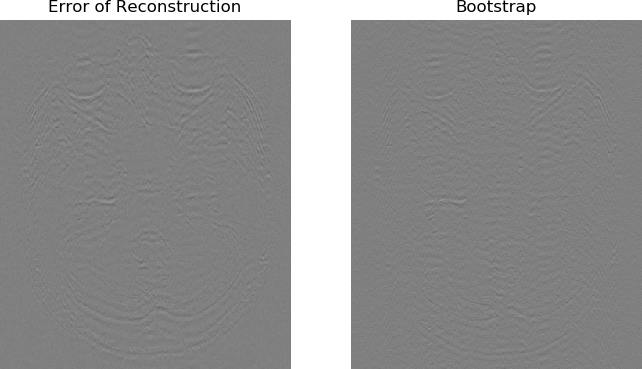}}

\end{centering}
\caption{$2\times$ horizontally retained sampling --- slice 5}
\end{figure}

\begin{figure}
\begin{centering}

\parbox{\imsizes}{\includegraphics[width=\imsizes]{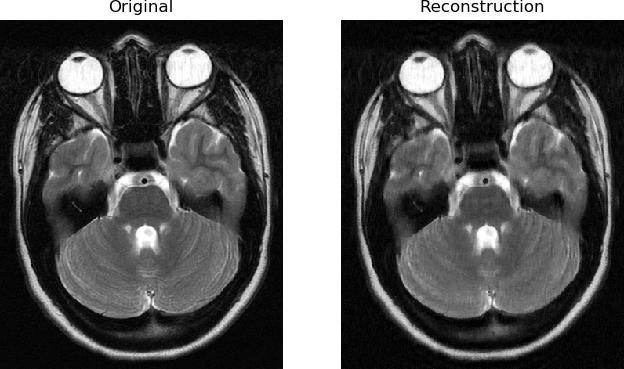}}

\vspace{\vertseps}

\parbox{\imsizes}{\includegraphics[width=\imsizes]{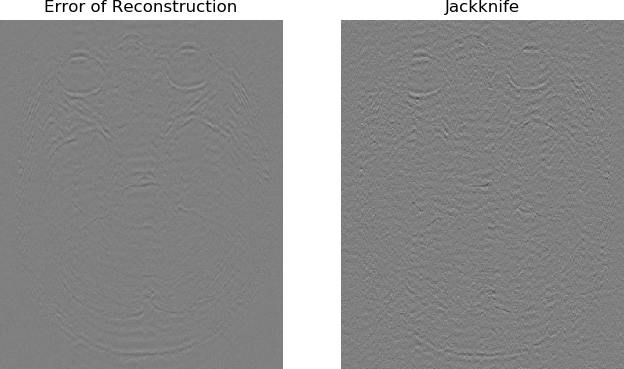}}
\hfill
\parbox{\imsizes}{\includegraphics[width=\imsizes]{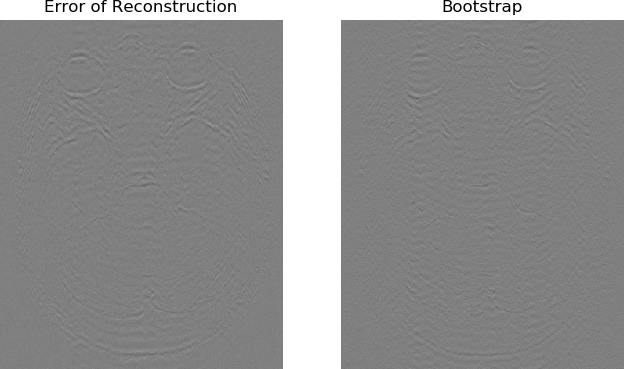}}

\end{centering}
\caption{$2\times$ horizontally retained sampling --- slice 6}
\end{figure}

\begin{figure}
\begin{centering}

\parbox{\imsizes}{\includegraphics[width=\imsizes]{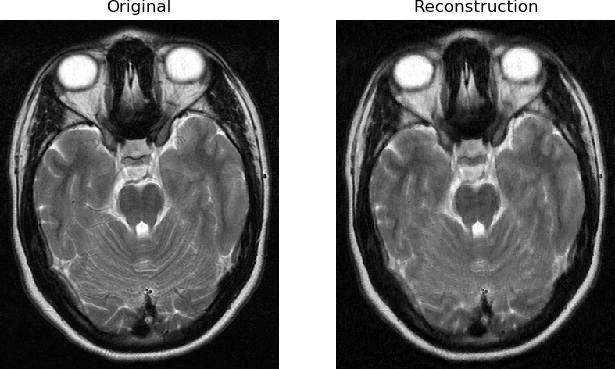}}

\vspace{\vertseps}

\parbox{\imsizes}{\includegraphics[width=\imsizes]{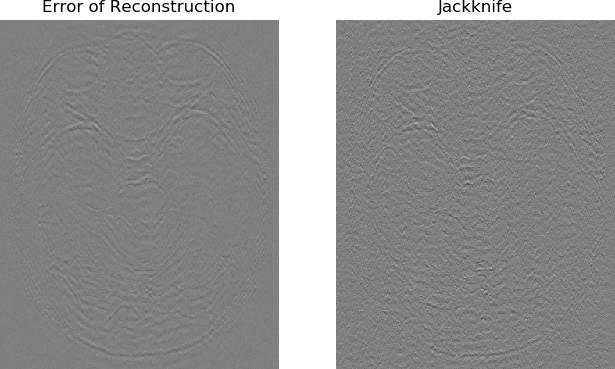}}
\hfill
\parbox{\imsizes}{\includegraphics[width=\imsizes]{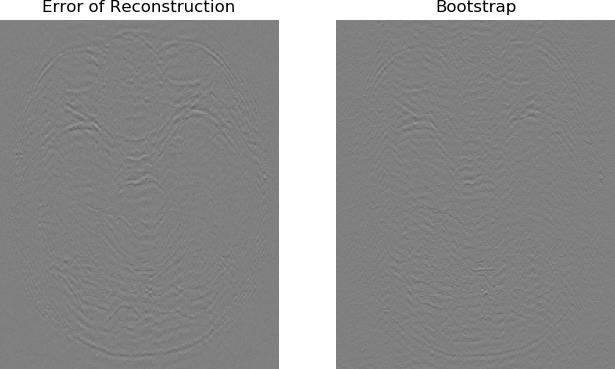}}

\end{centering}
\caption{$2\times$ horizontally retained sampling --- slice 7}
\end{figure}

\begin{figure}
\begin{centering}

\parbox{\imsizes}{\includegraphics[width=\imsizes]{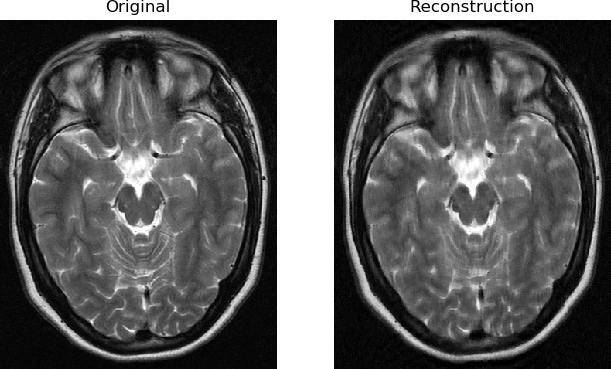}}

\vspace{\vertseps}

\parbox{\imsizes}{\includegraphics[width=\imsizes]{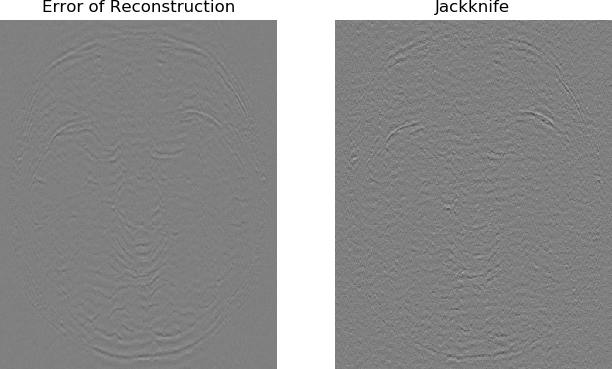}}
\hfill
\parbox{\imsizes}{\includegraphics[width=\imsizes]{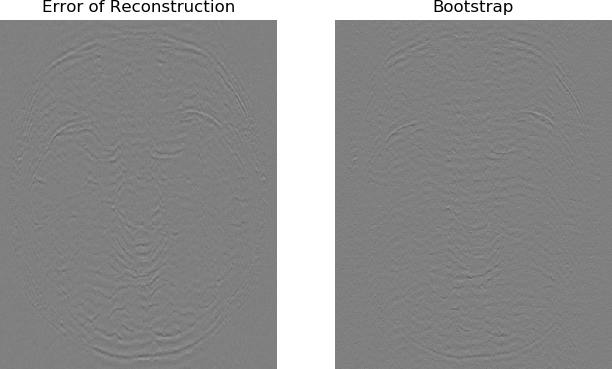}}

\end{centering}
\caption{$2\times$ horizontally retained sampling --- slice 8}
\end{figure}

\begin{figure}
\begin{centering}

\parbox{\imsizes}{\includegraphics[width=\imsizes]{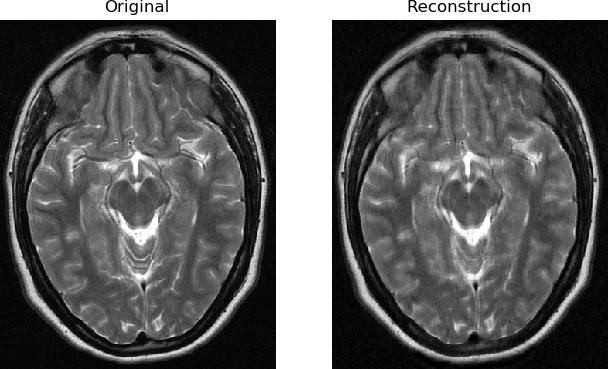}}

\vspace{\vertseps}

\parbox{\imsizes}{\includegraphics[width=\imsizes]{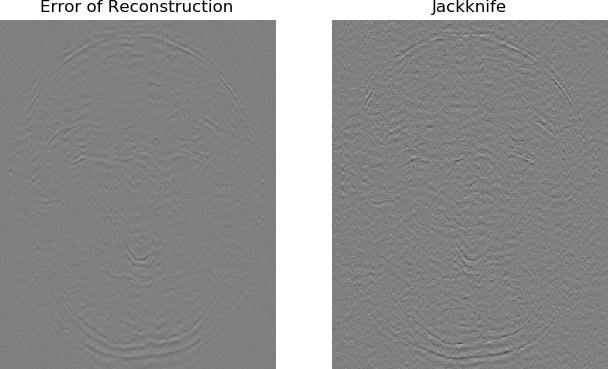}}
\hfill
\parbox{\imsizes}{\includegraphics[width=\imsizes]{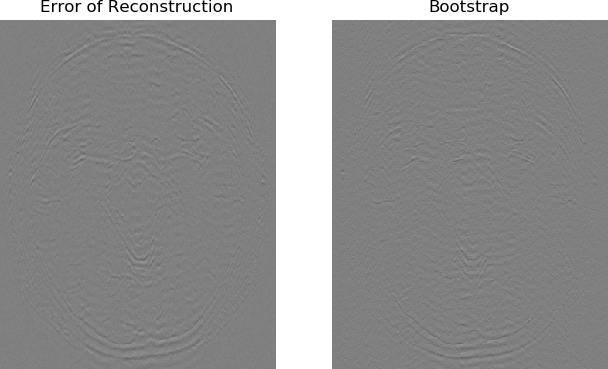}}

\end{centering}
\caption{$2\times$ horizontally retained sampling --- slice 9}
\end{figure}

\begin{figure}
\begin{centering}

\parbox{\imsizes}{\includegraphics[width=\imsizes]{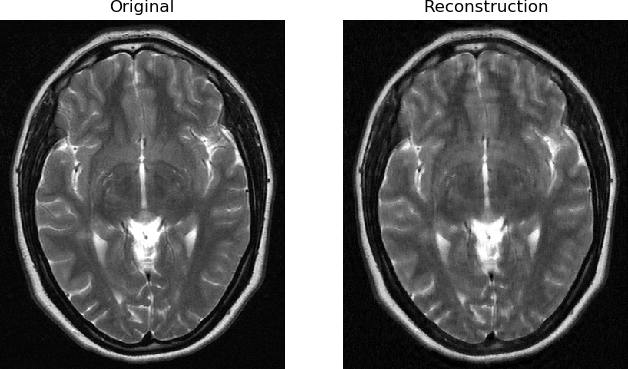}}

\vspace{\vertseps}

\parbox{\imsizes}{\includegraphics[width=\imsizes]{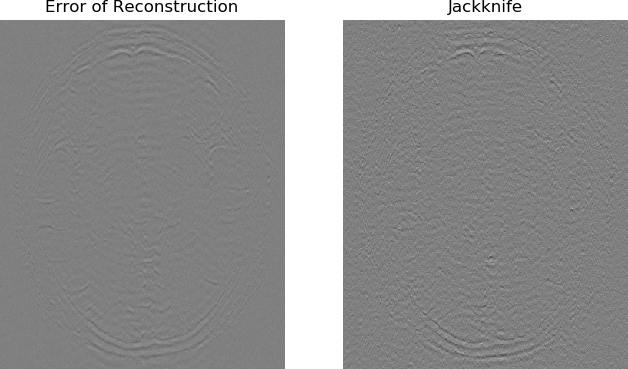}}
\hfill
\parbox{\imsizes}{\includegraphics[width=\imsizes]{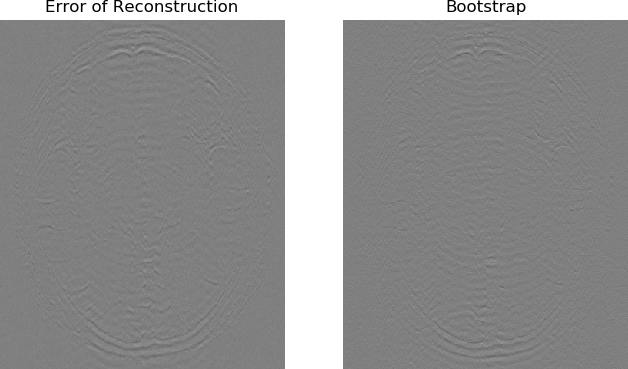}}

\end{centering}
\caption{$2\times$ horizontally retained sampling --- slice 10}
\end{figure}

\begin{figure}
\begin{centering}

\parbox{\imsizes}{\includegraphics[width=\imsizes]{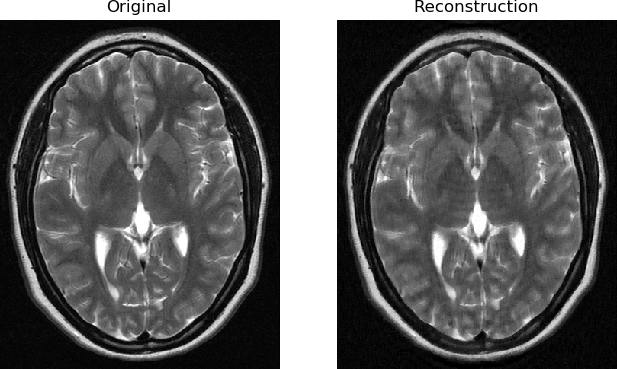}}

\vspace{\vertseps}

\parbox{\imsizes}{\includegraphics[width=\imsizes]{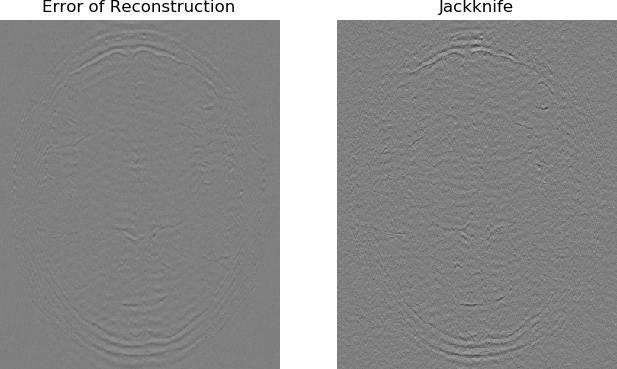}}
\hfill
\parbox{\imsizes}{\includegraphics[width=\imsizes]{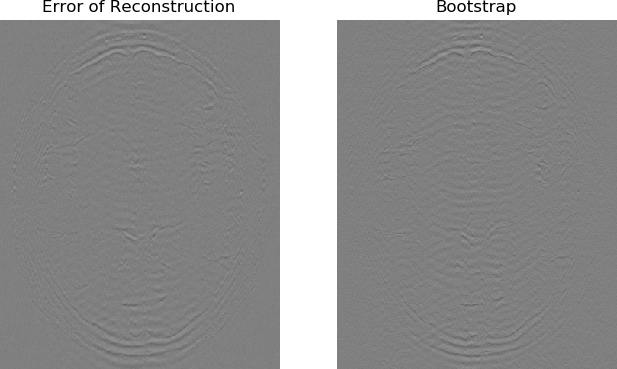}}

\end{centering}
\caption{$2\times$ horizontally retained sampling --- slice 11}
\end{figure}

\begin{figure}
\begin{centering}

\parbox{\imsizes}{\includegraphics[width=\imsizes]{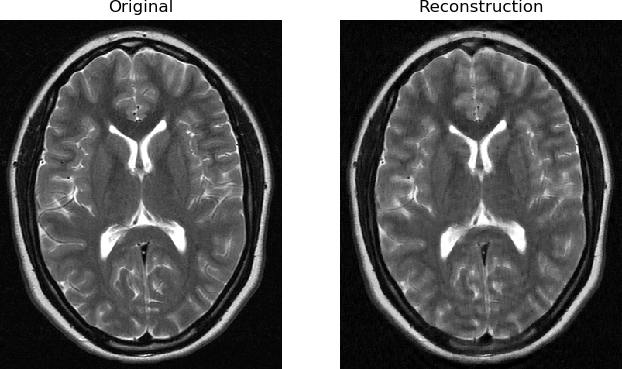}}

\vspace{\vertseps}

\parbox{\imsizes}{\includegraphics[width=\imsizes]{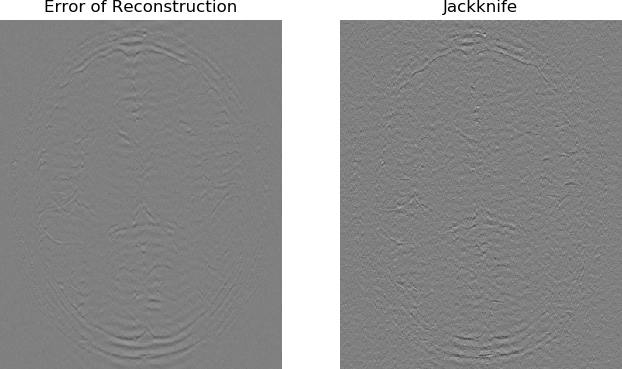}}
\hfill
\parbox{\imsizes}{\includegraphics[width=\imsizes]{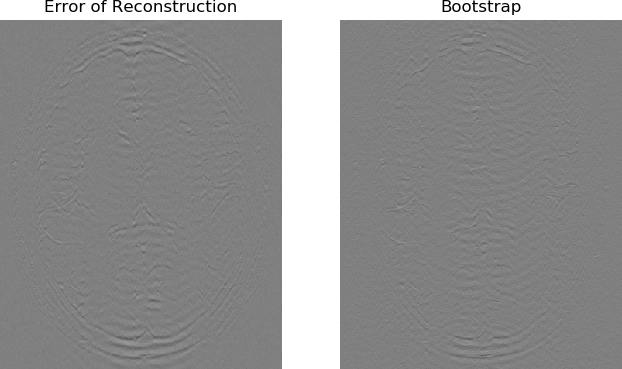}}

\end{centering}
\caption{$2\times$ horizontally retained sampling --- slice 12}
\end{figure}

\begin{figure}
\begin{centering}

\parbox{\imsizes}{\includegraphics[width=\imsizes]{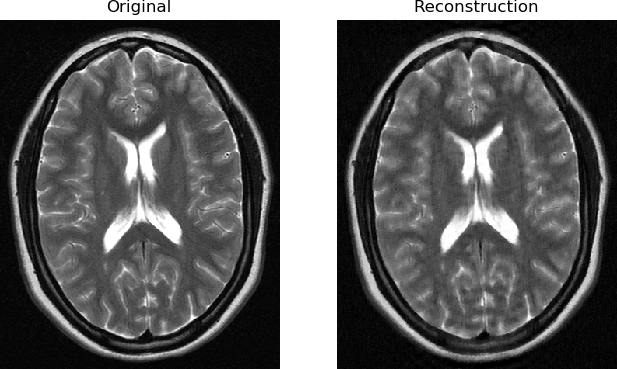}}

\vspace{\vertseps}

\parbox{\imsizes}{\includegraphics[width=\imsizes]{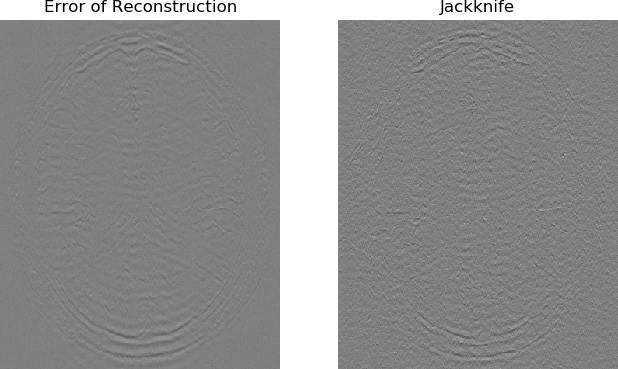}}
\hfill
\parbox{\imsizes}{\includegraphics[width=\imsizes]{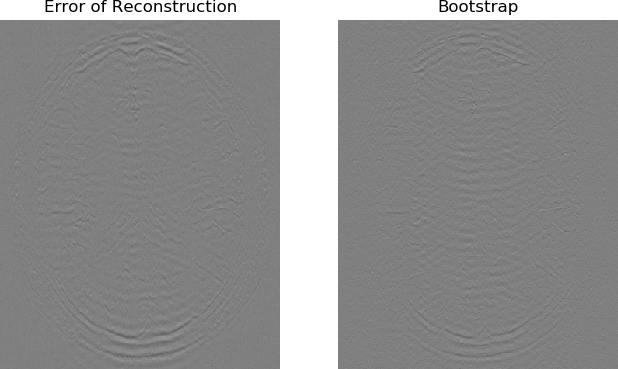}}

\end{centering}
\caption{$2\times$ horizontally retained sampling --- slice 13}
\end{figure}

\begin{figure}
\begin{centering}

\parbox{\imsizes}{\includegraphics[width=\imsizes]{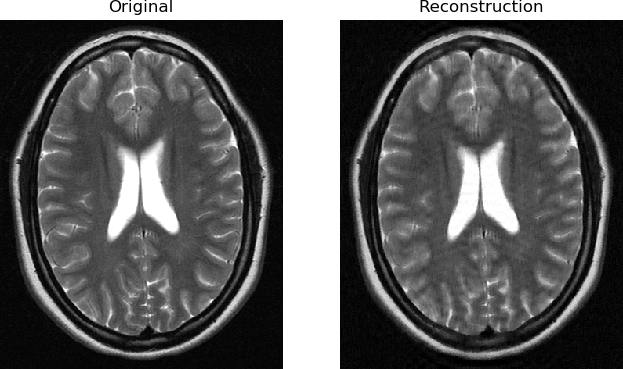}}

\vspace{\vertseps}

\parbox{\imsizes}{\includegraphics[width=\imsizes]{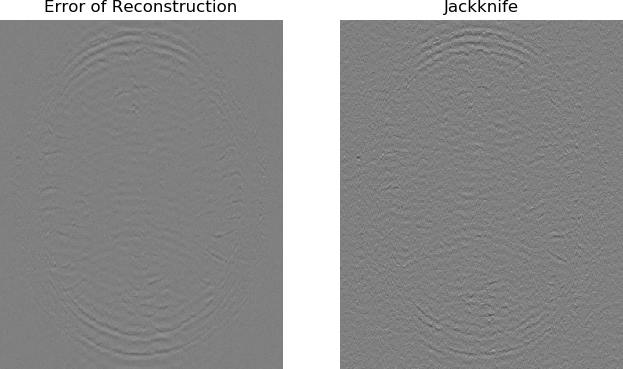}}
\hfill
\parbox{\imsizes}{\includegraphics[width=\imsizes]{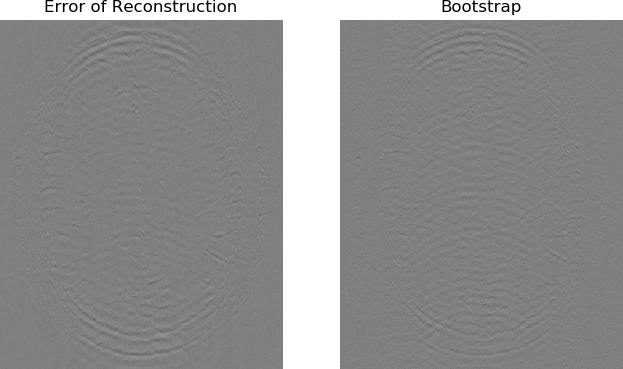}}

\end{centering}
\caption{$2\times$ horizontally retained sampling --- slice 14}
\end{figure}

\begin{figure}
\begin{centering}

\parbox{\imsizes}{\includegraphics[width=\imsizes]{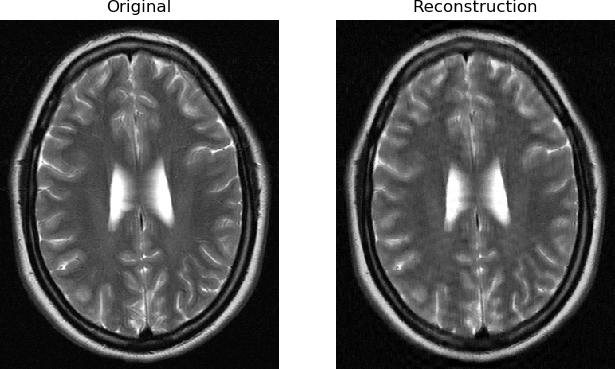}}

\vspace{\vertseps}

\parbox{\imsizes}{\includegraphics[width=\imsizes]{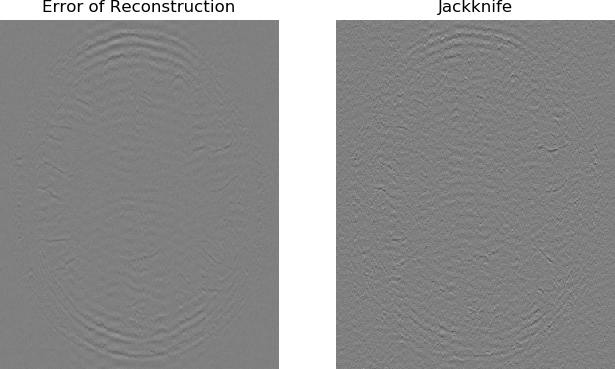}}
\hfill
\parbox{\imsizes}{\includegraphics[width=\imsizes]{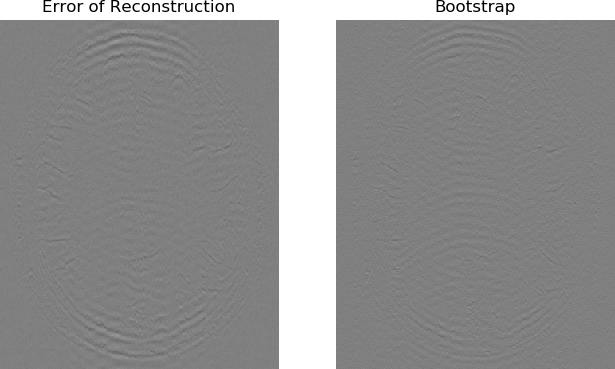}}

\end{centering}
\caption{$2\times$ horizontally retained sampling --- slice 15}
\end{figure}

\begin{figure}
\begin{centering}

\parbox{\imsizes}{\includegraphics[width=\imsizes]{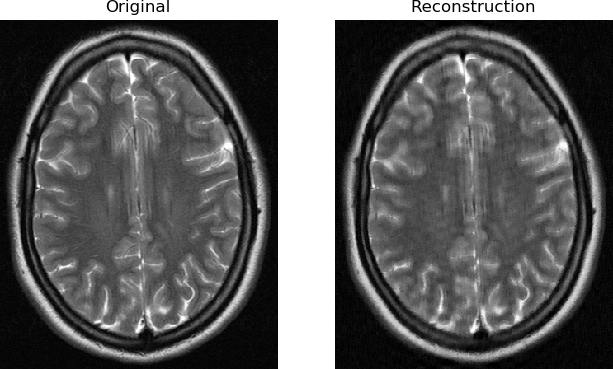}}

\vspace{\vertseps}

\parbox{\imsizes}{\includegraphics[width=\imsizes]{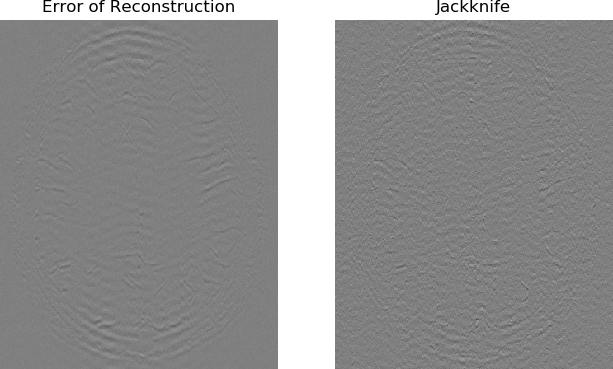}}
\hfill
\parbox{\imsizes}{\includegraphics[width=\imsizes]{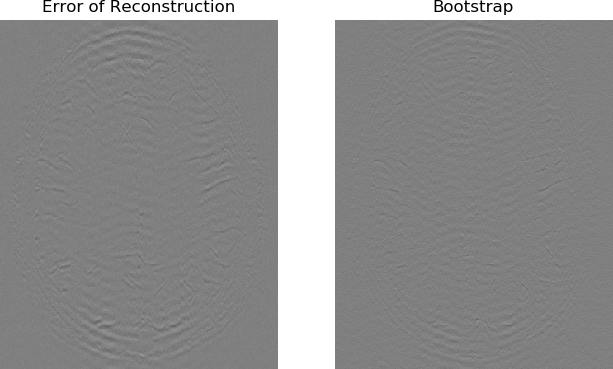}}

\end{centering}
\caption{$2\times$ horizontally retained sampling --- slice 16}
\end{figure}

\begin{figure}
\begin{centering}

\parbox{\imsizes}{\includegraphics[width=\imsizes]{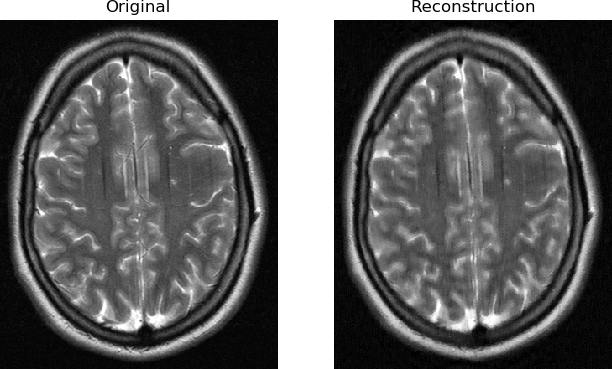}}

\vspace{\vertseps}

\parbox{\imsizes}{\includegraphics[width=\imsizes]{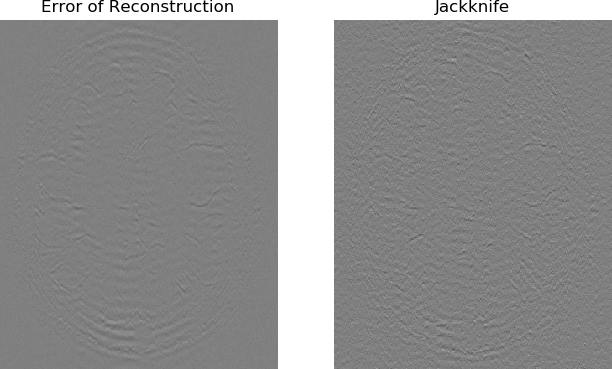}}
\hfill
\parbox{\imsizes}{\includegraphics[width=\imsizes]{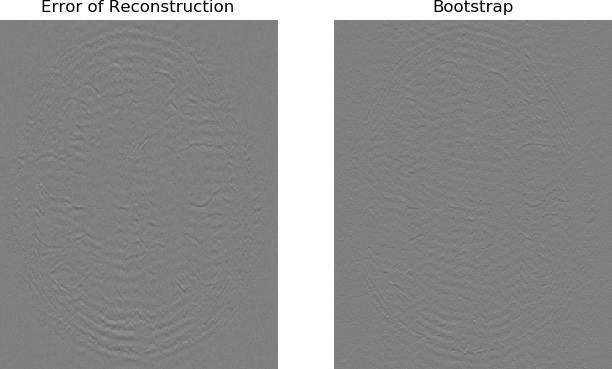}}

\end{centering}
\caption{$2\times$ horizontally retained sampling --- slice 17}
\end{figure}

\begin{figure}
\begin{centering}

\parbox{\imsizes}{\includegraphics[width=\imsizes]{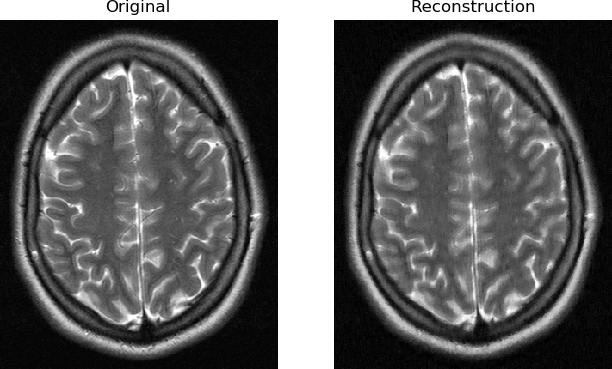}}

\vspace{\vertseps}

\parbox{\imsizes}{\includegraphics[width=\imsizes]{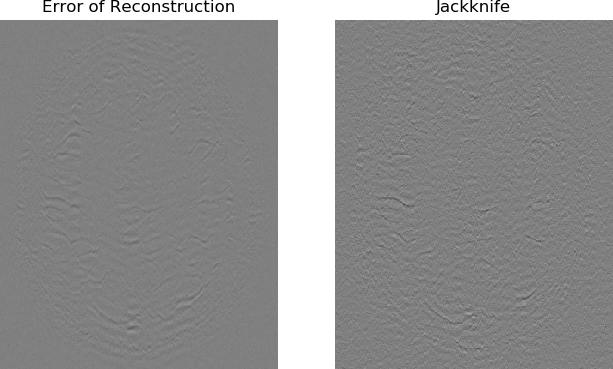}}
\hfill
\parbox{\imsizes}{\includegraphics[width=\imsizes]{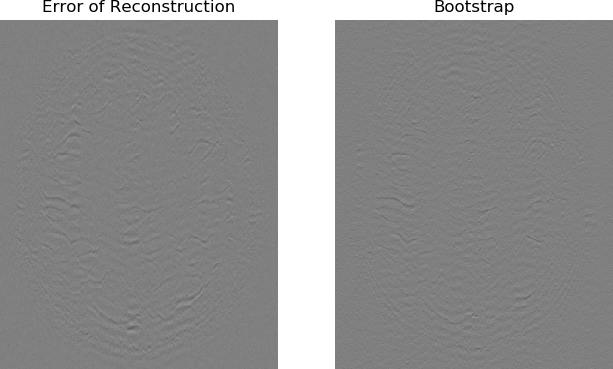}}

\end{centering}
\caption{$2\times$ horizontally retained sampling --- slice 18}
\end{figure}

\begin{figure}
\begin{centering}

\parbox{\imsizes}{\includegraphics[width=\imsizes]{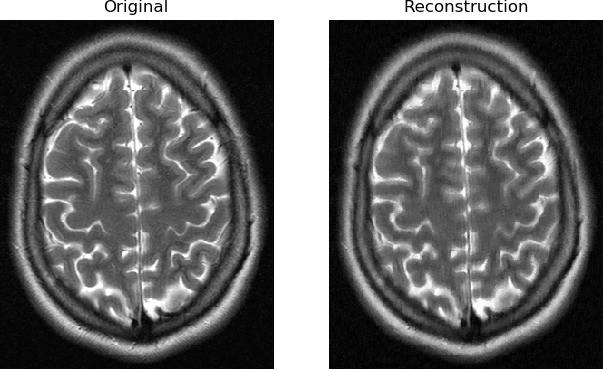}}

\vspace{\vertseps}

\parbox{\imsizes}{\includegraphics[width=\imsizes]{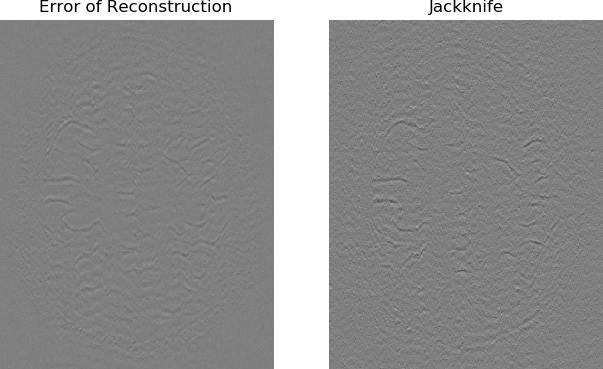}}
\hfill
\parbox{\imsizes}{\includegraphics[width=\imsizes]{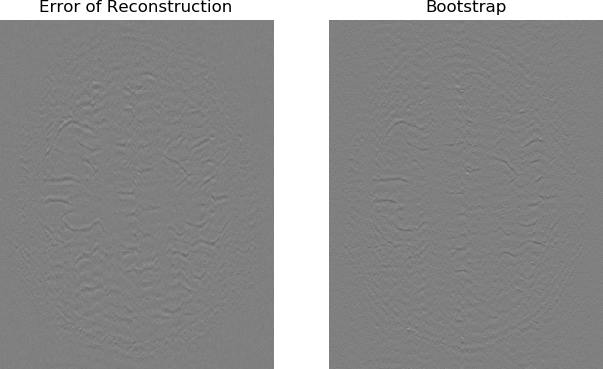}}

\end{centering}
\caption{$2\times$ horizontally retained sampling --- slice 19}
\end{figure}

\begin{figure}
\begin{centering}

\parbox{\imsizes}{\includegraphics[width=\imsizes]{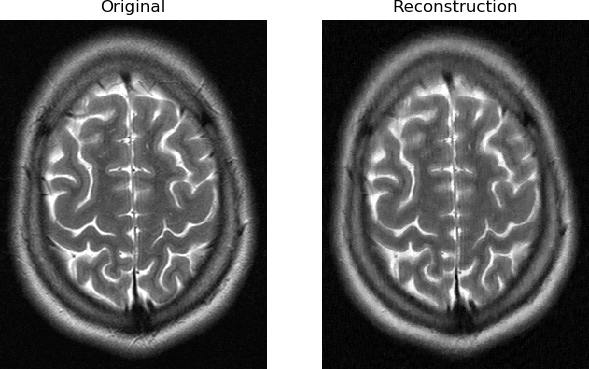}}

\vspace{\vertseps}

\parbox{\imsizes}{\includegraphics[width=\imsizes]{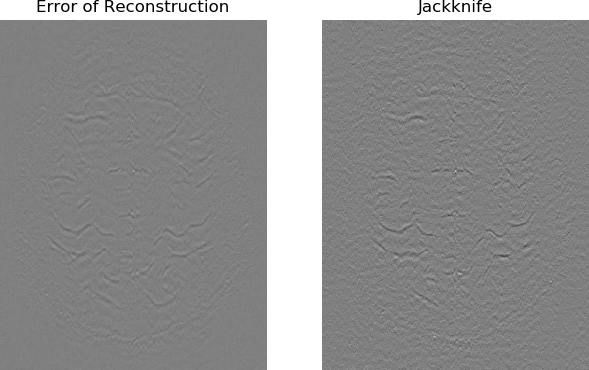}}
\hfill
\parbox{\imsizes}{\includegraphics[width=\imsizes]{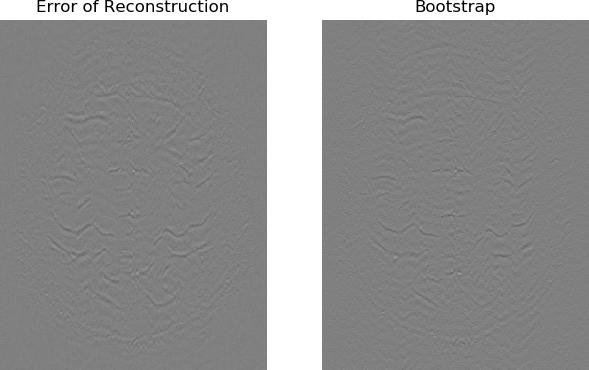}}

\end{centering}
\caption{$2\times$ horizontally retained sampling --- slice 20}
\end{figure}

\clearpage

\bibliography{mri}
\bibliographystyle{siam}

\end{document}